\documentclass[]{pasj01}
\usepackage{url}
\usepackage{color}

\draft

\Received{}
\Accepted{}
 
\begin{document} 
\title{ 
Revisiting Planetary Systems in Okayama Planet Search Program: A new long-period planet, RV astrometry joint analysis, and multiplicity-metallicity trend around evolved stars.}

\author{
Huan-Yu \textsc{Teng}\altaffilmark{1,*}, 
Bun'ei \textsc{Sato}\altaffilmark{1}, 
Masayuki \textsc{Kuzuhara}\altaffilmark{2}, 
Takuya \textsc{Takarada}\altaffilmark{2}, %
Masashi \textsc{Omiya}\altaffilmark{2,3}, %
Hiroki \textsc{Harakawa}\altaffilmark{4}, 
Hideyuki \textsc{Izumiura}\altaffilmark{5}, 
Eiji \textsc{Kambe}\altaffilmark{4}, 
Mesut \textsc{Yilmaz}\altaffilmark{6}, 
Ilfan \textsc{Bikmaev}\altaffilmark{7,8}, 
Selim O. \textsc{Selam}\altaffilmark{6}, 
Timothy D. \textsc{Brandt}\altaffilmark{9}, 
Guang-Yao \textsc{Xiao}\altaffilmark{10, 11}, 
Michitoshi \textsc{Yoshida}\altaffilmark{4}, %
Yoichi \textsc{Itoh}\altaffilmark{12}, 
Hiroyasu \textsc{Ando}\altaffilmark{3}, 
Eiichiro \textsc{Kokubo}\altaffilmark{3}, and 
Shigeru \textsc{Ida}\altaffilmark{13} 
}

\altaffiltext{1}{Department of Earth and Planetary Sciences, School of Science, Tokyo Institute of Technology, 2-12-1 Ookayama, Meguro-ku, Tokyo 152-8550, Japan}
\altaffiltext{2}{Astrobiology Center, National Institutes of Natural Sciences, 2-21-1 Osawa, Mitaka, Tokyo 181-8588, Japan}
\altaffiltext{3}{National Astronomical Observatory of Japan, National Institutes of Natural Sciences, 2-21-1 Osawa, Mitaka, Tokyo 181-8588, Japan}
\altaffiltext{4}{Subaru Telescope, National Astronomical Observatory of Japan, National Institutes of Natural Sciences, 650 North A’ohoku Pl., Hilo, HI, 96720, USA}
\altaffiltext{5}{Okayama Branch Office, Subaru Telescope, National Astronomical Observatory of Japan, National Institutes of Natural Sciences, Kamogata, Asakuchi, Okayama 719-0232, Japan}
\altaffiltext{6}{Ankara University, Dept. of Astronomy and Space Sciences, TR-06100 Besevler/Ankara, T\"URK\.IYE}
\altaffiltext{7}{Kazan Federal University, Kremlyevskaya Str. 18, Kazan, 420008, Russia} 
\altaffiltext{8}{Academy of Sciences of Tatarstan, Bauman Str. 20,  Kazan, 420111, Russia}
\altaffiltext{9}{Department of Physics, University of California, Santa Barbara, Santa Barbara, CA 93106, USA}
\altaffiltext{10}{CAS Key Laboratory of Optical Astronomy, National Astronomical Observatories, Chinese Academy of Sciences, Beijing 100101, China} 
\altaffiltext{11}{Department of Physics, College of Science, Tibet University, Lhasa 850000, China}
\altaffiltext{12}{Nishi-Harima Astronomical Observatory, Center for Astronomy, University of Hyogo, 407-2, Nishigaichi, Sayo, Hyogo 679-5313, Japan}
\altaffiltext{13}{Earth-Life Science Institute, Tokyo Institute of Technology, 2-12-1 Ookayama, Meguro-ku, Tokyo 152-8550, Japan}

\email{teng.h.aa@m.titech.ac.jp}


\maketitle
\begin{abstract}
In this study, we revisit 32 planetary systems around evolved stars observed within the framework of the Okayama Planet Search Program (OPSP) and its collaborative framework of the East Asian Planet Search Network (EAPS-Net) to search for additional companions and investigate the properties of stars and giant planets in multiple-planet systems.
With our latest radial velocities obtained from Okayama Astrophysical Observatory (OAO), we confirm an additional giant planet in the wide orbit of 75 Cet system ($P_{\rm{c}} = 2051.62_{-40.47}^{+45.98}\ \rm{d}$, $M_{\rm{c}}\sin i=0.912_{-0.090}^{+0.088}\ M_{\rm{J}}$, and $a_{\rm{c}}=3.929_{-0.058}^{+0.052}\ \rm{au}$), along with five stars exhibiting long-term radial velocity accelerations, which indicates massive companions in the wide orbits.
{We have also found that the radial velocity variations of several planet-harboring stars may indicate additional planet candidates, stellar activities, or other understudied sources. These stars include $\epsilon$ Tau, 11 Com, 24 Boo, 41 Lyn, 14 And, HD 32518, and $\omega$ Ser. }
We further constrain the orbital configuration of the HD 5608, HD 14067, HD 120084, and HD 175679 systems by combining radial velocities with astrometry, as their host central stars exhibit significant astrometric accelerations.
For other systems, we simply refine their orbital parameters.
{Moreover, our study indicates that the OPSP planet-harboring stars are more metal-poor compared to the currently known planet harboring stars, and this is likely due to the $B-V$ color upper limit at 1.0 for star selection in the beginning of the survey.}
{Finally, by investigating the less-massive giant planets ($< 5 M_{\rm{J}}$) around currently known planet-harboring evolved stars, we have found that metallicity positively correlates with the multiplicity and total planet mass of the system, which can be evidence for the core-accretion planet formation model.} 
\end{abstract}

\newpage
\section{Introduction}
So far, over 5000 extrasolar planets (exoplanets) having various orbital properties have been discovered around different types of stars. Most of these planets were known to orbit stars having solar mass or lower mass and living in their main-sequence stage.
Solar twins {($0.7 M_{\odot} \leq M_{\star} < 1.5 M_{\odot}$), e.g, G and K dwarfs}, are the optimal objects to search for ``solar systems'' in terms of giving an answer to the formation of solar systems. 
Low-mass stars {($M_{\star} < 0.7 M_{\odot}$), e.g., the M dwarfs} are the best targets to search for ``Earths'' in habitable zones in terms of terrestrial world and habitability. 
They are both befitting candidates for the most widely used observational techniques, i.e., the transit method and radial velocity (RV) method (or Doppler method).
On the contrary, those massive stars {($\gtsim 1.5 M_{\odot}$)}, e.g., intermediate-mass stars {($1.5 M_{\odot} \leq M_{\star} < 5 M_{\odot}$)}, are observationally poor candidates during the main-sequence phase. 
The large body of massive stars restricts the detectability of planets in transit, and the lack of absorption lines due to their high surface temperature and fast self-rotation \citep{Lagrange2009} confines RV measurement to bad accuracy.

However, planet surveys around massive stars are particularly valuable concerning the point that the properties of giant planets would constrain the time scale and the mechanism of planet formation.
Compared to less massive stars, both the lifetimes of the massive host stars and those of their surrounding disks (\cite{Haisch2001a, Haisch2001b, Andrews2013, Ribas2015}) can be as short as the critical time scales of core-accretion planet formation scenario (A few Myr; \cite{Pollack1996}), a popular and favorable solution to solve the puzzle. 
Another plausible refers to the disk-instability scenario \citep{Boss1997}, where planets can form in a much shorter period (as short as a few kyr) than in the core-accretion scenario.
Thus, more observational evidence of planets and their properties is welcome in understanding the formation of giant planets.

The evolved counterpart of massive stars, spectroscopically G- or K-type subgiants or giants, have lower temperatures and slower rotation, suggesting that these targets are more promising to RV measurements. 
From this, the evolved stars have been widely surveyed by a number of projects:
Lick \citep{Frink2001, Reffert2015},
Okayama Planet Search Program (OPSP: \cite{Sato2005}), with its extended Subaru program \citep{Sato2010}, and its collaborative survey ``EAPS-Net'' \citep{Izumiura2005},
the ESO planet search program \citep{Setiawan2003},
the Tautenburg Observatory Planet Search \citep{Hatzes2005, Dollinger2007},
the ``Retired A Stars and Their Companions'' \citep{Johnson2007},
Penn State-Toru{\'n} Planet Search \citet{Niedzielski2007}, including its follow-up program ``Tracking Advanced Planetary systems'' \citet{Niedzielski2015a},
the BOAO K-giant survey \citep{Han2010},
the Pan-Pacific Planet Search (PPPS, \cite{Wittenmyer2011}),
the EXoPlanet aRound Evolved StarS project (EXPRESS: \cite{Jones2011}).
Based on large surveys, over 150 planetary systems around evolved stars have been discovered\footnote{We defined evolved stars as stars having surface gravity of $\log g < 3.5$. Here we do not strictly account for samples by intermediate mass ($1.5M_{\odot} < M_{\star} < 5M_{\odot}$), thus some less-massive stars are included.}.

Among these systems, $\sim 15\%$ of them are known to be multiple-planet systems. Thanks to  long-baseline RV surveys, these systems can be more widely discovered. 
These multiple-planet systems are of primary focus and a particular interest in terms of the formation and evolution of planetary systems. 
Further, these multiple-planet systems around evolved stars are mostly in the pattern of ``planet pairs'', and many of them have small orbital separations or they are in mean-motion resonance (MMR).
Some typical systems with small orbital separation include:
BD+20 2457 \citep{Niedzielski2009}, a metal-poor K2-giant with $\rm{[Fe/H]}=-1.0$ orbited by two brown dwarf-mass bodies with near 3:2 MMR; 
HD 33844 \citep{Wittenmyer2016a}, a two-giant-planets in 5:3 resonance;
7 CMa, HD 5319, HD 99706, HD 102329, HD 116029, and HD 200964 \citep{Wittenmyer2011,Johnson2011,Giguere2015,Bryan2016,Luque2019}, harboring giant planet pairs with a period ratio equal to or close to 4:3. 
Some typical systems with large orbital separation include: 
HD 4732 \citep{Sato2013a}, a subgiant star with two giant planets at 1.19 and 4.60 au, respectively;
$\nu$ Oph \citep{Quirrenbach2011, Sato2012, Quirrenbach2019}, a $M_{\star}=2.7 M_{\odot}$ giant star having two brown dwarfs in 6:1 resonance at 1.8 and 6.0 au respectively.

{In addition, precise astrometry helps much more in planet detection and characterization recently than before, and it is especially sensitive for massive companions in wide orbits. Precise astrometry} can be effectively combined with RV measurements to distinguish and constrain the masses of giant planets, brown dwarfs, and low-mass stellar companions, as well as to search for wide-orbit planets suggested by RV long-term acceleration.
Several groups of researchers have been working on joint constraints on RV-detected planets.
\citet{Li2021} used the Keplerian orbital code \texttt{orvara} \citep{Brandt2021} to fit the published RVs and Hipparcos-Gaia Catalog of Accelerations (HGCA: \cite{Brandt2018, Brandt2021}; Gaia EDR3 version) of nine single and massive RV-detected exoplanets and obtained an accurate determination of their masses.
\citet{Kiefer2021} investigated the nature of hundreds of RV-detected exoplanets with \texttt{GASTON} \citep{Kiefer2019}.
\citet{Feng2022} identified 167 cold giants and 68 other types of companions by combining RVs and astrometric data with \texttt{pexo} \citep{Feng2019}.
\citet{Delisle2022} analyzed three systems on the basis of Hipparcos data, together with on-ground RVs, with their newly-developed \texttt{kepmodel} code.
More recently, \citet{Xiao2023} independently measured the masses of 115 RV-detected substellar companions with \texttt{orvara} and \texttt{HGCA}.

The giant planet-metallicity relation has been studied for years.
Many studies have suggested a positive planet-metallicity correlation that giant planets are preferentially found around more metal-rich stars (e.g., \cite{Fischer2005}; \cite{Reffert2015}; \cite{Jones2016}; \cite{Wittenmyer2017}; \cite{Wolthoff2022}), but some others (e.g., \cite{Jofre2015}) have conservatively said no clear correlation.
Yet, a relation between stellar metallicity and the multiplicity of planets has not been well established or quantified.
\citet{Fischer2005} have pointed out that the presence of multiple-planet systems appeared to correlate with high stellar metallicity. 
\citet{Jofre2015} have also shown that evolved stars hosting multiple planets appear to be more metal-rich compared with evolved stars hosting single planets.
These trends follow the study on main-sequence stars, given by \citet{Wright2009}.

In this study, we revisit 32 planetary systems which are involved in the Okayama Planet Search Program (OPSP). 
These systems are released no later than 2018 with limited RV measurements. 
We here extend their time baseline of RV monitoring to hunt for extra companions and update the orbital parameters of the system as well as their host star. 
We also investigate the correlation between orbital properties and host star properties, especially for multiple-planet systems.
In section \ref{sec:opsp}, we revise the general information and 20-year operation of OPSP, and we update the properties of stars involved in this study.
In section \ref{sec:observations}, we describe the observations and data we have used in this study, and we provide the analysis of RV data.
In section \ref{sec:res}, we release the latest RV results of the 32 planetary systems, including the newly discovered 75 Cet c.
In section \ref{sec:interest}, we present other seven stars showing {noteworthy} RV variability.
In section \ref{sec:astrometry}, we present RV astrometry joint fitting results of four systems.
In section \ref{sec:discussion}, we discuss the properties of planet-harboring evolved stars and their planetary systems.  
Finally, in section \ref{sec:summary}, we summarize the paper.

\newpage
\section{Okayama Planet Search Program}\label{sec:opsp}
Okayama Planet Search Program (OPSP) surveys late-G (including early-K) giant stars in radial velocity, aiming to search for planets around intermediate-mass stars in their evolved stages.
OPSP selected 300 late-G giant stars and began the survey in 2001. 

To extend the OPSP program, an international network was established among Japanese, Korean, and Chinese researchers using three telescopes in the 2 m class in 2005 (East-Asian Planet Search Network: \cite{Izumiura2005}), and the Subaru Planet Search Program was started in 2006.
Subaru program targets at $\sim$ 300 Late-G giant stars, which could be barely followed up with 2 m class telescopes after a quick identification of stars exhibiting large RV variation.
The survey at $\rm{T\ddot{U}B\dot{I}TAK}$ National Observatory (TUG) is another extension of OPSP within the framework of an international collaboration between $\rm{T\ddot{u}rkiye}$, Russia, and Japan.

The targeted stars in the aforementioned surveys were selected from the Hipparcos catalog \citep{ESA1997} and they generally obey the following criteria: stars with 
1) $V < 7$ to attain a sufficient signal-to-noise (S/N) ratio,
2) a color index of $0.6 \lesssim B-V \lesssim 1.0$ to achieve intrinsic radial velocity stability to a level of $\sigma \lesssim 20\ \rm{m\ s^{-1}}$,
3) an absolute magnitude of $-3 \lesssim M_{V} \lesssim 2$ to include stars with masses of 1.5–5 $M_{\odot}$.

By the end of 2021, OPSP and its extensions have discovered 36 planetary systems, including systems with substellar companions, 4 candidates ($\epsilon$ Psc, HD 40956, HD 111591, and HD 175679), and independently confirmed 3 systems ($\kappa$ CrB, $\nu$ Oph, and HD 210702).
In 2022, OPSP confirmed another two systems around two red giant stars (HD 167768 and HD 184010)

In this study, we revisited planetary systems, which were observed within OPSP and its extensions and were released before 2019. 
But we did not include stars involved in the Korean-Japanese planet search program (\cite{Omiya2009}: HD 119445, \cite{Omiya2012}: HD 100655, \cite{Jeong2018}: HD 40956 and HD 111591). The update on these stars will appear in a forthcoming paper.

\begin{figure}
\centering
\includegraphics[scale=0.5]{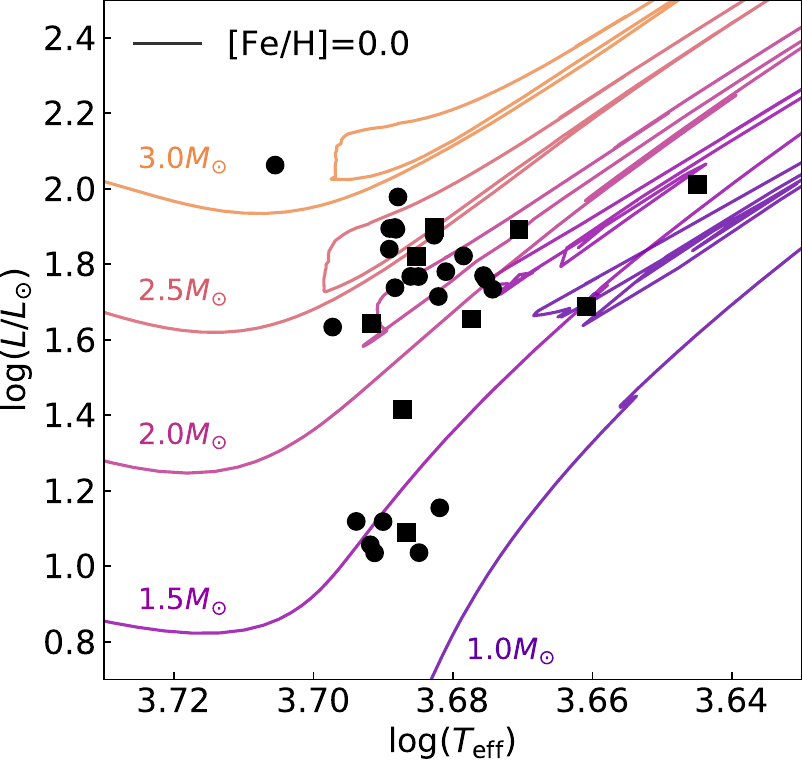}
\caption{
HR diagram of the 32 OPSP planet-harboring stars in this study. The stars included in \citet{Takeda2008} are marked by circles, while other stars are marked by squares. The colored lines represent evolutionary tracks for solar-metallicity stars of masses between 1.0 $M_{\odot}$ and 3.0 $M_{\odot}$.
}\label{fig:HR}
\end{figure}

The stellar parameters in this study, {as listed in Table \ref{tab:star_pars},} were obtained in two ways, which were discriminatively illustrated with different markers in Figure \ref{fig:HR}.
For stars involved in \citet{Takeda2008}, we followed the same method in \citet{Teng2022b} and re-estimated their stellar parameters by using \texttt{isoclassify} \citep{Huber2017} with the atmospheric parameters given by \citet{Takeda2008}. 
For other stars, we directly adopted the stellar parameters given in the planet detection paper, {as indicated in Table \ref{tab:star_ref}}.

\begin{table*}[p]
\begin{center}
\rotatebox{90}{\begin{minipage}{1.0\vsize}
\tbl{Stellar parameters.\footnotemark[$*$]}{
\scriptsize
\begin{tabular}{lcccccccccccccc}\hline\hline
\ & $\pi$ & $V$ & $B-V$ & Spec. type  & $T_{\mathrm{eff,sp}}$  & $\mathrm{[Fe/H]}_{\mathrm{sp}}$  & $\log g_{\star,sp}$  & $T_{\mathrm{eff}}$  & ${\mathrm{[Fe/H]}}$  & $\log g_{\star}$  & $L_{\star}$  & $M_{\star}$  & $R_{\star}$  &  \\  
\ & $(\rm{mas})$ & (mag) & (mag) & - & $({\mathrm{K}})$  & $\mathrm{(dex)}$  & $({\mathrm{cgs}})$  & $({\mathrm{K}})$  & ${\mathrm{(dex)}}$  & $({\mathrm{cgs}})$  & $(L_{\odot})$  & $ (M_{\odot})$  & $(R_{\odot})$  &  \\  
\hline
HD 2952 & $9.1205$  & $5.92$  & $1.04$  & K0 III  & $4844$  & $0.00$  & $2.67$  & $4853$  & $-0.00_{-0.10}^{+0.10}$  & $2.65_{-0.08}^{+0.09}$  & $58.59_{-6.43}^{+8.12}$  & $1.91_{-0.22}^{+0.42}$  & $10.90_{-0.81}^{+0.57}$  &  \\  
HD 4732 & $18.1441$  & $5.90$  & $0.94$  & K0 IV  & $4959$  & $0.01$  & $3.16$  & $4898$  & $-0.01_{-0.11}^{+0.05}$  & $3.17_{-0.06}^{+0.05}$  & $13.13_{-0.78}^{+0.81}$  & $1.39_{-0.17}^{+0.15}$  & $5.03_{-0.13}^{+0.12}$  &  \\  
HD 5608 & $17.0697$  & $5.98$  & $1.00$  & K0 IV  & $4854$  & $0.06$  & $3.03$  & $4807$  & $0.04_{-0.10}^{+0.07}$  & $3.06_{-0.08}^{+0.08}$  & $14.28_{-3.26}^{+4.07}$  & $1.29_{-0.23}^{+0.23}$  & $5.43_{-0.62}^{+0.70}$  &  \\  
HD 14067 & $7.1092$  & $6.51$  & $1.03$  & G9 III  &  -  &  -  &  -  & $4815$  & $-0.10_{-0.08}^{+0.08}$  & $2.61_{-0.10}^{+0.10}$  & $79$  & $2.40_{-0.20}^{+0.20}$  & $12.40_{-1.10}^{+1.10}$  &  \\  
75 Cet & $12.1717$  & $5.36$  & $1.00$  & G3 III  & $4846$  & $0.00$  & $2.63$  & $4809$  & $0.10_{-0.05}^{+0.08}$  & $2.69_{-0.04}^{+0.03}$  & $51.82_{-2.45}^{+1.46}$  & $1.92_{-0.08}^{+0.07}$  & $10.38_{-0.26}^{+0.15}$  &  \\  
81 Cet & $9.5199$  & $5.65$  & $1.02$  & G5 III  & $4785$  & $-0.06$  & $2.35$  & $4734$  & $-0.11_{-0.06}^{+0.10}$  & $2.41_{-0.06}^{+0.07}$  & $57.58_{-5.14}^{+8.43}$  & $1.23_{-0.25}^{+0.30}$  & $11.21_{-0.35}^{+0.81}$  &  \\  
$\epsilon$ Tau & $22.3654$  & $3.53$  & $1.10$  & K0 III  & $4883$  & $0.13$  & $2.57$  & $4877$  & $0.10_{-0.10}^{+0.11}$  & $2.66_{-0.05}^{+0.03}$  & $78.11_{-5.82}^{+5.68}$  & $2.57_{-0.25}^{+0.17}$  & $12.35_{-0.36}^{+0.41}$  &  \\  
HD 32518 & $8.2192$  & $6.42$  & $1.11$  & K1 III  &  -  &  -  &  -  & $4580$  & $-0.15_{-0.04}^{+0.04}$  & $2.10_{-0.15}^{+0.15}$  & $49$  & $1.13_{-0.18}^{+0.18}$  & $10.22_{-0.87}^{+0.87}$  &  \\  
6 Lyn & $18.2183$  & $5.86$  & $0.93$  & K0 IV  & $4978$  & $-0.13$  & $3.16$  & $4941$  & $-0.17_{-0.07}^{+0.11}$  & $3.16_{-0.08}^{+0.08}$  & $13.14_{-2.66}^{+3.23}$  & $1.32_{-0.21}^{+0.20}$  & $4.94_{-0.50}^{+0.55}$  &  \\  
HD 47366 & $11.9187$  & $6.11$  & $0.99$  & K1 III  &  -  &  -  &  -  & $4866$  & $-0.02_{-0.09}^{+0.09}$  & $2.97_{-0.06}^{+0.06}$  & $26$  & $1.81_{-0.13}^{+0.13}$  & $7.30_{-0.33}^{+0.33}$  &  \\  
o Uma & $17.9335$  & $3.42$  & $0.86$  & G4 II-III  & $5242$  & $-0.09$  & $2.64$  & $5075$  & $-0.16_{-0.10}^{+0.08}$  & $2.58_{-0.05}^{+0.06}$  & $115.52_{-17.61}^{+20.53}$  & $2.72_{-0.16}^{+0.14}$  & $13.84_{-1.04}^{+1.19}$  &  \\  
4 Uma & $13.2035$  & $4.61$  & $1.17$  & K1 III  &  -  &  -  &  -  & $4415$  & $-0.25_{-0.04}^{+0.04}$  & $1.80_{-0.15}^{+0.15}$  & $103$  & $1.23_{-0.15}^{+0.15}$  & $18.11_{-1.47}^{+1.47}$  &  \\  
41 Lyn & $11.8163$  & $5.39$  & $0.99$  & K0 III-IV  & $4771$  & $-0.34$  & $2.26$  & $4797$  & $-0.32_{-0.10}^{+0.10}$  & $2.37_{-0.07}^{+0.05}$  & $60.30_{-7.39}^{+14.76}$  & $1.07_{-0.16}^{+0.27}$  & $11.13_{-0.68}^{+1.58}$  &  \\  
HD 104985 & $9.9280$  & $5.78$  & $1.03$  & G9 III  & $4679$  & $-0.35$  & $2.47$  & $4738$  & $-0.32_{-0.10}^{+0.10}$  & $2.42_{-0.06}^{+0.07}$  & $58.87_{-11.20}^{+11.69}$  & $1.22_{-0.23}^{+0.29}$  & $11.12_{-0.74}^{+1.23}$  &  \\  
11 Com & $10.1677$  & $4.74$  & $1.00$  & G8 III  & $4841$  & $-0.28$  & $2.51$  & $4874$  & $-0.26_{-0.10}^{+0.10}$  & $2.45_{-0.08}^{+0.08}$  & $95.11_{-29.11}^{+48.85}$  & $2.09_{-0.63}^{+0.64}$  & $13.76_{-2.45}^{+2.85}$  &  \\  
HD 120084 & $9.6277$  & $5.90$  & $1.00$  & G7 III  & $4892$  & $0.09$  & $2.71$  & $4879$  & $0.10_{-0.10}^{+0.08}$  & $2.74_{-0.09}^{+0.07}$  & $54.68_{-7.44}^{+9.23}$  & $2.16_{-0.26}^{+0.31}$  & $10.37_{-0.80}^{+0.82}$  &  \\  
24 Boo & $10.5895$  & $5.58$  & $0.86$  & G3 IV  & $4893$  & $-0.77$  & $2.21$  & $4816$  & $-0.78_{-0.10}^{+0.10}$  & $2.28_{-0.07}^{+0.09}$  & $75.32_{-14.85}^{+17.34}$  & $1.05_{-0.18}^{+0.27}$  & $12.19_{-1.28}^{+1.45}$  &  \\  
o CrB & $11.9113$  & $5.51$  & $1.02$  & K0 III  & $4749$  & $-0.29$  & $2.34$  & $4769$  & $-0.27_{-0.10}^{+0.09}$  & $2.36_{-0.12}^{+0.07}$  & $66.37_{-14.04}^{+82.64}$  & $1.30_{-0.36}^{+0.95}$  & $11.87_{-1.50}^{+6.20}$  &  \\  
$\gamma$ Lib & $20.2593$  & $3.91$  & $1.01$  & K0 III  & $4822$  & $-0.30$  & $2.56$  & $4879$  & $-0.27_{-0.10}^{+0.08}$  & $2.50_{-0.07}^{+0.06}$  & $78.94_{-4.49}^{+4.18}$  & $1.78_{-0.24}^{+0.24}$  & $12.38_{-0.24}^{+0.30}$  &  \\  
$\omega$ Ser & $13.1026$  & $5.22$  & $1.02$  & G8 III  & $4770$  & $-0.24$  & $2.21$  & $4724$  & $-0.12_{-0.03}^{+0.01}$  & $2.37_{-0.04}^{+0.02}$  & $54.18_{-2.23}^{+1.50}$  & $1.01_{-0.08}^{+0.07}$  & $10.99_{-0.24}^{+0.15}$  &  \\  
$\kappa$ CrB & $33.3433$  & $4.82$  & $1.00$  & K0 III-IV  & $4877$  & $0.10$  & $3.21$  & $4840$  & $0.09_{-0.10}^{+0.07}$  & $3.20_{-0.10}^{+0.10}$  & $10.86_{-3.18}^{+4.38}$  & $1.33_{-0.25}^{+0.24}$  & $4.68_{-0.71}^{+0.84}$  &  \\  
HD 145457 & $7.3867$  & $6.57$  & $1.04$  & K0 III  &  -  &  -  &  -  & $4757$  & $-0.14_{-0.09}^{+0.09}$  & $2.77_{-0.10}^{+0.10}$  & $45$  & $1.90_{-0.30}^{+0.30}$  & $9.90_{-0.50}^{+0.50}$  &  \\  
$\nu$ Oph & $22.9031$  & $5.44$  & $0.99$  & K0 III  & $4928$  & $0.13$  & $2.63$  & $4887$  & $0.10_{-0.10}^{+0.10}$  & $2.64_{-0.10}^{+0.10}$  & $78.48_{-23.33}^{+34.05}$  & $2.61_{-0.57}^{+0.32}$  & $12.24_{-1.76}^{+2.55}$  &  \\  
HD 167042 & $20.1313$  & $5.97$  & $0.94$  & K1 IV  & $4943$  & $0.00$  & $3.28$  & $4919$  & $-0.01_{-0.11}^{+0.10}$  & $3.23_{-0.07}^{+0.07}$  & $11.40_{-1.98}^{+2.30}$  & $1.39_{-0.19}^{+0.17}$  & $4.63_{-0.40}^{+0.43}$  &  \\  
HD 173416 & $7.5363$  & $6.04$  & $1.06$  & G8  &  -  &  -  &  -  & $4683$  & $-0.22_{-0.09}^{+0.09}$  & $2.48_{-0.10}^{+0.10}$  & $78$  & $2.00_{-0.30}^{+0.30}$  & $13.50_{-0.90}^{+0.90}$  &  \\  
HD 175679 & $5.9475$  & $6.13$  & $0.96$  & G8 III  &  -  &  -  &  -  & $4844$  & $-0.14_{-0.10}^{+0.10}$  & $2.59_{-0.10}^{+0.10}$  & $66$  & $2.70_{-0.30}^{+0.30}$  & $11.60_{-1.60}^{+1.60}$  &  \\  
HD 180314 & $8.1261$  & $6.61$  & $1.02$  & K0  &  -  &  -  &  -  & $4917$  & $0.20_{-0.09}^{+0.09}$  & $2.98_{-0.10}^{+0.10}$  & $44$  & $2.60_{-0.30}^{+0.30}$  & $9.20_{-0.10}^{+0.40}$  &  \\  
$\xi$ Aql & $17.5183$  & $4.71$  & $1.02$  & K0 III  & $4802$  & $-0.18$  & $2.72$  & $4841$  & $-0.11_{-0.10}^{+0.06}$  & $2.61_{-0.08}^{+0.08}$  & $58.54_{-12.37}^{+6.02}$  & $1.74_{-0.28}^{+0.21}$  & $10.86_{-1.05}^{+0.44}$  &  \\  
18 Del & $13.3021$  & $5.51$  & $0.93$  & G6 III  & $4985$  & $-0.05$  & $2.84$  & $4980$  & $-0.06_{-0.10}^{+0.10}$  & $2.87_{-0.05}^{+0.03}$  & $43.02_{-4.04}^{+6.49}$  & $2.10_{-0.09}^{+0.13}$  & $8.80_{-0.40}^{+0.61}$  &  \\  
HD 208897 & $14.9591$  & $6.51$  & $1.01$  & K0  &  -  &  -  &  -  & $4860$  & $0.21_{-0.15}^{+0.15}$  & $3.13_{-0.14}^{+0.14}$  & $12$  & $1.25_{-0.11}^{+0.11}$  & $4.98_{-0.20}^{+0.20}$  &  \\  
HD 210702 & $18.4593$  & $5.93$  & $0.95$  & K1 IV  & $4967$  & $0.01$  & $3.19$  & $4912$  & $-0.02_{-0.10}^{+0.07}$  & $3.23_{-0.07}^{+0.08}$  & $10.85_{-2.14}^{+2.56}$  & $1.33_{-0.20}^{+0.19}$  & $4.54_{-0.45}^{+0.49}$  &  \\  
14 And & $13.1681$  & $5.22$  & $1.03$  & K0 III  & $4813$  & $-0.24$  & $2.63$  & $4888$  & $-0.21_{-0.10}^{+0.10}$  & $2.55_{-0.07}^{+0.06}$  & $69.17_{-7.51}^{+14.25}$  & $1.78_{-0.29}^{+0.43}$  & $11.55_{-0.51}^{+1.12}$  &  \\
\hline 
Source & Gaia & HIP & HIP & HIP & T08 & T08 & T08 & This work & This work & This work & This work & This work & This work \\
\hline 
\end{tabular}
}
\begin{tabnote}
\hangindent6pt\noindent
\hbox to6pt{\footnotemark[$*$]\hss}\unskip%
Stellar parameters of the 32 OPSP planet-harboring stars in this study. 
Gaia refers to Gaia EDR3, HIP refers to Hipparcos, and T08 refers to \citet{Takeda2008}.
The stellar parameters of stars included in T08 are re-estimated by using the same methodology in \citet{Teng2022a}. 
The subscript ``sp'' refers to spectral determination in T08. 
\end{tabnote}
\label{tab:star_pars}
\end{minipage}}
\end{center}
\end{table*}

\newpage
\section{Observations and analyses}\label{sec:observations}
\begin{table*}[p]
\footnotesize
\begin{center}
\begin{tabular}{llccc}\hline\hline
Star name & HD number & Data acquisition outside this work & Stellar parameters \\  
\hline
HD 2952 & HD 2952 & \citet{Sato2013b} & This work \\  
HD 4732 & HD 4732 & \citet{Sato2013a} & This work \\  
HD 5608 & HD 5608 & \citet{Sato2012} & This work \\  
HD 14067 & HD 14067 & \citet{Wang2014} & \citet{Wang2014} \\  
75 Cet & HD 15779 & \citet{Sato2012} & This work \\  
81 Cet & HD 16400 & \citet{Sato2008b} & This work \\  
$\epsilon$ Tau & HD 28305 & \citet{Sato2007}  & This work \\  
HD 32518 & HD 32518 & \citet{Dollinger2009} & \citet{Dollinger2009} \\  
6 Lyn & HD 45410 & \citet{Sato2008b}  & This work \\  
HD 47366 & HD 47366 & \citet{Sato2016} & \citet{Sato2016} \\  
o Uma & HD 71369 & \citet{Sato2012} & This work  \\  
4 Uma & HD 73108 & --- & \citet{Dollinger2007} \\  
41 Lyn & HD 81688 & \citet{Sato2008a} & This work  \\  
HD 104985 & HD 104985 & \citet{Sato2003}, \citet{Sato2008a} & This work  \\  
11 Com & HD 107383 & \citet{Liu2008} & This work  \\  
HD 120084 & HD 120084 & \citet{Sato2013b} & This work \\  
24 Boo & HD 127243 & \citet{Takarada2018} & This work  \\  
o CrB & HD 136512 & \citet{Sato2013b} & This work  \\  
$\gamma$ Lib & HD 138905 & \citet{Takarada2018} & This work  \\  
$\omega$ Ser & HD 141680 & \citet{Sato2013b} & This work   \\  
$\kappa$ CrB & HD 142091 & \citet{Bowler2010}, \citet{Sato2012} & This work  \\  
HD 145457 & HD 145457 & \citet{Sato2010} & \citet{Sato2010}  \\  
$\nu$ Oph & HD 163917 & \citet{Sato2012}, \citet{Quirrenbach2019} & This work \\  
HD 167042 & HD 167042 & \citet{Sato2008b}, \citet{Bowler2010} & This work\\  
HD 173416 & HD 173416 & \citet{Liu2009}  & \citet{Liu2009}\\  
HD 175679 & HD 175679 & \citet{Wang2012} & \citet{Wang2012} \\  
HD 180314 & HD 180314 & \citet{Sato2010} & \citet{Sato2010} \\  
$\xi$ Aql & HD 188310 & \citet{Sato2008a} & This work \\  
18 Del & HD 199665 & \citet{Sato2008a}, \citet{Ryu2016} & This work\\  
HD 208897 & HD 208897 & \citet{Yilmaz2017} & \citet{Yilmaz2017}  \\  
HD 210702 & HD 210702 & \citet{Bowler2010}, \citet{Sato2012} & This work \\  
14 And & HD 221345  & \citet{Sato2008b} & This work \\
\hline 
\end{tabular}
\end{center}
\caption{
Data acquisition details for each planet-harboring star included in the Okayama Planet Search Program (OPSP) studied in this study. 
The RVs for stars previously studied in OPSP papers were re-extracted from 1-D spectra, and may differ from those reported in previous studies.
RV data of 6 Lyn in \citet{Bowler2010} and HD 73108 in \citet{Dollinger2007} are not provided, 
while the RV data of $\nu$ Oph in \citet{Quirrenbach2019} include the RVs reported in the original planet discovery paper \citep{Quirrenbach2011}.
}\label{tab:star_ref}
\end{table*}

\subsection{OAO Observations}
The RV data in this study were mainly extracted from the spectra that were obtained by the 1.88-m reflector with HIgh Dispersion Echelle Spectrograph (HIDES: \cite{Izumiura1999}) at OAO from 2001 to 2021.
The instrument was equipped with an iodine cell in the HIDES optical path to HIDES, which provided numerous iodine absorption lines in the range of 5000-5800 \AA ~as a reference for precise radial velocity measurements. 

HIDES has experienced several upgrades.
At first, HIDES was set to cover a wavelength range of 5000-6100 \AA ~with one $2 \mathrm{K} \times 4 \mathrm{K}$ CCD.
In December 2007, HIDES was upgraded from a single CCD to a mosaic of three.
This widened the wavelength region to 3700-7500 \AA ~(3700-5000 \AA, ~5000-5800 \AA ~and ~5800-7500 \AA ~for each respectively) and enabled us to simultaneously measure the level of stellar activities (e.g. Ca \emissiontype{II} HK lines) and line profiles as well as radial velocities.
In 2010, HIDES was equipped with a new high-efficiency fiber-link system with its own iodine cell installed in the optical path.
This greatly enhanced the overall throughput \citep{Kambe2013} and efficiently shortened the exposure time of our targets. 
In 2018, HIDES was upgraded with its slit mode elements fully removed from the HIDES optical path, 
and with the optical instruments of fiber-link mode re-arranged on a new stabilized platform in the precise temperature-controlled Coud\'e room. 
Since the optical path was re-arranged in the upgrade, fiber-link modes pre-/post-upgrade are considered to be two independent instruments. 

In this study, the HIDES RVs were extracted by spectra from both conventional slit mode (HIDES-S) and fiber mode pre- and post-upgrade (HIDES-F1 and HIDES-F2). 

\subsection{Other Observations}
The RV data obtained from sites other than OAO were mainly obtained from published papers. 
These include data obtained from 
the High Dispersion Spectrograph (HDS: \cite{Noguchi2002}) mounted on Subaru Telescope at Mauna Kea,
the Coud\'{e} Echelle Spectrograph (CES\footnote{The CES pre-upgrade and post-upgrade are respectively marked as CES-O and CES-N.}: \cite{Zhao2001}) and High-Resolution Spectrograph (HRS) on 2.16m telescope at Xinglong Observatory,
the UCLES echelle spectrograph \citep{Diego1990} on Anglo-Australian Telescope (AAT) at Siding Spring Observatory,
the Hamilton Echelle Spectrograph on Coud\'{e} Auxiliary Telescope at Lick Observatory (Lick),
the coud\'{e} echelle spectrograph on 2m Alfred Jensch Telescope (AJT) at Thuringia State Observatory,
and Coude Echelle Spectrograph on 1.5m Russian-Turkish Telescope at TUG.
Furthermore, we obtained new RV data from TUG after the last data release in 2017 January \citep{Yilmaz2017}.
A listing of data acquisition for each star is given in Table \ref{tab:star_ref}.

\subsection{Radial Velocity from HIDES spectra}
We extracted RVs with the same methodology given in \citet{Teng2022b} and \citet{Teng2022c}. Here, we briefly review the flow.

The reduction of echell{\'e} spectra taken by HIDES were performed with IRAF\footnote{IRAF is distributed by the National Optical Astronomy Observatories, which is operated by the Association of Universities for Research in Astronomy, Inc. under a cooperative agreement with the National Science Foundation, USA.} packages in a standard way. 
To make RV measurement, we used spectra covering 5000\AA\ to 5800\AA\ in which $\rm{I_{2}}$ absorption lines are superimposed by an $\rm{I_{2}}$ cell. 
Following the procedure in \citet{Sato2002} and \citet{Sato2012}, a method based on \citet{Butler1996}, we modeled the spectra ($\rm{I_{2}}$ superposed stellar spectra) by using the stellar template and high-resolution $\rm{I_{2}}$ spectra convolved with the instrumental profile (IP) of the spectrograph. 
The stellar template spectra used for HIDES-S were obtained by deconvolving a pure stellar spectrum with the IP estimated from $\rm{I_{2}}$ superposed B type star or flat spectra.
If a star has spectra taken by HIDES-F1 high-resolution mode (R $\sim$ 100000) without $\rm{I_{2}}$-cell in its optical path, the stellar template spectra used for HIDES-F1 and -F2 were obtained by its high-resolution spectra. Otherwise, we use the same deconvolved spectra for HIDES-S. 
The final RV values and their measurement errors were taken from the average of measurement in each segment set in the $\rm{I_{2}}$ absorption region for each instrument.

\subsection{Period search and Keplerian orbital fit}\label{sec:orbfit}
For a given RV time series, Generalized Lomb-Scargle periodogram (hereafter GLS: \cite{Zechmeister2009}) was performed 
to search for periodicity caused by planets in the time series.
To assess the significance of the periodicity, False Alarm Probability (FAP) was applied with the approximation method developed by \citet{Baluev2008} in the same package. {A peak higher than 0.1\% threshold, i.e. with its FAP value significantly lower than 0.1\%, i.e., was recognized as a credible signal.}

The Keplerian model was generated with free parameters including Keplerian orbital elements (orbital period $P$, RV semi-amplitude $K$, the combination of eccentricity $e$ and argument of periastron $\omega$, $\sqrt{e} \cos \omega$ and $\sqrt{e} \sin \omega$, and time of inferior conjunction\footnote{The periastron passage $T_{\rm{p}}$ is converted from inferior conjunction $T_{\rm{c}}$ after parameter fitting.} $T_{\rm{c}}$), extra Gaussian noise $s_{\rm{inst}}$, RV offset of different instruments $\gamma$, and long term trends at highest of 2nd order.
\begin{equation}
V_{\mathrm{r}} = \sum_k^{N_{\rm pl}} V_{r,k}(t | P, K, e, \omega, T_{\mathrm{p}} ) + \gamma + \dot{\gamma}(t - t_0) + \ddot{\gamma}(t - t_0)^2,
\label{eqn:vr}
\end{equation}
where $N_{\rm pl}$ is the total number of planets in the system, $\dot{\gamma}$ indicates the linear trend term, $\ddot{\gamma}$ indicates the quadratic trend term, and $t_0$ is the midpoint of the full time span. 
{Since the main instrument, HIDES, is confirmed to be stable in 20 years without long-term RV drifts (Appendix \ref{sec:inst_stability}), the trend terms in Equation \ref{eqn:vr} are used to fit the long-term RV variations, which are not induced by the instrumental systematics.} 

The best-fit Keplerian orbits and the uncertainties were derived by maximum a posteriori (MAP) method from Markov Chain Monte Carlo (MCMC) sampling.
The MCMC starting point was generated from likelihood maxima, which were determined by a truncated Newton \textsf{TNC} method and a \textsf{Nelder-Mead} method \citep{Nelder1965}.

The long-term trend was determined by model comparison. The goodness of fit was indicated by reduced Chi-square $\chi^{2}_{\rm{red}}$ ($\chi^{2}_{\mathrm{red}} = \chi^{2} / N_{\mathrm{DoF}}$, where $N_{\mathrm{DoF}}$ is the degree of freedom), and the best model was determined by Bayesian Information Criteria (BIC: \cite{Schwarz1978}; $\mathrm{BIC} = -2\ln{(\hat{L})} + n_{\mathrm{pars}} \ln{(n_{\mathrm{data}})} $, where $\hat{L}$ is the maximized value of the likelihood function, $n_{\mathrm{pars}}$ is the number of parameters, and $n_{\rm{data}}$ is the number of data). Fittings with $\chi^{2}_{\rm{red}}$ far from 1 was considered as not well fitted. Models with absolute value of $\Delta\mathrm{BIC}$ less than 10 were not considered to be distinct \citep{Kass1996}. {For comparable two models with different long-term trend, we chose simpler one with fewer free parameters, i.e., the model described by a simpler long-term trend. However, only if two models are comparable in BIC but simultaneously varied in long-term trend and the Keplerian orbit, e.g. the orbit solutions of HD 14067 b in \citet{Wang2014}, we provide both models since because the current observations cannot determine the true scenario.} 

The determination of a possible extra planet was in a similar way. 
We compared Keplerian models with different planet numbers with $\chi^{2}_{\rm{red}}$ and BIC if an extra periodic signal appeared in the periodogram. 
Similarly, we applied a model with possible extra planets only if this model is significantly better, i.e. $\chi^{2}_{\rm{red}}\sim 1$, and a lower BIC value with a difference greater than 10.

\subsection{Chromospheric activity and Line profile analysis of HIDES spectra}
Spectral line profile deformation can lead to wavelength shifts and mimic planetary signals, and stellar chromospheric activity can also result in RV variations. Thus line profile analyses are necessary in order to eliminate any fake planetary signal. 
In this study, these analyses were performed the same as \citet{Teng2022a}.

Bisector inverse span {(BIS: \cite{Queloz2001})} was adopted as an indicator of line profile asymmetry. The spectra in the iodine-free region (4000-5000 \AA) were taken for measurement. The BIS was defined as the offset of averaged velocity between the cross-correlation function (CCF\footnote{CCF is constructed by shifting the mask as a function of Doppler velocity, and its weight is determined by the depth of the spectral line. The numerical mask is generated from \texttt{SPECTRUM} \citep{Gray1994} for a G-type giant star with approximately 800 lines.}: \cite{Baranne1996,Pepe2002})  upper region (5\%--15\% from the continuum of the CCF) and lower region (85\%--95\% from the continuum of the CCF). 
The mean removed BIS, $\rm{BIS}^{\prime}=\rm{BIS}-\overline{BIS}$, was calculated in order to suppress the offset between instruments. Notably, while the amplitude of BIS measurements might be expected to change with different instruments, only the offset (without scaling) was taken into account in this analysis.

The line cores of Ca \emissiontype{II} H\&K lines are widely used in the visible band to trace chromospheric activity \citep{Duncan1991}. 
In this study, the Ca \emissiontype{II} H index $S_{\rm{H}}$, defined in \citet{Sato2013b}, was applied to quantify the strength of chromospheric activity. However, Ca \emissiontype{II} K lines were not included since the signal-to-noise ratio for the spectra in this region was not high enough for measurement. 
If a stellar chromospheric activity causes an apparent RV variation, a clear correlation between RV and $S_{\rm{H}}$ can be detected (e.g., Figure 9 in \cite{Sato2013b}). Additionally, it should be noticed that Ca \emissiontype{II} H\&K lines are located in the order overlapping regions in HIDES-F1 and -F2 spectra, therefore $S_{\rm{H}}$ cannot be measured with HIDES-F1 and -F2 spectra.

\newpage
\section{Fitting Results and discovery of a new planet}\label{sec:res}
\subsection{Refined orbital parameters}
\begin{table*}[p]
\rotatebox{90}{\begin{minipage}{1.0\vsize}
\begin{center}
\tbl{Orbital parameters.\footnotemark[$*$]}{
\scriptsize
\begin{tabular}{llccccccccc}\hline\hline
Parameters  & $P$  & $K$  & $e$  & $\omega$  & $T_{\rm{p}}$ & $M_{\rm{p}}\sin i$  & $a$  & $\dot{\gamma}$  & $\ddot{\gamma}$\\  
Unit  & $(\rm{d})$  & $(\rm{m\ s^{-1}})$  &   & $\rm{(rad)}$  & (JD$-2450000$)  & $(M_{\rm{J}})$  & $(\rm{au})$  & $(\rm{m\ s^{-1}\ yr^{-1}})$  & $(\rm{m\ s^{-1}\ yr^{-2}})$  \\  
\hline  
HD 2952 b & $312.08_{-1.69}^{+1.36}$  & $18.42_{-2.43}^{+2.53}$  & $0.269_{-0.150}^{+0.165}$  & $-2.715_{-0.629}^{+0.600}$  & $4896.3_{-48.0}^{+21.6}$  & $0.896_{-0.125}^{+0.133}$  & $1.117_{-0.004}^{+0.003}$  &   &   \\  
HD 4732 b & $371.74_{-1.86}^{+0.72}$  & $46.80_{-4.07}^{+4.78}$  & $0.266_{-0.068}^{+0.081}$  & $1.093_{-0.193}^{+0.169}$  & $5660.0_{-14.5}^{+12.8}$  & $1.986_{-0.154}^{+0.156}$  & $1.129_{-0.004}^{+0.001}$  &   &   \\  
HD 4732 c & $3184.20_{-57.60}^{+58.12}$  & $38.03_{-3.15}^{+1.30}$  & $0.187_{-0.066}^{+0.034}$  & $1.893_{-0.231}^{+0.289}$  & $9459.3_{-125.9}^{+138.8}$  & $3.365_{-0.269}^{+0.116}$  & $4.725_{-0.057}^{+0.057}$  &   \\  
HD 5608 b & $768.70_{-1.67}^{+4.72}$  & $21.95_{-1.14}^{+1.17}$  & $0.110_{-0.080}^{+0.029}$  & $-1.604_{-1.206}^{+0.268}$  & $3936.4_{-721.7}^{+27.8}$  & $1.168_{-0.057}^{+0.062}$  & $1.790_{-0.003}^{+0.007}$  & $-5.138_{-0.594}^{+0.167}$  &   \\  
HD 14067 b & $2725.81_{-7.50}^{+13.12}$  & $100.17_{-5.60}^{+4.17}$  & $0.701_{-0.040}^{+0.023}$  & $1.787_{-0.086}^{+0.090}$  & $5797.5_{-11.8}^{+8.8}$  & $8.806_{-0.387}^{+0.421}$  & $5.113_{-0.009}^{+0.016}$  &   &   \\  
75 Cet b & $696.62_{-1.69}^{+1.33}$  & $36.92_{-1.35}^{+1.09}$  & $0.093_{-0.042}^{+0.026}$  & $-2.750_{-0.184}^{+5.513}$  & $5056.9_{-60.5}^{+34.6}$  & $2.479_{-0.090}^{+0.074}$  & $1.912_{-0.003}^{+0.002}$  &   &   \\  
\textbf{75 Cet c} & $2051.62_{-40.47}^{+45.98}$  & $9.44_{-1.39}^{+1.04}$  & $0.023_{-0.003}^{+0.191}$  & $2.147_{-0.890}^{+2.737}$  & $4284.9_{-765.7}^{+474.0}$  & $0.912_{-0.143}^{+0.088}$  & $3.929_{-0.052}^{+0.058}$  &   \\  
81 Cet b & $1005.57_{-1.94}^{+1.84}$  & $58.51_{-1.20}^{+1.39}$  & $0.037_{-0.025}^{+0.015}$  & $-2.909_{-0.628}^{+0.950}$  & $3458.4_{-180.6}^{+106.0}$  & $3.307_{-0.067}^{+0.078}$  & $2.104_{-0.003}^{+0.003}$  &   &   \\  
$\epsilon$ Tau b & $585.82_{-0.33}^{+0.26}$  & $93.24_{-0.73}^{+0.74}$  & $0.076_{-0.008}^{+0.009}$  & $1.883_{-0.106}^{+0.119}$  & $3492.3_{-10.0}^{+11.3}$  & $7.190_{-0.056}^{+0.056}$  & $1.878_{-0.001}^{+0.001}$  &   &   \\  
HD 32518 b & $157.35_{-0.08}^{+0.10}$  & $98.88_{-5.88}^{+5.60}$  & $0.028_{-0.019}^{+0.034}$  & $-0.618_{-1.244}^{+1.876}$  & $4375.2_{-6.5}^{+108.3}$  & $2.849_{-0.171}^{+0.160}$  & $0.594_{-0.000}^{+0.000}$  &   &   \\  
6 Lyn b & $919.86_{-3.71}^{+1.92}$  & $32.76_{-1.52}^{+0.67}$  & $0.042_{-0.032}^{+0.024}$  & $2.111_{-1.077}^{+1.540}$  & $3537.3_{-224.7}^{+121.2}$  & $1.881_{-0.089}^{+0.038}$  & $2.028_{-0.005}^{+0.003}$  &   &   \\  
HD 47366 b & $360.08_{-0.72}^{+0.27}$  & $40.15_{-2.20}^{+1.97}$  & $0.071_{-0.046}^{+0.030}$  & $-2.654_{-0.143}^{+5.231}$  & $5308.7_{-68.8}^{+24.4}$  & $2.082_{-0.113}^{+0.099}$  & $1.207_{-0.002}^{+0.001}$  &   &   \\  
HD 47366 c & $677.53_{-1.16}^{+1.29}$  & $24.02_{-1.08}^{+1.18}$  & $0.161_{-0.041}^{+0.045}$  & $2.856_{-0.525}^{+0.163}$  & $3950.3_{-54.0}^{+16.5}$  & $1.522_{-0.070}^{+0.072}$  & $1.840_{-0.002}^{+0.002}$  &   \\  
o Uma b & $1569.81_{-13.59}^{+12.76}$  & $33.11_{-2.77}^{+1.64}$  & $0.149_{-0.079}^{+0.044}$  & $0.712_{-0.374}^{+0.539}$  & $3348.8_{-84.1}^{+132.6}$  & $3.651_{-0.298}^{+0.176}$  & $3.691_{-0.021}^{+0.020}$  &   &   \\  
4 Uma b & $270.27_{-0.09}^{+0.24}$  & $242.24_{-14.31}^{+7.45}$  & $0.453_{-0.036}^{+0.021}$  & $0.499_{-0.094}^{+0.055}$  & $5958.1_{-4.1}^{+2.3}$  & $7.902_{-0.378}^{+0.199}$  & $0.877_{-0.000}^{+0.001}$  &   &   \\  
41 Lyn b & $183.93_{-0.09}^{+0.09}$  & $56.42_{-1.88}^{+1.87}$  & $0.040_{-0.031}^{+0.022}$  & $-0.252_{-1.099}^{+1.158}$  & $4354.4_{-16.9}^{+80.2}$  & $1.654_{-0.055}^{+0.054}$  & $0.648_{-0.000}^{+0.000}$  &   &   \\  
HD 104985 b & $199.68_{-0.07}^{+0.09}$  & $169.06_{-3.74}^{+3.33}$  & $0.047_{-0.028}^{+0.015}$  & $2.336_{-0.488}^{+0.581}$  & $5083.5_{-16.1}^{+17.7}$  & $5.559_{-0.122}^{+0.110}$  & $0.715_{-0.000}^{+0.000}$  &   &   \\  
11 Com b & $323.21_{-0.05}^{+0.06}$  & $288.63_{-2.37}^{+2.39}$  & $0.238_{-0.007}^{+0.007}$  & $1.594_{-0.031}^{+0.030}$  & $4519.4_{-1.6}^{+1.5}$  & $15.464_{-0.125}^{+0.123}$  & $1.178_{-0.000}^{+0.000}$  &   &   \\  
HD 120084 b & $2134.79_{-13.31}^{+8.08}$  & $40.50_{-2.04}^{+1.88}$  & $0.478_{-0.034}^{+0.035}$  & $1.902_{-0.092}^{+0.130}$  & $4965.7_{-19.3}^{+29.1}$  & $3.762_{-0.188}^{+0.156}$  & $4.192_{-0.017}^{+0.011}$  &   &   \\  
24 Boo b & $30.33_{-0.01}^{+0.00}$  & $55.67_{-2.66}^{+2.31}$  & $0.032_{-0.023}^{+0.039}$  & $-3.128_{-1.666}^{+1.559}$  & $5129.2_{-15.4}^{+3.0}$  & $0.883_{-0.043}^{+0.036}$  & $0.194_{-0.000}^{+0.000}$  &   &   \\  
o CrB b & $187.01_{-0.16}^{+0.15}$  & $33.56_{-2.16}^{+1.71}$  & $0.105_{-0.079}^{+0.036}$  & $0.771_{-0.837}^{+0.742}$  & $4443.7_{-18.9}^{+27.8}$  & $1.120_{-0.071}^{+0.057}$  & $0.699_{-0.000}^{+0.000}$  &   &   \\  
$\gamma$ Lib b & $414.88_{-1.47}^{+1.24}$  & $21.17_{-2.01}^{+1.94}$  & $0.160_{-0.107}^{+0.071}$  & $-2.962_{-0.086}^{+5.754}$  & $5020.3_{-86.3}^{+23.7}$  & $1.126_{-0.108}^{+0.101}$  & $1.319_{-0.003}^{+0.003}$  &   &   \\  
$\gamma$ Lib c & $966.49_{-1.96}^{+2.27}$  & $73.38_{-2.02}^{+1.75}$  & $0.065_{-0.022}^{+0.033}$  & $2.663_{-0.366}^{+0.453}$  & $5563.8_{-56.7}^{+67.6}$  & $5.231_{-0.147}^{+0.120}$  & $2.318_{-0.003}^{+0.004}$  &   \\  
$\omega$ Ser b & $278.59_{-0.61}^{+0.80}$  & $23.11_{-3.38}^{+1.94}$  & $0.180_{-0.127}^{+0.066}$  & $2.499_{-1.272}^{+0.600}$  & $5022.3_{-63.7}^{+20.1}$  & $0.736_{-0.106}^{+0.060}$  & $0.838_{-0.001}^{+0.002}$  &   &   \\  
$\kappa$ CrB b & $1253.68_{-6.48}^{+3.07}$  & $23.78_{-0.99}^{+0.52}$  & $0.040_{-0.032}^{+0.013}$  & $3.099_{-7.392}^{+4.617}$  & $1743.5_{-633.1}^{+181.3}$  & $1.522_{-0.065}^{+0.033}$  & $2.500_{-0.009}^{+0.004}$  & $0.518_{-0.316}^{+0.033}$  &   \\  
HD 145457 b & $176.13_{-0.20}^{+0.18}$  & $66.46_{-2.65}^{+3.12}$  & $0.111_{-0.040}^{+0.039}$  & $-0.848_{-0.490}^{+0.516}$  & $3347.1_{-10.2}^{+19.7}$  & $2.794_{-0.117}^{+0.132}$  & $0.762_{-0.001}^{+0.001}$  &   &   \\  
$\nu$ Oph b & $529.93_{-0.07}^{+0.08}$  & $287.60_{-0.74}^{+1.19}$  & $0.126_{-0.004}^{+0.003}$  & $0.164_{-0.021}^{+0.028}$  & $1506.0_{-1.8}^{+2.3}$  & $21.548_{-0.056}^{+0.092}$  & $1.765_{-0.000}^{+0.000}$  &   &   \\  
$\nu$ Oph c & $3180.60_{-4.52}^{+5.01}$  & $175.47_{-1.02}^{+1.57}$  & $0.179_{-0.005}^{+0.007}$  & $0.149_{-0.034}^{+0.033}$  & $-114.1_{-17.6}^{+16.6}$  & $23.693_{-0.139}^{+0.205}$  & $5.829_{-0.006}^{+0.006}$  &   \\  
HD 167042 b & $417.67_{-0.41}^{+0.43}$  & $30.52_{-0.95}^{+0.70}$  & $0.010_{-0.004}^{+0.033}$  & $0.950_{-2.358}^{+1.213}$  & $2935.5_{-65.6}^{+136.9}$  & $1.397_{-0.044}^{+0.031}$  & $1.220_{-0.001}^{+0.001}$  & $1.361_{-0.291}^{+0.230}$  &   \\  
HD 173416 b & $322.03_{-1.52}^{+1.33}$  & $35.33_{-4.46}^{+4.07}$  & $0.208_{-0.094}^{+0.089}$  & $-2.058_{-0.578}^{+0.589}$  & $4735.0_{-253.1}^{+23.4}$  & $1.841_{-0.225}^{+0.206}$  & $1.158_{-0.004}^{+0.003}$  &   &   \\  
HD 175679 b & $1357.64_{-2.04}^{+1.74}$  & $378.37_{-5.65}^{+3.51}$  & $0.386_{-0.007}^{+0.011}$  & $-0.222_{-0.031}^{+0.019}$  & $3274.6_{-6.1}^{+4.4}$  & $36.868_{-0.626}^{+0.337}$  & $3.341_{-0.003}^{+0.003}$  &   &   \\  
HD 180314 b & $394.73_{-0.13}^{+0.13}$  & $343.39_{-3.92}^{+3.83}$  & $0.247_{-0.009}^{+0.009}$  & $-0.982_{-0.044}^{+0.044}$  & $3570.6_{-2.4}^{+2.4}$  & $22.707_{-0.255}^{+0.252}$  & $1.448_{-0.000}^{+0.000}$  &   &   \\  
$\xi$ Aql b & $136.97_{-0.18}^{+0.11}$  & $53.07_{-4.72}^{+2.69}$  & $0.059_{-0.037}^{+0.087}$  & $-0.000_{-0.040}^{+1.857}$  & $3001.0_{-2.1}^{+46.7}$  & $1.941_{-0.179}^{+0.091}$  & $0.625_{-0.001}^{+0.000}$  &   &   \\  
18 Del b & $982.85_{-0.92}^{+1.06}$  & $114.96_{-0.96}^{+1.98}$  & $0.024_{-0.018}^{+0.007}$  & $1.119_{-1.012}^{+0.503}$  & $4356.1_{-118.2}^{+102.9}$  & $9.207_{-0.077}^{+0.160}$  & $2.476_{-0.002}^{+0.002}$  & $-1.291_{-0.473}^{+0.377}$  & $0.177_{-0.069}^{+0.052}$  \\  
HD 208897 b & $358.27_{-1.50}^{+0.87}$  & $29.45_{-1.04}^{+1.25}$  & $0.020_{-0.010}^{+0.049}$  & $-1.981_{-0.084}^{+2.815}$  & $6178.2_{-311.4}^{+4.7}$  & $1.194_{-0.044}^{+0.049}$  & $1.063_{-0.003}^{+0.002}$  & $4.073_{-0.825}^{+0.441}$  &   \\  
HD 210702 b & $355.50_{-0.61}^{+0.38}$  & $36.11_{-1.27}^{+1.46}$  & $0.071_{-0.051}^{+0.019}$  & $1.962_{-0.447}^{+0.858}$  & $2189.9_{-28.5}^{+40.9}$  & $1.515_{-0.053}^{+0.062}$  & $1.079_{-0.001}^{+0.001}$  &   &   \\  
14 And b & $186.76_{-0.12}^{+0.11}$  & $86.08_{-2.95}^{+2.76}$  & 0 (fixed)  & 0 (fixed)  & $2853.0_{-2.1}^{+2.1}$  & $3.559_{-0.122}^{+0.114}$  & $0.775_{-0.000}^{+0.000}$  &   &   \\  
\hline 
\end{tabular}
}
\begin{tabnote}
\hangindent6pt\noindent
\hbox to6pt{\footnotemark[$*$]\hss}\unskip%
These parameters are derived with only RV data. $\dot{\gamma}$ $(\rm{m\ s^{-1}\ yr^{-1}})$ and $\ddot{\gamma}$ $(\rm{m\ s^{-1}\ yr^{-2}})$ are linear trend and quadratic trend, respectively.
The newly confirmed planet (75 Cet c) is marked in bold text.
\end{tabnote}
\label{tab:orbpar_orbits}
\end{center}
\end{minipage}}
\end{table*}

Following the methodology outlined in Section \ref{sec:orbfit}, we refined the orbital parameters of the 32 revisited systems. 
Our analysis revealed an additional planet, 75 Cet c, in one of the systems, and five systems with significant long-term trends: HD 5608, $\kappa$ CrB, HD 167042, 18 Del, and HD 208897.

This manuscript provides a comprehensive listing of the best-fit orbital parameters in Table \ref{tab:orbpar_orbits}. 
The online materials contain not only the orbital parameters but also additional parameters such as extra jitter and instrumental offsets.
Furthermore, the online materials include a compilation of all RV and other spectral measurements.

\subsection{A new planet: 75 Cet c}
75 Cet (HD 15779, HIP 11791) is a G3 III star and reported to harbor a giant planet in \citet{Sato2012}. We started to observe 75 Cet in 2002 February at OAO with HIDES-S. After 10-year observation, we released a giant planet of $M_{\rm{p}}\sin i=3.0\ M_{\rm{J}}$ at $a = 2.1$ au, and possible planet(s) at wider orbit(s). We continued monitoring the star with HIDES-S, -F1, and -F2 at OAO until 2020 December and totally obtained 147 spectra with RVs.

\begin{figure}
\begin{center}
\includegraphics[scale=0.45]{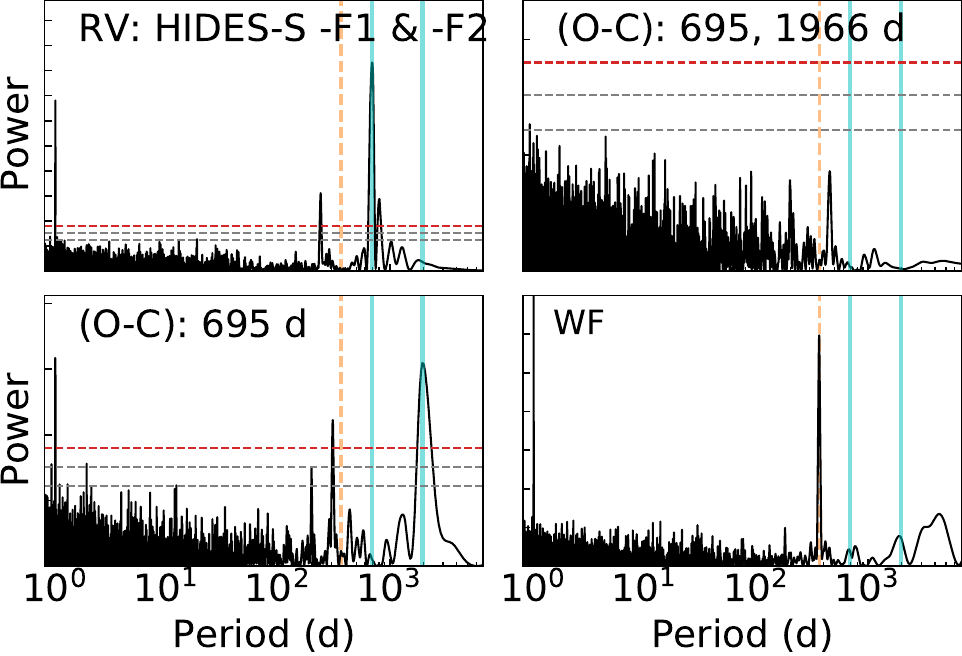}
\end{center}
\caption{
GLS periodograms for 75 Cet. Top left: The observed RVs. Bottom left: The RV residuals to the single Keplerian fitting of the inner planet. Top right: The RV residuals to the 2-Keplerian fitting. Bottom right: The window function of the observed RVs. 
The horizontal lines represent 10\%, 1\%, and 0.1\% FAP levels from bottom to top. The vertical cyan solid lines indicate the planetary signals, and the vertical orange dashed line indicates 1 year. 
}\label{fig:HD15779_resid_ls}
\end{figure}

We fitted the latest data with a single Keplerian model, and it yielded the orbital parameters for the known companion of $P = 696.26\ \rm{d}$, $K = 35.63\ \rm{m}\ \rm{s}^{-1}$, and $e = 0.05$, with an rms scatter of $10.45\ \rm{m}\ \rm{s}^{-1}$.
We then performed a periodogram analysis of the residuals and found possible peaks at round 307 d and 1966 d (Figure \ref{fig:HD15779_resid_ls}), which were close to those reported in \citet{Sato2012}. 
Since the observation had a strong sampling window at one year, and two periodic signals had a relation of $1/307 \sim 1/365 + 1/1966$, we realized the 307 d signal could be the alias of the 1966 d signal. 

\begin{figure}
\begin{center}
\includegraphics[scale=0.25]{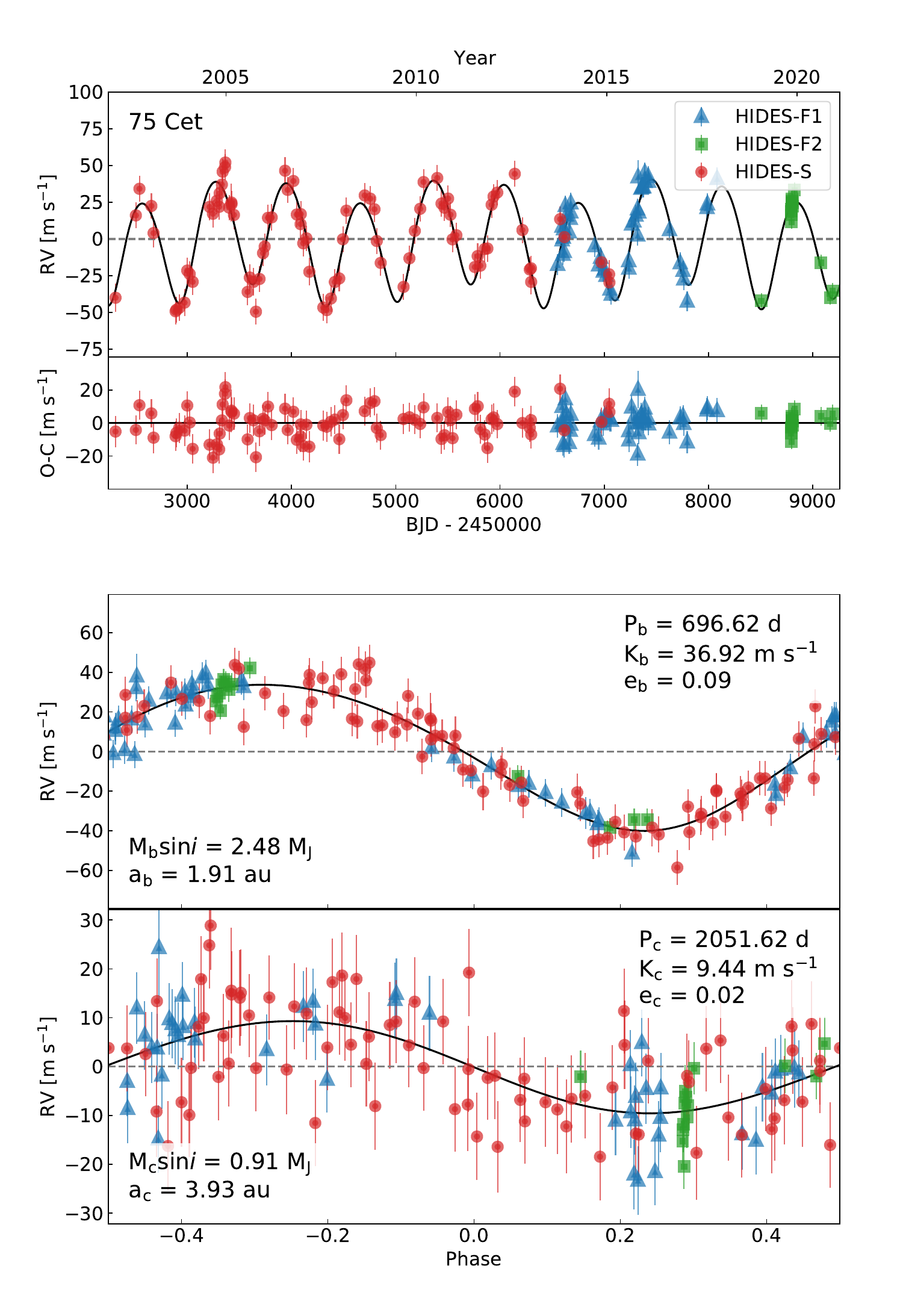}
\end{center}
\caption{
A double Keplerian solution to 75 Cet.
Top: Best fit double Keplerian curve in the full observation span, including fitted RV offsets between instruments and jitters that are included in the error bars. Mid: Residuals of the RVs with respect to the best-fit model. Bottom: phase-folded orbits of planet b and c. Data taken by HIDES-S, -F1, and -F2 using 1.88m Telescope at OAO are shown by red circles, blue triangles, and green squares, respectively.
}\label{fig:HD15779_phase}
\end{figure}

We thus applied a double Keplerian model, and the orbit fitting yielded $P_{\rm{b}} = 696.62_{-1.69}^{+1.33}\ \rm{d}$, $K_{\rm{b}} = 36.92_{-1.35}^{+1.09}\ \rm{m}\ \rm{s}^{-1}$, and $e_{\rm{b}} = 0.093_{-0.042}^{+0.026}$ for the inner known planet, and $P_{\rm{c}} = 2051.62_{-40.47}^{+45.98}\ \rm{d}$, $K_{\rm{c}} = 9.44_{-1.39}^{+1.04}\ \rm{m}\ \rm{s}^{-1}$, and $e_{\rm{c}} = 0.023_{-0.003}^{+0.191}$.
Adopting a stellar mass of $M_{\star}=1.92\ M_{\odot}$, we obtained the minimum masses and semimajor axes $M_{\rm{b}}\sin i=2.479\ M_{\rm{J}}$, $a_{\rm{b}}=1.912\ \rm{au}$, and $M_{\rm{c}}\sin i=0.912\ M_{\rm{J}}$, $a_{\rm{c}}=3.929\ \rm{au}$ respectively (Figure \ref{fig:HD15779_phase}). The rms scatter of residuals of the double Keplerian model was {suppressed to $10.45\ \rm{m}\ \rm{s}^{-1}$}, and the periodogram did not show any significant signal (Figure \ref{fig:HD15779_resid_ls}). 

We used the scaling relation in \citet{Kjeldsen1995} to estimate the expected p-mode oscillation, based on the stellar parameters derived in this study. The amplitude obtained was approximately $6.4\rm{m}\ \rm{s}^{-1}$, which provides a plausible explanation for the rms of the residuals to the double Keplerian fitting. 
Additionally, we searched for regular variability corresponding to 696 d and 2051 d in BIS,  which indicates Line profile, and Ca \emissiontype{II} index $S_{\rm{H}}$, but did not find any. Thus we confirm the presence of an outer planet in the 75 Cet system.

Interestingly, only considering the gas giant planets, the configuration of 75 Cet system is similar to our solar system. 
The mass ratio of the two gas giants in the 75 Cet system, assuming co-planarity, is $M_{\rm{b}}/M_{\rm{c}} = 2.72$, while their semimajor axis ratio is $a_{\rm{b}}/a_{\rm{S}} = 0.487$. Notably, these values are comparable to those of Jupiter and Saturn in our own solar system, with a mass ratio of $M_{\rm{J}}/M_{\rm{S}} = 3.34$ and a semimajor axis ratio of $a_{\rm{J}}/a_{\rm{S}} = 0.544$.

{75 Cet c, defined as a planet around evolved star, having a mass of approximately $1\ M_{\rm{J}}$ and an orbital separation of around $\gtrsim 4$ au, adds to a population of planets of which we currently have only two other examples with comparable properties:} BD+49 828 b, which orbits a giant star with a mass of $1.6 M_{\odot}$ at a distance of $4.2$ au and has a minimum mass of $1.6 M_{\rm{J}}$ \citep{Niedzielski2015b}, and HD 25723 c, which orbits a giant star with a mass of $2.12 M_{\odot}$ at a distance of $4.35$ au and has a minimum mass of $1.33 M_{\rm{J}}$ \citep{TalaPinto2020}. 
{The shortage of planets like 75 Cet c is currently unclear, and this might due to the selection effect because} these planets are always hard to be confirmed due to the necessity of decades-long observational baseline.

\subsection{RV long-term trend}
\begin{figure*}
\begin{center}
\includegraphics[scale=0.25]{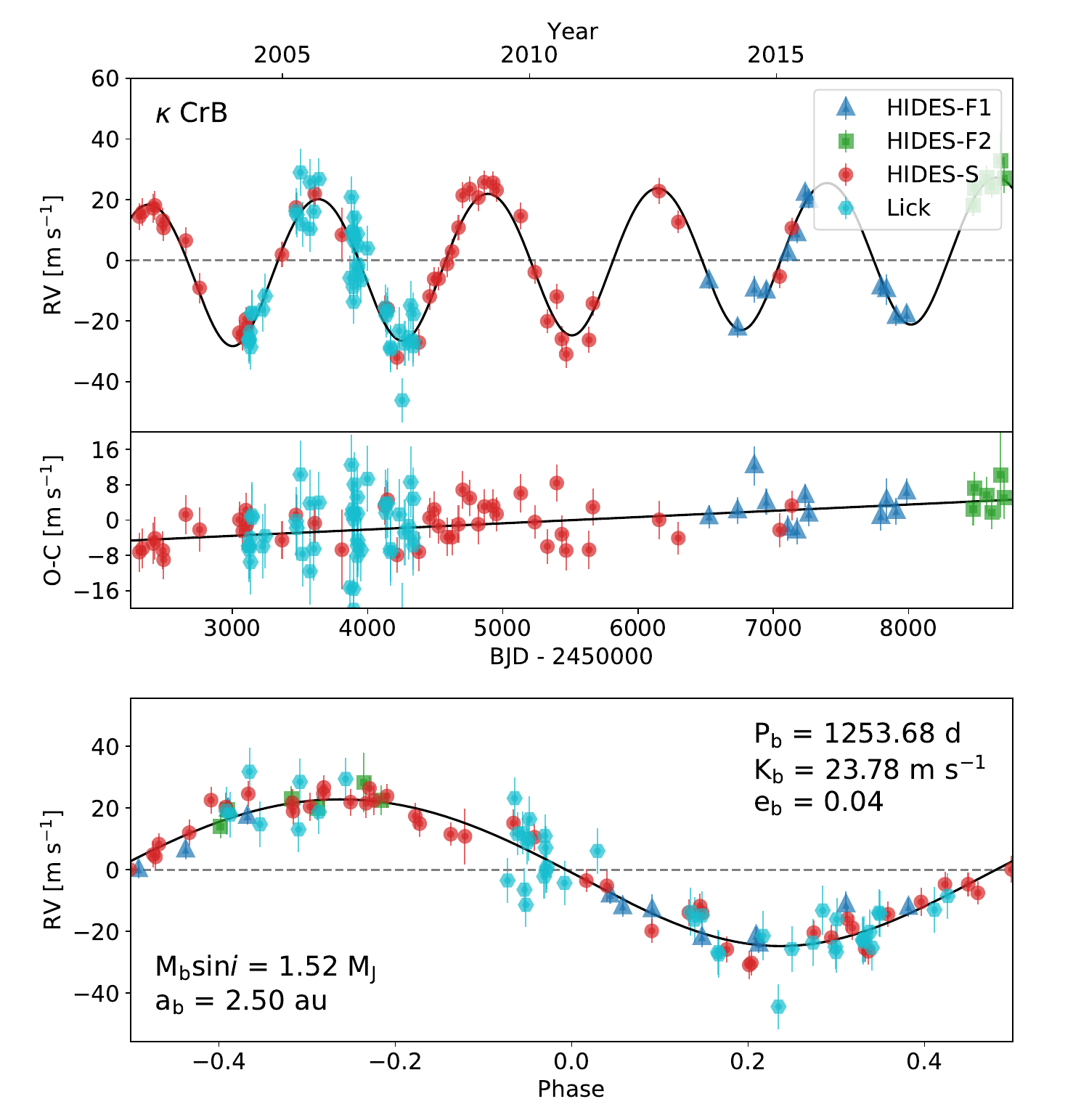}
\includegraphics[scale=0.25]{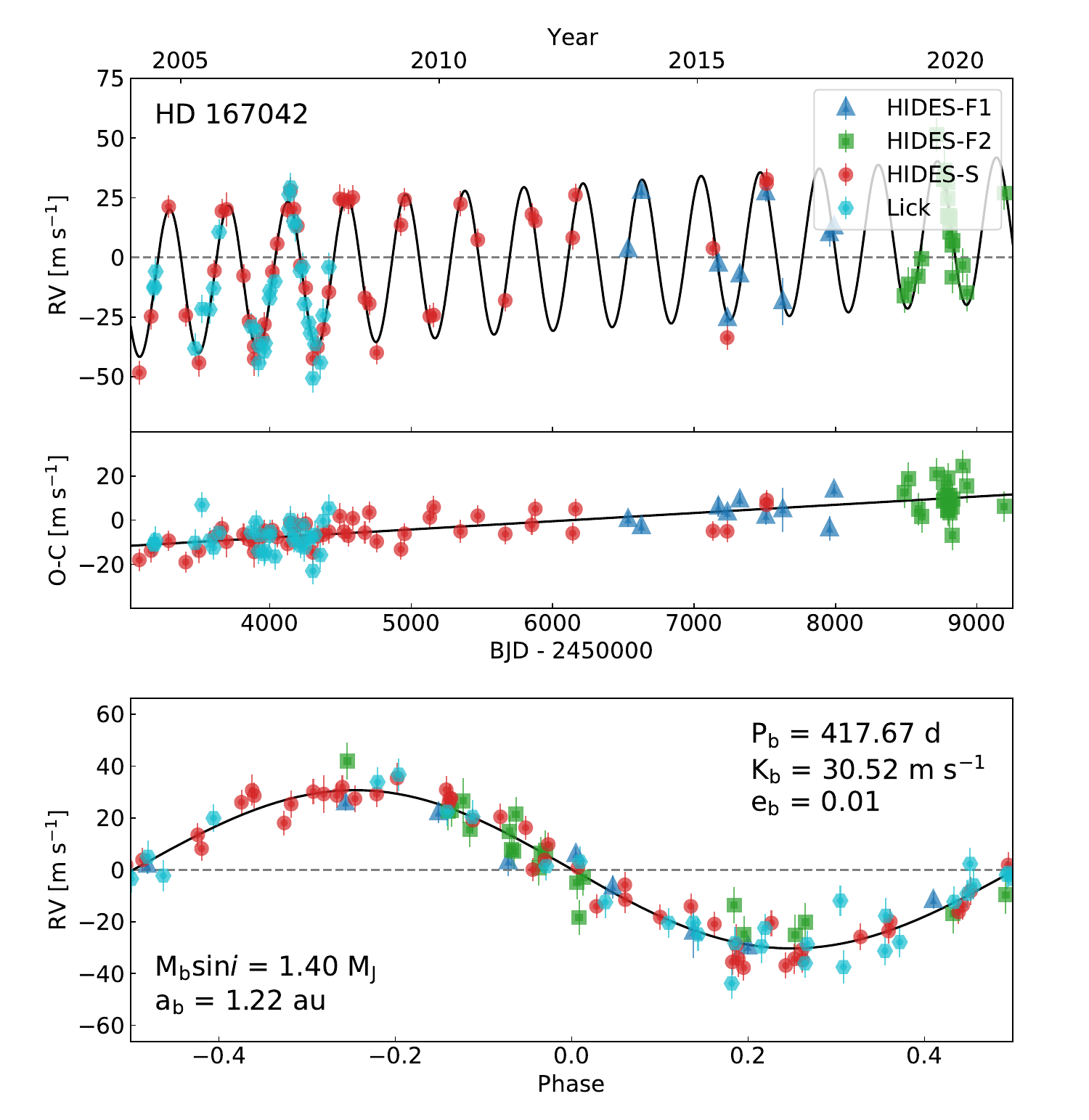}
\includegraphics[scale=0.25]{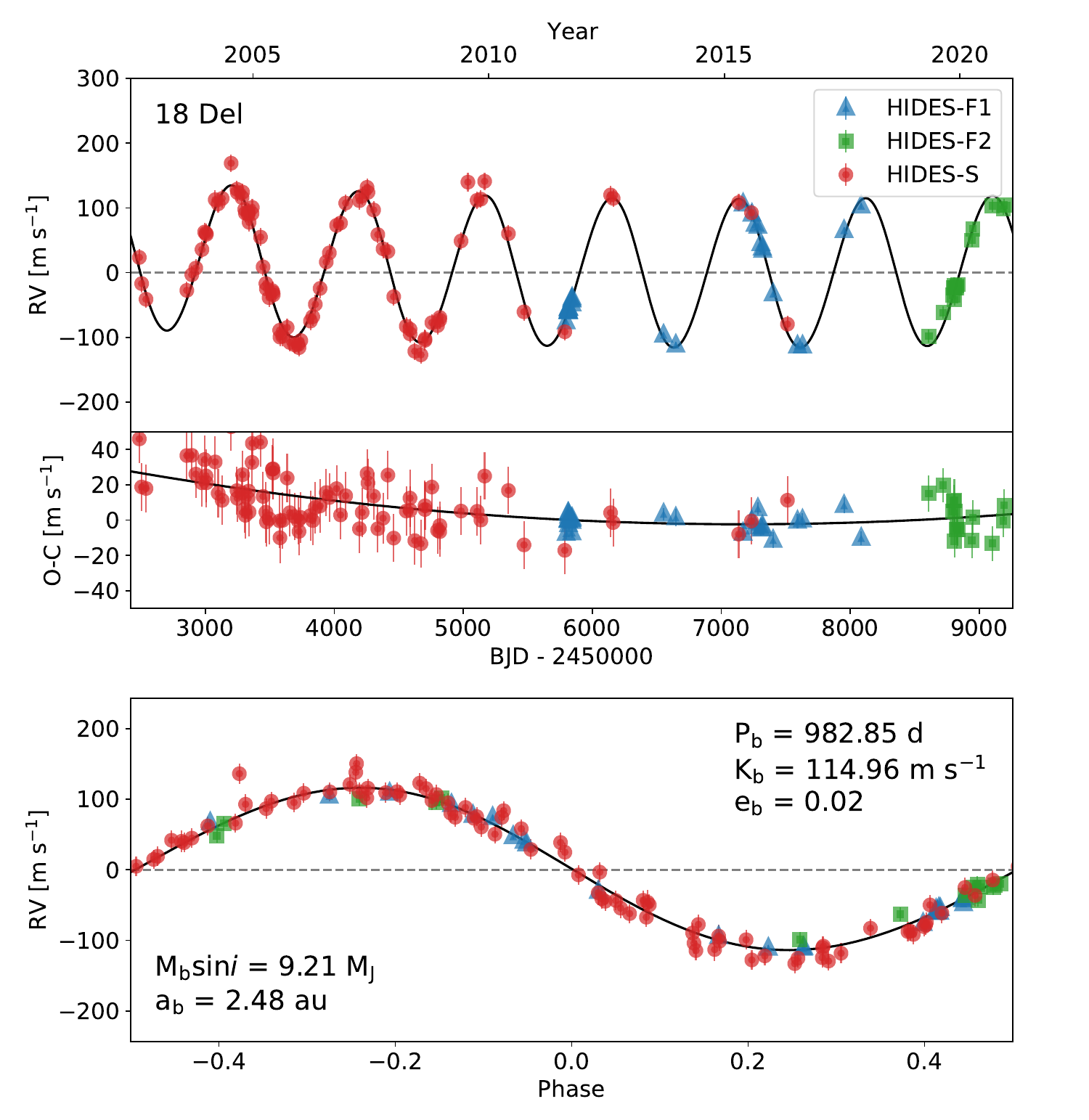}
\includegraphics[scale=0.25]{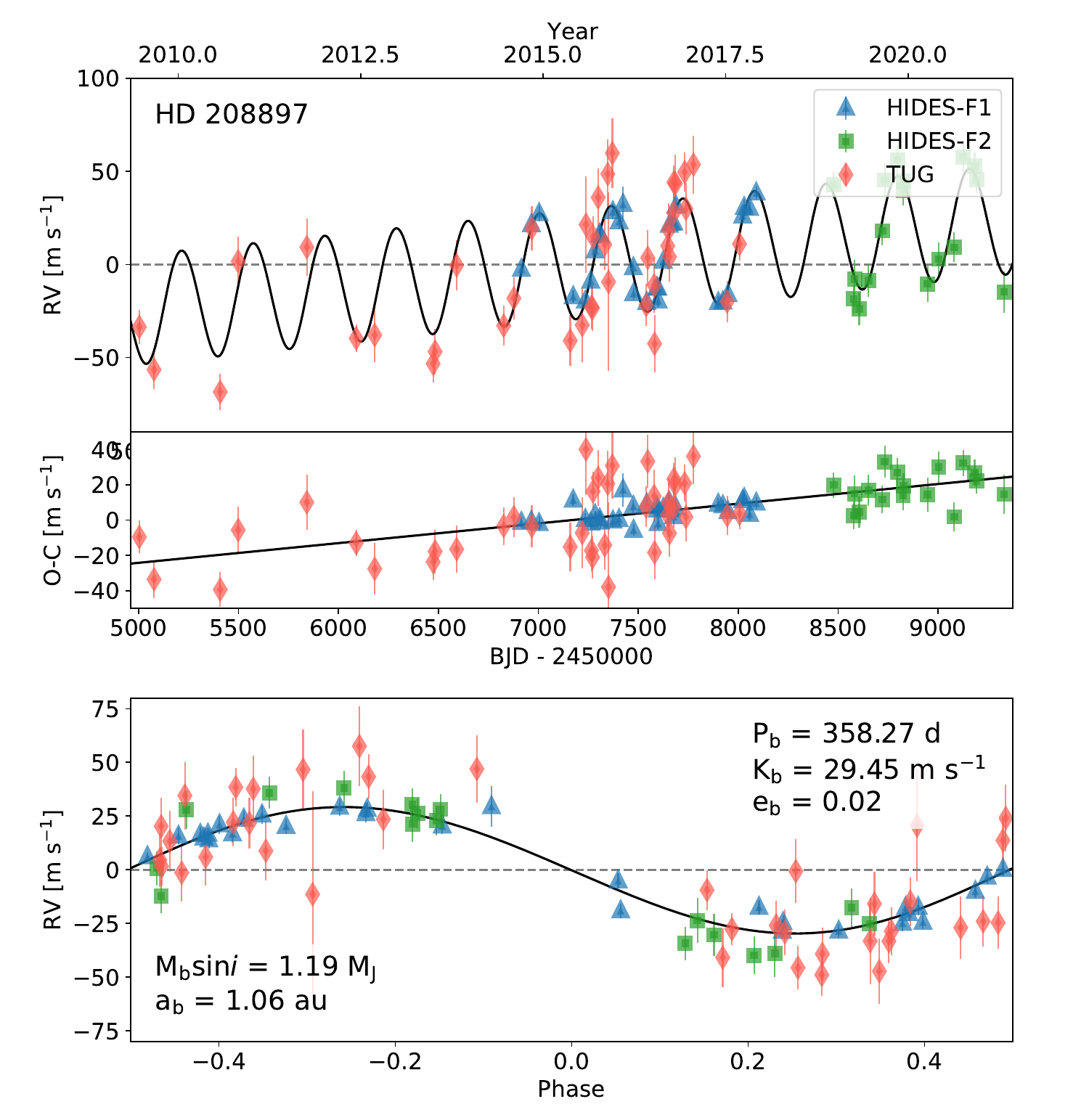}
\end{center}
\caption{
Keplerian orbital fit to $\kappa$ CrB, HD 167042, 18 Del, and HD 208897. 
(In each subplots) Top: Best fit 1-Keplerian curve in the full observation span, including fitted RV offsets between instruments and jitters that are included in the error bars. Mid: Residuals of the RVs with respect to the best-fit model. Bottom: phase-folded orbit of the planet. 
Data taken by HIDES-S, -F1, and -F2 using 1.88m Telescope at OAO are shown by red circles, blue triangles, and green squares, respectively.  
Data taken by the Hamilton Echelle Spectrograph on Coud\'{e} Auxiliary Telescope at Lick Observatory (Lick) are shown by cyan hexagons, 
while the data taken by the Coude Echelle Spectrograph on 1.5m Russian-Turkish Telescope at TUG are shown by bright red diamonds.
}\label{fig:HD142091_phase}
\end{figure*}

Our best fits showed that five stars have significant RV accelerations over decades, i.e., the RV long-term trend.
Four of the five, including HD 5608, $\kappa$ CrB, HD 167042, and HD 208897, showed a linear trend, {and their best-fit orbits are shown shown in Table \ref{tab:orbpar_orbits} and Figure \ref{fig:HD142091_phase})}. 
The slope of the RV trend can be used for estimating the minimum of dynamical mass of the outer body \citep{Winn2009}:
\begin{equation}
\label{eq:dvdt}
\frac{M_{\rm c}\sin i_{\rm c}}{a_{\rm c}^{2}} \sim \frac{\dot{\gamma}}{G} 
\end{equation}
where $\dot{\gamma}$ is the RV acceleration, $M_{\rm c}$ is the mass of the outer companion, $i_{\rm c}$ is the orbital inclination, $G$ is the gravitational constant, and $a_{\rm c}$ is the semimajor axis of the outer companion. 
Assuming (near) circular orbits for the outer companions, the RV linear acceleration trend could be assumed to be at least half of the orbit.
Our observation baselines are approximately 20 years, and the stars are more massive than 1 $M_{\odot}$. Thus, the prospective outer planets should reside over $\sim10$ au away from the central star. 

\begin{figure}
\begin{center}
\includegraphics[scale=0.25]{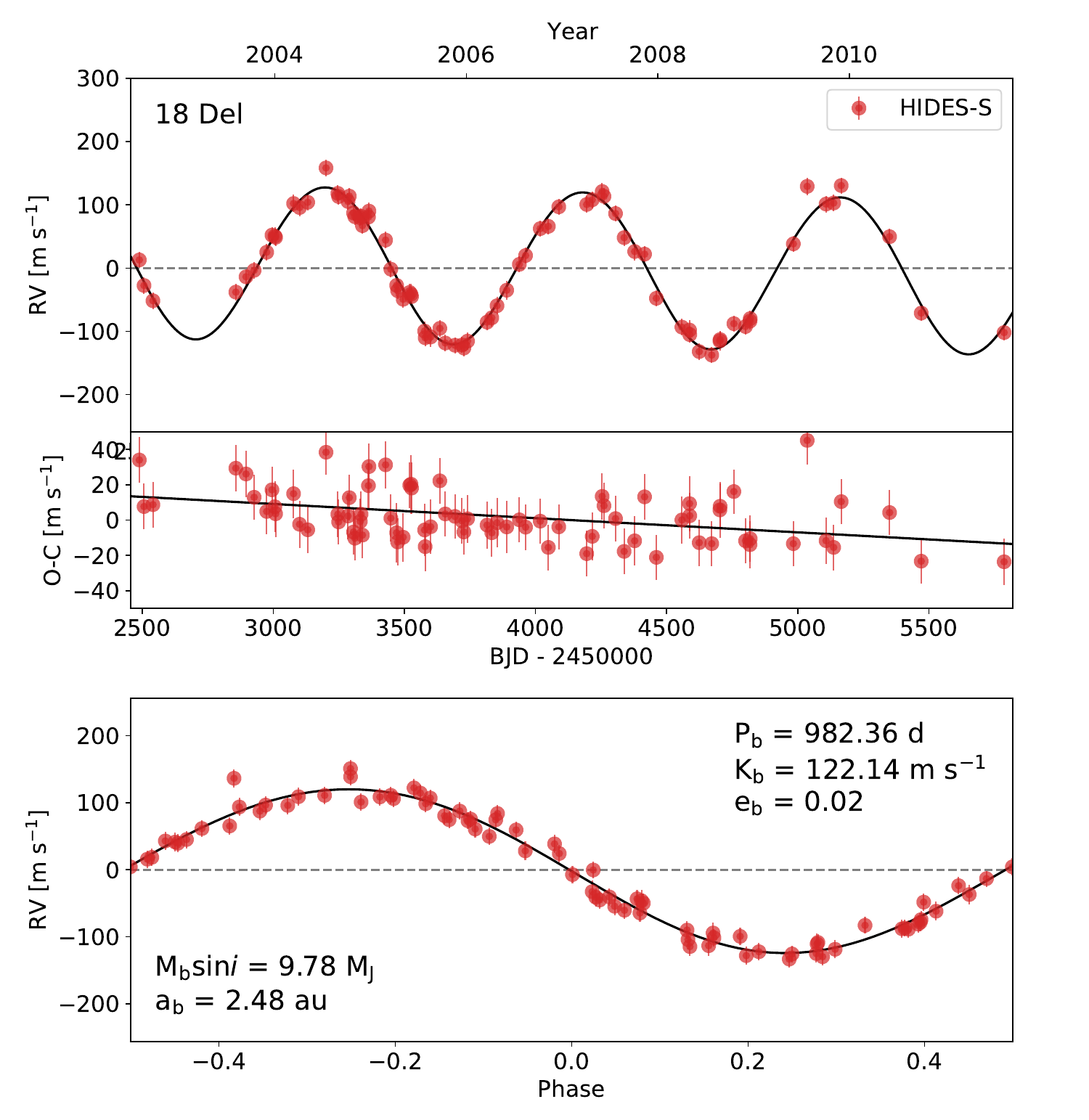}
\end{center}
\caption{
Single Keplerian solution to 18 Del by using RVs obtained by HIDES-S earlier than JD 2456000.
Top: Best fit single Keplerian curve
Mid: Residuals of the RVs with respect to the best-fit model.
Bottom: phase-folded orbit of the planet.
}\label{fig:HD199665_phase_hidess}
\end{figure}

Another one of the five, 18 Del, displayed a significant turning point {at around JD 2457000 in the RV curve, and we derived a best-fit orbit is shown in Table \ref{tab:orbpar_orbits} and Figure \ref{fig:HD142091_phase}}.
To estimate the mass of a prospective outer companion, a linear acceleration or deceleration should be determined. Thus, we divided the RV time series into acceleration and deceleration sides.
Notably, the deceleration side provided a longer baseline and contained more of data points and predominantly taken by HIDES-S.
{We performed a Keplerian orbital fit with HIDES-S data showing deceleration (observation earlier than JD 2456000), and it consequently yielded a value of $\dot{\gamma}=-2.922\ \rm{m\ s^{-1}\ yr^{-1}}$ (Figure \ref{fig:HD199665_phase_hidess})}.

\begin{figure}
\begin{center}
\includegraphics[scale=0.6]{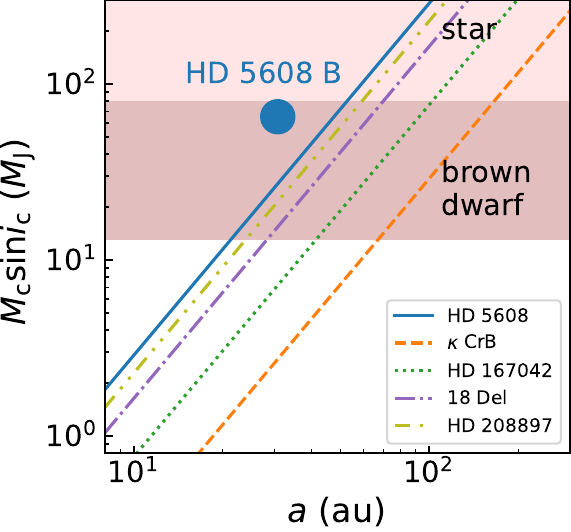}
\end{center}
\caption{
The mass of prospective outer companion as a function of the semimajor axis. HD 5608, $\kappa$ Crb, HD 167042, and HD 208897 are represented by different line styles, and specifically, we mark the confirmed companion HD 5608 B in this figure. The upper light shaded area indicates stellar regime, while the middle dark shaded area indicates brown dwarf regime. 
}\label{fig:rvtrend}
\end{figure}

Given the slopes, we could derive the $(M_{\rm c}\sin i_{\rm c})/{a_{\rm c}^{2}}$ are $-0.029$, $0.003$, $0.008$, $-0.016$, and $0.023$ for HD 5608, $\kappa$ Crb, HD 167042, 18 Del, and HD 208897, respectively. 
We present the correlation between mass and semimajor axis in Figure \ref{fig:rvtrend}. 

Among the five stars, HD 5608 has been confirmed to possess a low-mass stellar companion through imaging detection, as reported by \citet{Ryu2016}. 
Due to our intention to present a joint fit using radial velocity, relative astrometry, and direct imaging for this system in Section \ref{sec:astrometry}, we will not elaborate on it in this section.
{Conversely, the remaining four systems were not reported to have outer companions}, with $\kappa$ CrB and 18 Del subjected to imaging attempts by \citet{Bonsor2013} and \citet{Ryu2016}, respectively.

Despite the non-detection, some estimates on $\kappa$ CrB and 18 Del were given according to the observations.
In particular, \citet{Bonsor2013} used adaptive optics imaging and RV trend to impose a mass-separation constraint and put forth three possible scenarios for $\kappa$ CrB.
Moreover, they formulated three disk scenarios based on their constraint and resolved images from the \textit{Hershel} spacecraft.
However, it is possible that the previously established constraints for the potential outer companion were overestimated.
By incorporating our latest RV trend and updated stellar properties, we revised the estimation, which may in turn require modifications to the three proposed scenarios for $\kappa$ CrB:
\begin{itemize}
    \item All three scenarios require an outer companion with a minimum mass of 0.63 $M_{\rm{J}}$ and a semimajor axis wider than 14.7 AU.
    \item In scenario 1, which involves a wide belt ranging from 20 AU to 220 AU, the presence of an outer companion can only be supported within the narrow range of 14.7 to 20 AU.
    \item In scenario 2, which consists of two narrow belts at 40 AU and 165 AU, the survival of the outer companion depends on its location relative to the two belts. If it exists inside the inner belt, it must be situated between 14.7 and 40 AU. If it exists between the two belts, it must have a mass of at least 4.9 $M_{\rm{J}}$. Additionally, based on the RV trend and assuming an upper limit of 40 $M_{\rm{J}}$ for the outer companion's mass beyond 70 AU, we estimate that the outer companion must be situated within approximately 125 AU to avoid affecting the stability of the inner belt.
    \item In scenario 3, which involves a stirred disk with dust peaking at 83 AU, the disk can be stirred out to 53 AU by the inner body. We also apply equation 15 from \citet{Mustill2009} to demonstrate that the outer companion's influence on the disk is negligible if it is located very close to the host star, with a minimum mass of 0.63 $M_{\rm{J}}$ and a radial distance of 14.7 AU, and an orbit characterized by an extremely low eccentricity of less than 0.003. Although we cannot exclude this unlikely situation, the probability of it occurring is relatively low in this scenario.
\end{itemize}

18 Del was reported to harbor a distant companion 18 Del B \citep{Mugrauer2014} outside the field of view of HiCIAO frame \citep{Ryu2016}. 
The projected separation of 18 Del B is 2199 au and its mass is 0.19 $M_{\odot}$.
Based on the mass separation relation shown in Figure \ref{fig:rvtrend}, it can be confidently concluded that 18 Del B does not contribute to the observed trend.
An upper mass limit as a function of separation with an inner orbital boundary was obtained from imaging observation by \citep{Ryu2016}, 
Their results indicated that, if present, the companion should have a maximum mass of approximately $300 M_{\rm{J}}$ and a minimum mass of approximately $4 M_{\rm{J}}$ at 10 au for the inner most, and $90 M_{\rm{J}}$ at 50 au for the outer most, suggesting that it could be a high-mass planet, brown dwarf, or low-mass stellar companion.

\newpage
\section{Noteworthy systems}\label{sec:interest}
\subsection{$\epsilon$ Tau: Additional RV variations}
$\epsilon$ Tau (HD 28305, HR 1409, HIP 20889) is a K0 giant star in the nearest open cluster, the Hyades cluster, and its planet is the first one ever discovered in an open cluster. We started to collect data in 2003 December, and we released the discovery of the planet after 20 data in \citet{Sato2007}. 
We continued the follow-up observation with HIDES-S, -F1, and -F2 until 2021 March and totally obtained 386 spectra with RVs. 

We fitted the data with a single Keplerian model, and it yielded the orbital parameters for the companion of $P = 585.82_{-0.33}^{+0.26}\ \rm{d}$, $K = 93.24_{-0.73}^{+0.74}\ \rm{m}\ \rm{s}^{-1}$, and $e = 0.076_{-0.008}^{+0.009}$.
Adopting a stellar mass of $M_{\star}=2.57\ M_{\odot}$, we obtained the companion mass of $M_{\rm{p}}\sin i=7.194\ M_{\rm{J}}$ and semimajor-axis of $a=1.878\ \rm{au}$. 
The rms scatter of the single Keplerian fitting was $9.33\ \rm{m}\ \rm{s}^{-1}$.

\begin{figure*}
\begin{center}
\includegraphics[scale=0.5]{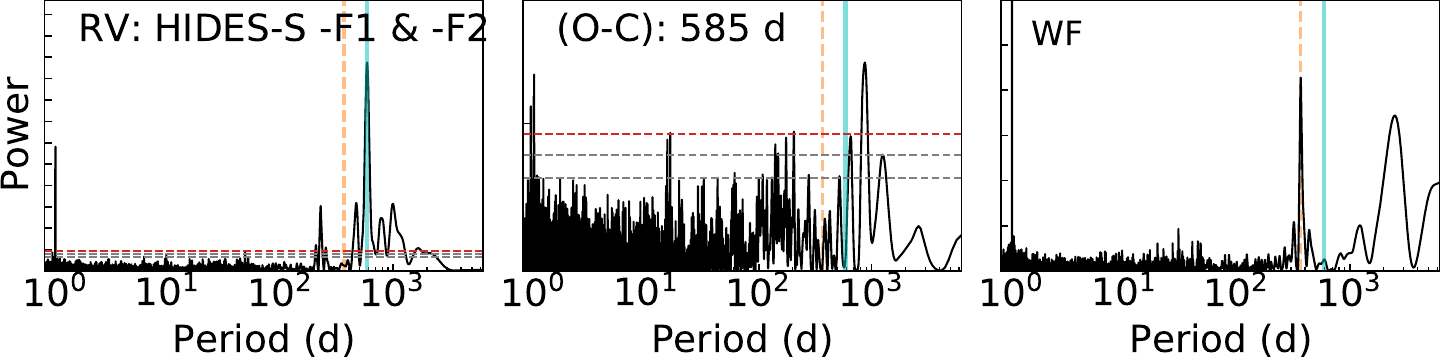}
\end{center}
\caption{
GLS periodograms for $\epsilon$ Tau. From left to right: The observed RVs, the RV residuals to the single Keplerian fitting of the known planet, and the window function of the observed RVs. 
The horizontal lines represent 10\%, 1\%, and 0.1\% FAP levels from bottom to top. The vertical cyan solid lines indicate the planetary signal and the vertical orange dashed line indicates 1 year. 
}\label{fig:HD28305_resid_ls}
\end{figure*}

\begin{figure}
\begin{center}
\includegraphics[scale=0.25]{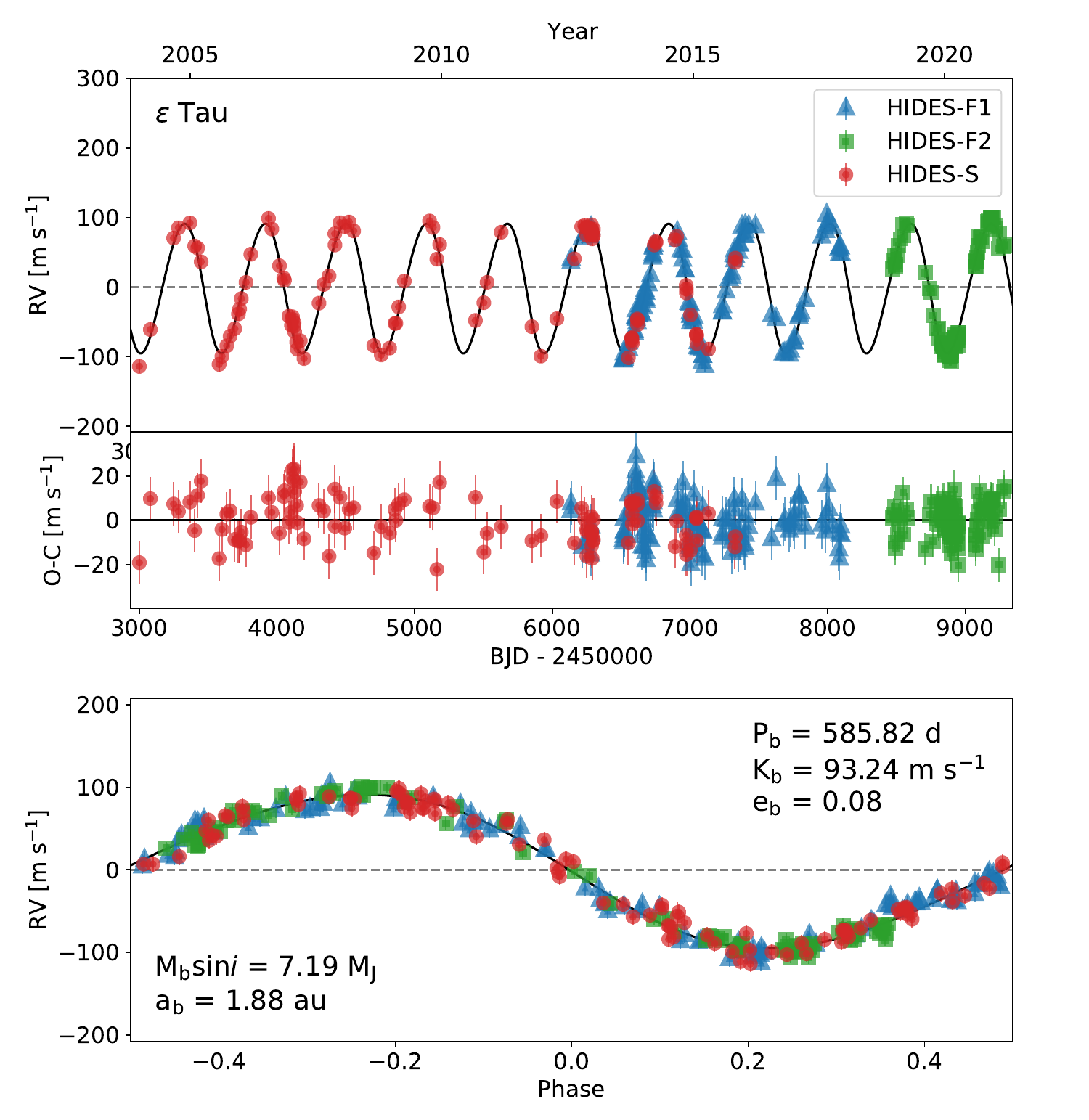}
\end{center}
\caption{
Single Keplerian solution to $\epsilon$ Tau.
Top: Best fit single Keplerian curve in the full observation span, including fitted RV offsets between instruments and jitters that are included in the error bars. Mid: Residuals of the RVs with respect to the best-fit model. Bottom: phase-folded orbit of the planet. Data taken by HIDES-S, -F1, and -F2 using 1.88m Telescope at OAO are shown by red circles, blue triangles, and green squares, respectively.
}\label{fig:HD28305_1pl_phase}
\end{figure}

\begin{figure}
\begin{center}
\includegraphics[scale=0.45]{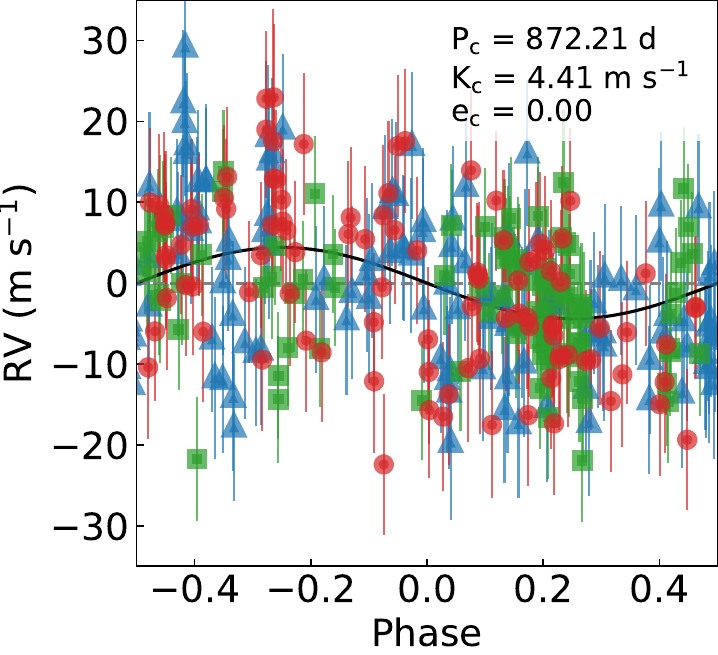}
\end{center}
\caption{
The phase-folded curve of the regular 872 d variation in $\epsilon$ Tau RVs. The symbols are the same as those in Figure \ref{fig:HD28305_1pl_phase}.
}\label{fig:HD28305_2pl_phase}
\end{figure}

As seen in Figure \ref{fig:HD28305_resid_ls}, the RV residuals to the single Keplerian fitting (Figure \ref{fig:HD28305_1pl_phase}) exhibit a distinct regular variation at 872 d with FAP apparently lower than 0.1\%, suggesting a possible outer companion in the system.
We thus performed a double Keplerian fit to the RV system, with the eccentricity of possible outer companion equal to 0. 
Consequently, we obtained $P = 872.21 \rm{d}$ and $K = 4.41\ \rm{m}\ \rm{s}^{-1}$ (Figure \ref{fig:HD28305_2pl_phase}), referring to minimum mass of $M_{\rm{p}}\sin i=0.326\ M_{\rm{J}}$ and semimajor-axis of $a = 2.28\ \rm{au}$. 
The rms scatter of 2-Keplerian model was {suppressed to $\sim 8.69\ \rm{m}\ \rm{s}^{-1}$ (decreased by $\sim 0.64\ \rm{m}\ \rm{s}^{-1}$)}. 

\begin{figure}
\begin{center}
\includegraphics[scale=0.45]{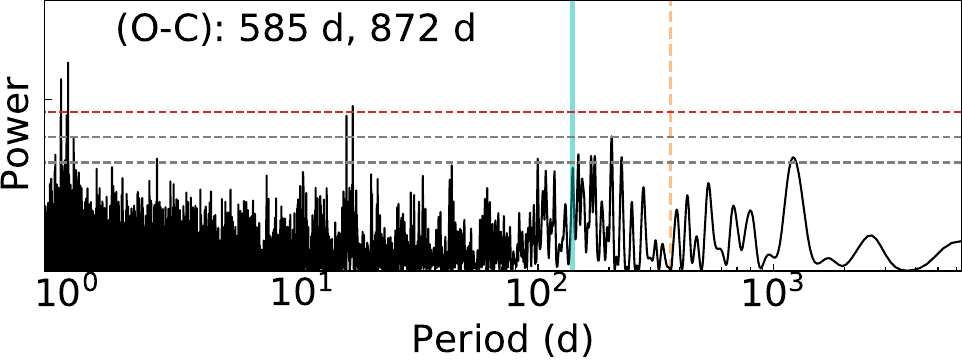}
\end{center}
\caption{
GLS periodograms of $\epsilon$ Tau RVs with the removal of the known planet and regular 872 d variation. 
The horizontal lines represent 10\%, 1\%, and 0.1\% FAP levels from bottom to top.
The vertical cyan solid line indicates the planetary signal, and the vertical orange dashed line indicates 1 year. 
}\label{fig:HD28305_2pl_resid_ls}
\end{figure}

After the subtraction of 872 d variation, another two independent signals, respectively at periods of 16 d and 15 d, remained in the residuals with FAP slightly lower than 0.1\% (Figure \ref{fig:HD28305_2pl_resid_ls}). 
They can be interpreted by sinusoidal curves with period of $P = 16.31\ \rm{d}$ and $P = 15.34\ \rm{d}$, and semi-amplitude of $K = 3.70\ \rm{m}\ \rm{s}^{-1}$ and $K = 3.33\ \rm{m}\ \rm{s}^{-1}$. 
With the subtraction of 16 d and 15 d variations, the rms scatter was further {suppressed to $\sim 7.83\ \rm{m}\ \rm{s}^{-1}$ (by $\sim 0.86\ \rm{m}\ \rm{s}^{-1}$)} compared to the double Keplerian model with $P = 585\ \rm{d}$ and $P = 872\ \rm{d}$.

The solar-like oscillation and instrumental jitter could be responsible for the remaining rms scatter. According to the scaling relation in \citet{Kjeldsen1995}, the oscillation amplitude for $\epsilon$ Tau is approximately $7\ \rm{m}\ \rm{s}^{-1}$. 
\citet{Ando2010} first detected oscillations in this star based on a few days of RV observations, \citet{Stello2017} then obtained consistent results with their data. 
\citet{Arentoft2019} reported intensive observations of the solar-like oscillation with over 4800 spectra. 
The filtered RV data in \citet{Arentoft2019} has rms scatter of $5.38\ \rm{m}\ \rm{s}^{-1}$. Considering the instrumental jitter of HIDES ($\lesssim 4.5\ \rm{m}\ \rm{s}^{-1}$: \cite{Teng2022a}; and this work), solar-like oscillation and instrumental jitter could be a plausible explanation to rms scatter.

\begin{figure*}
\begin{center}
\includegraphics[scale=0.45]{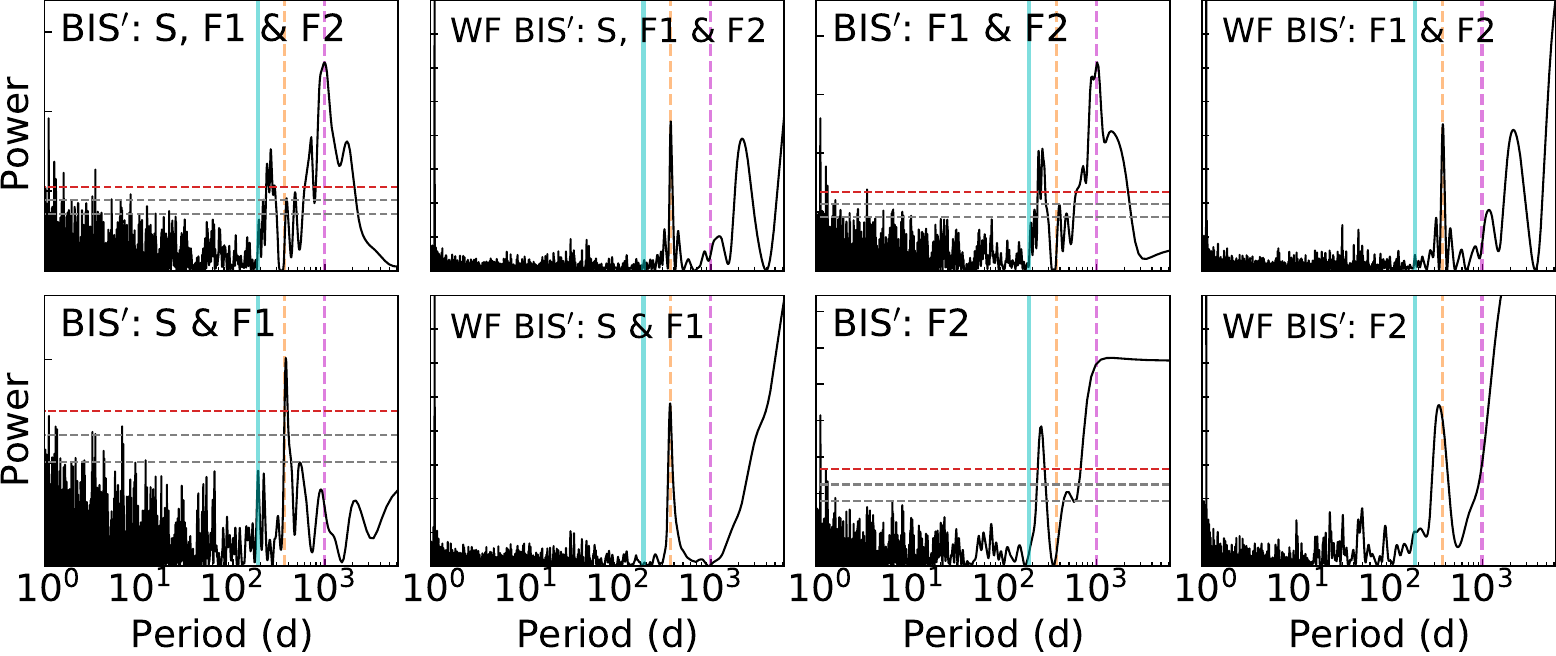}
\end{center}
\caption{
GLS periodograms of BIS$^{\prime}$ of $\epsilon$ Tau (on left panels) and the observational window functions (on right panels). BIS$^{\prime}$ calculated by data taken from HIDES-S, -F1, and -F2 are represented by S, F1, and F2, respectively. 
The horizontal lines represent 10\%, 1\%, and 0.1\% FAP level from bottom to top. The vertical cyan solid line indicates the planetary signal, the vertical orange dashed line indicates 1 year, and the vertical magenta dashed line indicates the BIS$^{\prime}$ signal introduced by HIDES-F2. 
}\label{fig:HD28305_bis_ls}
\end{figure*}

In order to investigate the true scenario for additional regular variations in the RV time series, we calculated the GLS periodogram to the BIS, BIS$^{\prime}$, and Ca \emissiontype{II} H index $S_{\rm{H}}$. 
The periodogram to BIS$^{\prime}$ exhibited a strong 990 d signal, which was then confirmed to be caused by HIDES-F2 (Figure \ref{fig:HD28305_bis_ls}). 
The periodogram to BIS$^{\prime}$ of HIDES-S and HIDES-F1 did not illustrate periodicity relevant to periods of RV variations. 
Therefore, 872 d variation, which exhibits the strong power in the periodogram to the 1-Keplerian model, can be considered a dubious planetary signal.
Yet, for the 16 d and 15 d variations, which show relatively weak power in the periodogram of the 2-Keplerian model, we prefer they are more likely to be instrumental or intrinsically from the star.

\begin{figure}
\begin{center}
\includegraphics[scale=0.45]{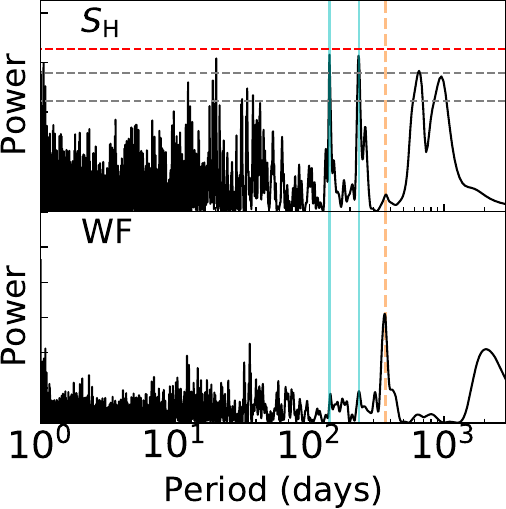}
\end{center}
\caption{
GLS periodograms of $S_{\rm{H}}$ of $\epsilon$ Tau. 
The horizontal lines represent 10\%, 1\%, and 0.1\% FAP level from bottom to top. The vertical cyan solid lines indicate the focused signals respectively at 141 d and 232 d in $S_{\rm{H}}$ time series, and the vertical orange dashed line indicates 1 year.
}\label{fig:HD28305_cahk_gls}
\end{figure}

The periodogram to $S_{\rm{H}}$ exhibited two possible periods with almost the same power, respectively, at 141 d and 232 d (Figure \ref{fig:HD28305_cahk_gls}).
Considering the strong window function effect at 1 yr, one should be the true signal and possibly refer to the rotation of the star, and the other one should be the alias ($1/141 - 1/232 \simeq 1/365$). 
Adopting the projected rotational velocity of $v \sin i = 3.52 \rm{km}\ \rm{s}^{-1}$ \citep{Takeda2008}, the maximum rotational period should be around 177 d. This suggests 141 d is more likely to be the true rotational period.

\begin{figure}
\begin{center}
\includegraphics[scale=0.45]{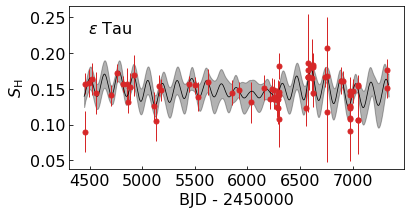}
\end{center}
\caption{
Gaussian Process (GP) results of $S_{\rm{H}}$ of $\epsilon$ Tau. The red circles with errorbars are $S_{\rm{H}}$ measurements taken by HIDES-S. The solid black line is the GP prediction, with its error shown by gray shade.
}\label{fig:HD28305_cahk_gp}
\end{figure}

\citet{Schroeder2020} investigated the magnetic activity of four giant stars in the Hyades cluster by Ca \emissiontype{II} H\&K emission. 
They estimated the rotational periods of three of them ($\gamma$ Tau, $\delta$ Tau and $\theta^{1}$ Tau) using Gaussian Process and adopted the rotational periods around 140 d for these three giants. 
However, they did not determine the  rotational period of $\epsilon$ Tau. 

In our work, we performed Gaussian Process (GP) to HIDES $S_{\rm{H}}$ data of $\epsilon$ Tau. 
We adopted the same GP Kernel in \citet{Schroeder2020}, {the \textit{celerite} approximation \citep{Foreman-Mackey2017},} but set a uniform prior between 100 d and 200 d to the period. 
{The formulation is given in the Equation \ref{eq:gp_rotation_kernel}}:
\begin{equation}
\label{eq:gp_rotation_kernel}
k(\tau) = \frac{B}{2+C} e^{-\tau/L} \left[ \cos\left(\frac{2 \pi \tau}{P_{\rm{rot}}}\right) + (1 + C) \right],
\end{equation}
where, $k$ is the covariance matrix, $\tau$ is the time difference, $B$ and $C$ are normalization constants, $L$ describes the lifetime of the features, $P_{\rm{rot}}$ is the desired rotational period.

Considering the inactiveness of Ca \emissiontype{II} H\&K emission with a low amplitude of $S_{\rm{H}}$ variation, {we by-eye set factor $C$ in the GP Kernel equal to 0}. 
As a result, we obtained a rotational period of 141.1 d (Figure \ref{fig:HD28305_cahk_gp}), which is close to the rotational periods of the other three giant stars, i.e., $\gamma$ Tau, $\delta$ Tau and $\theta^{1}$ Tau. 
Here, we also note that either 141 d or 232 d revealed by periodogram of $S_{\rm{H}}$ is contradictory to \citet{Schroeder2020}'s estimation of a minimal rotational period of 350 d based on scaling relation in \citet{Mittag2018}.

\begin{figure}
\begin{center}
\includegraphics[scale=0.45]{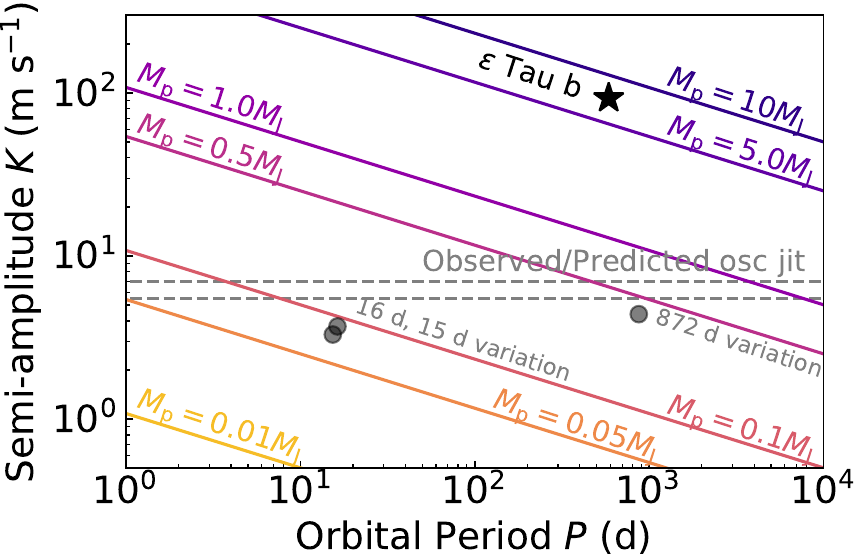}
\end{center}
\caption{
The detectability of a planet orbiting $\epsilon$ Tau. The colored solid lines indicate the orbital period against RV semi-amplitude of planets having different masses. The known planet is marked by a black star, and three regular variations in the RV residuals are marked by gray circles. The RV semi-amplitudes of observed and predicted oscillation jitter of $\epsilon$ Tau are marked by dashed gray lines. 
}\label{fig:HD28305_detect_limit}
\end{figure}

The long-baseline monitoring of the host star aims at searching for an additional companion. 
In Figure \ref{fig:HD28305_detect_limit}, we present RV semi-amplitude against the orbital period with hypothetical planets of different masses. 
Based on existing data, regular variation with semi-amplitude of $\sim 4\ \rm{m}\ \rm{s}^{-1}$ can be detected, and there is no long-term trend in the RV time series. 
Figure \ref{fig:HD28305_detect_limit} also suggests that we may not find any Jupiter-mass planet surviving around $\epsilon$ Tau by RVs. 
However, the instrumental jitter of HIDES is close to $4\ \rm{m}\ \rm{s}^{-1}$, therefore less-massive close-in planets, if they exist, are probably hidden in the jitter. 
In addition, for a giant star like $\epsilon$ Tau with $R_{\star} > 10 R_{\odot}$, photometry is also hard to detect a close-in planet, since the transit depth for a Jupiter-like planet would be lower than 0.0001. 

\newpage
\subsection{11 Com: Additional regular RV variations}
11 Com (HD 107383, HR 4697, HIP 60202) is a G8 III star and reported to harbor a substellar companion in \citet{Liu2008}. 
We started observations of 11 Com at OAO with HIDES-S in 2003 December. 
After its RV variations were detected, we then started observation of this star with CES-O at Xinglong in 2005 to confirm the variations of the star independently. 
We continued monitoring the star after the report of the companion at OAO with HIDES-S, -F1, and F2 until 2021 June and totally obtained 415 spectra with RVs.

We fitted the data with a single Keplerian model, and it yielded the orbital parameters for the companion of $P = 323.21_{-0.05}^{+0.06}\ \rm{d}$, $K = 288.63_{-2.37}^{+2.39}\ \rm{m}\ \rm{s}^{-1}$, and $e = 0.238_{-0.007}^{+0.007}$.
Adopting a stellar mass of $M_{\star}=2.09\ M_{\odot}$, we obtained the companion mass of $M_{\rm{p}}\sin i=15.464\ M_{\rm{J}}$ and semimajor-axis of $a=1.178\ \rm{au}$. 
The rms scatter of the single Keplerian fitting was $24.70\ \rm{m}\ \rm{s}^{-1}$. 

\begin{figure*}
\begin{center}
\includegraphics[scale=0.5]{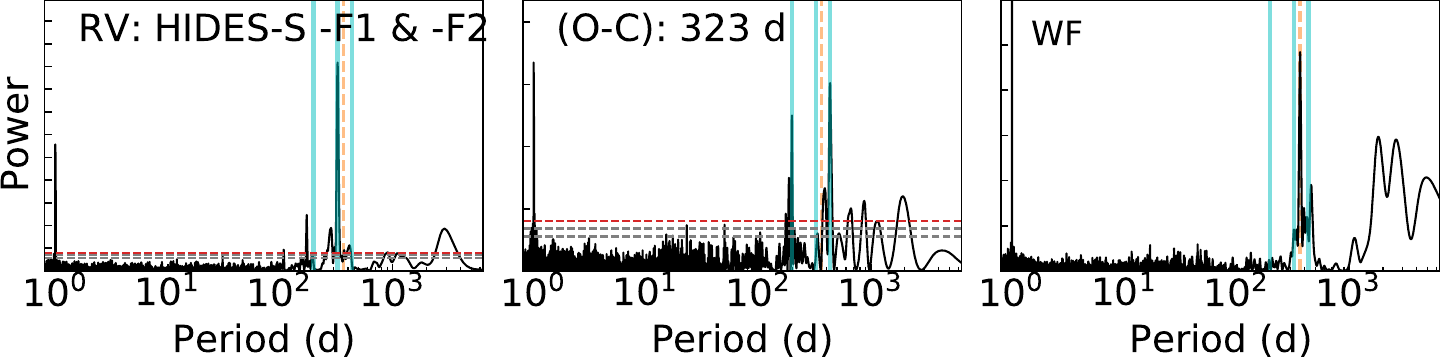}
\end{center}
\caption{
GLS periodograms for 11 Com. From left to right: The observed RVs, the RV residuals to the single Keplerian fitting of the known substellar companion, and the window function of the observed RVs. 
The horizontal lines represent 10\%, 1\%, and 0.1\% FAP levels from bottom to top. The vertical cyan solid line indicates the focused periodic signal, including a planetary signal at 323 d and two significant signals in the residuals at 198 d and 434 d  respectively, and the vertical orange dashed line indicates 1 year. 
}\label{fig:HD107383_resid_ls}
\end{figure*}

\begin{figure}
\begin{center}
\includegraphics[scale=0.25]{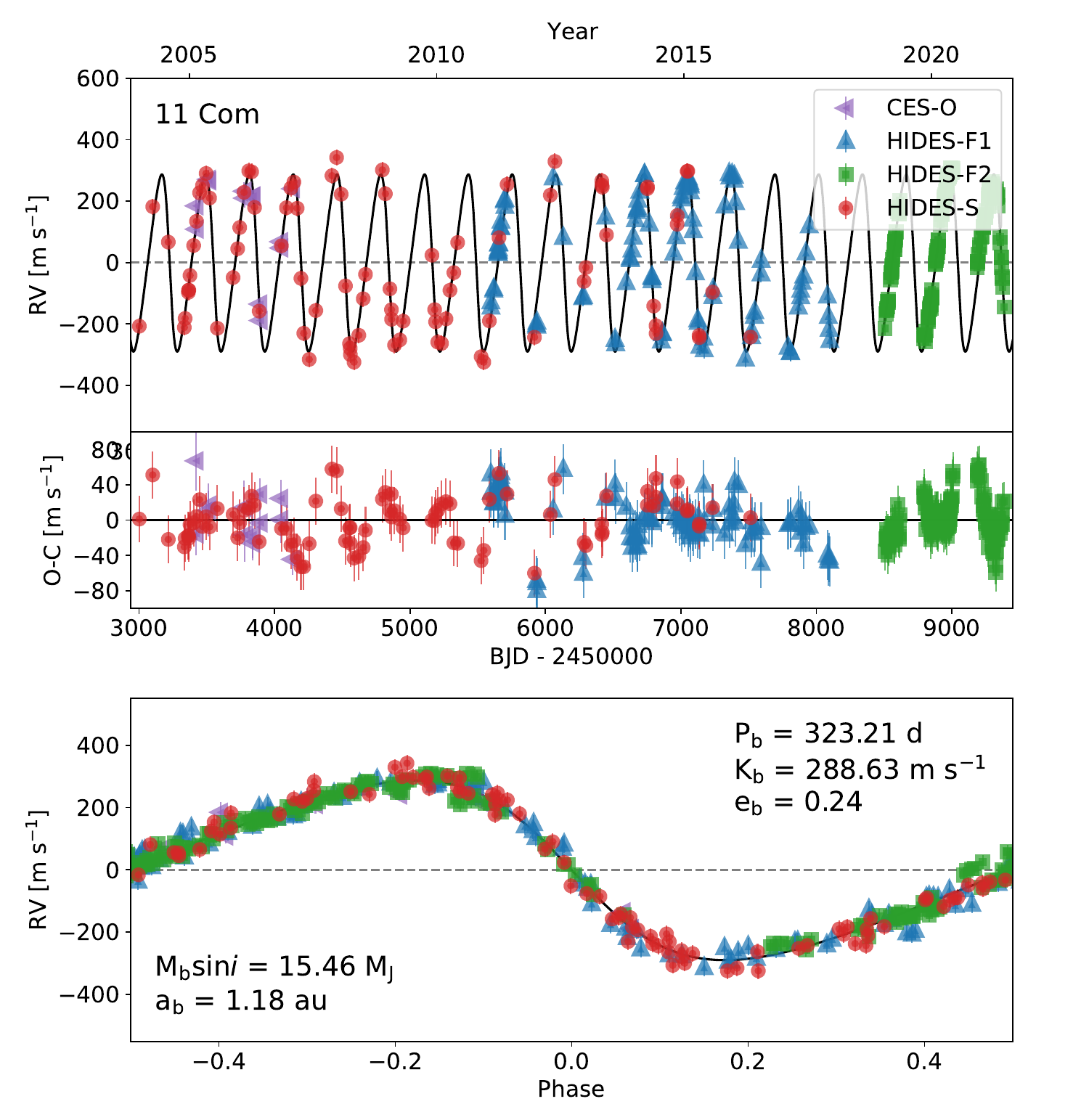}
\end{center}
\caption{
Single Keplerian solution to 11 Com.
Top: Best fit 1-Keplerian curve in the full observation span, including fitted RV offsets between instruments and jitters that are included in the error bars. Mid: Residuals of the RVs with respect to the best-fit model. Bottom: phase-folded orbit of the planet. Data taken by HIDES-S, -F1, and -F2 using 1.88m Telescope at OAO are shown by red circles, blue triangles, and green squares, respectively, and data taken by the CES pre-upgrade (CES-O) on 2.16m telescope at Xinglong Observatory are shown by purple triangles.
}\label{fig:HD107383_1pl_phase}
\end{figure}

\begin{figure}
\begin{center}
\includegraphics[scale=0.45]{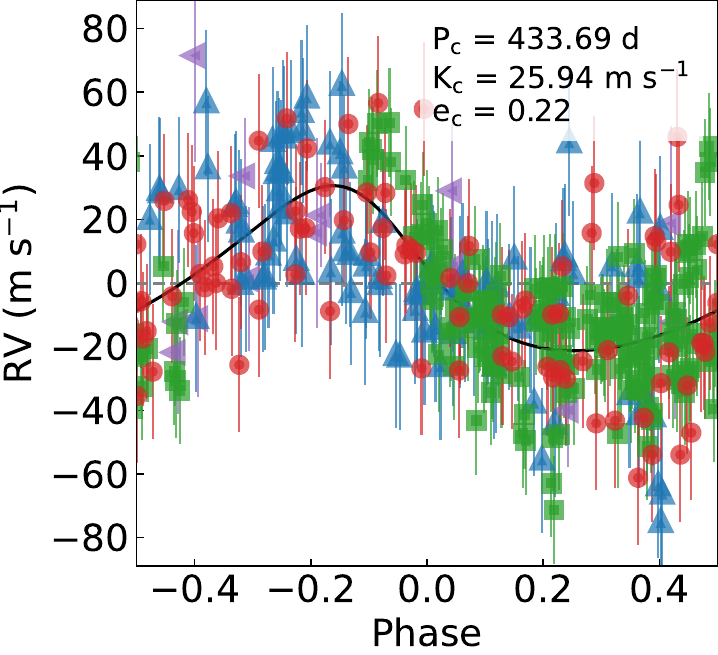}
\end{center}
\caption{
The phase-folded curve of the regular 434 d variation in 11 Com RVs. The symbols are the same as those in Figure \ref{fig:HD107383_1pl_phase}.
}\label{fig:HD107383_resid_phase}
\end{figure}

\begin{figure}
\begin{center}
\includegraphics[scale=0.5]{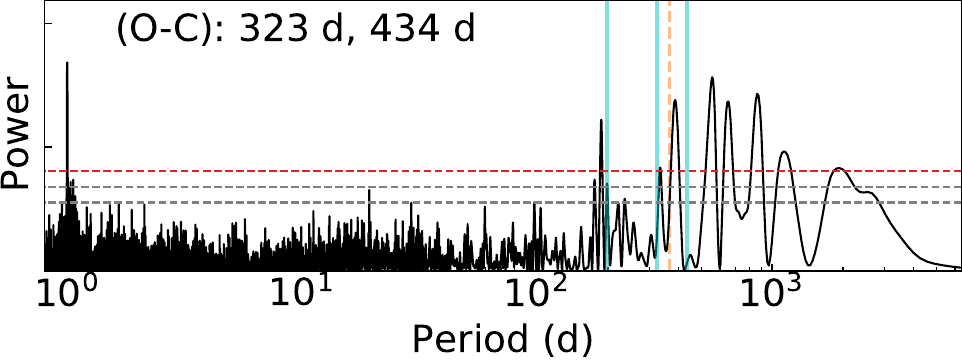}
\end{center}
\caption{
GLS periodograms of 11 Com RVs with removal of the known substellar companion and regular 434 d variation. 
The horizontal lines represent 10\%, 1\%, and 0.1\% FAP level from bottom to top.
The vertical cyan solid line indicates the focused signals including a planetary signal at 323 d and two significant signals in the residuals respectively at 198 d and 434 d, and the vertical orange dashed line indicates 1 year. 
}\label{fig:HD107383_2pl_resid_ls}
\end{figure}

As illustrated in Figure \ref{fig:HD107383_resid_ls}, the RV residuals to the single Keplerian fitting (Figure \ref{fig:HD107383_1pl_phase}) exhibited periodic variations at 198 d and 434 d, suggesting possible inner or outer companion(s). 
Considering the strong power at 1-year in the window function, 198 d signal could be the alias of 434 d signal ($1/198 \simeq 1/434 + 1/365$). 
We thus applied a double Keplerian fit with the known companion fixed to its best-fit orbital parameters. 
Consequently, the rms scatter was {suppressed to $20.21\ \rm{m}\ \rm{s}^{-1}$ (decreased by $4.49\ \rm{m}\ \rm{s}^{-1}$)}, and we obtained $P = 433.69\ \rm{d}$, $K = 25.94\ \rm{m}\ \rm{s}^{-1}$, $e = 0.220$, $M_{\rm{p}}\sin i=1.513\ M_{\rm{J}}$ and $a=1.43\ \rm{au}$ (Figure \ref{fig:HD107383_resid_phase}). 
Nevertheless, prominent signals persist in the periodogram of the residuals at periods of several hundred days (Figure \ref{fig:HD107383_2pl_resid_ls}).

However, the double planet scenario can be hardly feasible since the strong interaction between the two companions could make the system dynamically unstable. 
We tested the dynamical stability with \texttt{MERCURY6} integrator code  \citep{Chambers1999}. 
The integrator was initiated using the RV-derived orbital parameters of the inner companion, under the assumption of the co-planarity and prograde motion of the outer companion.
Consequently, the system always dissipated quickly and never kept stable from the RV-derived initials.
The inner companion (11 Com b) is highly massive which could clear up the neighboring area so that the existence of either an outer planet or an inner companion with short separation should be doubtful.
To comprehensively investigate the stability of the system, it is necessary to perform simulations across a wider parameter space. However, this analysis will be presented in a subsequent publication.

\begin{figure}
\centering
\includegraphics[scale=0.45]{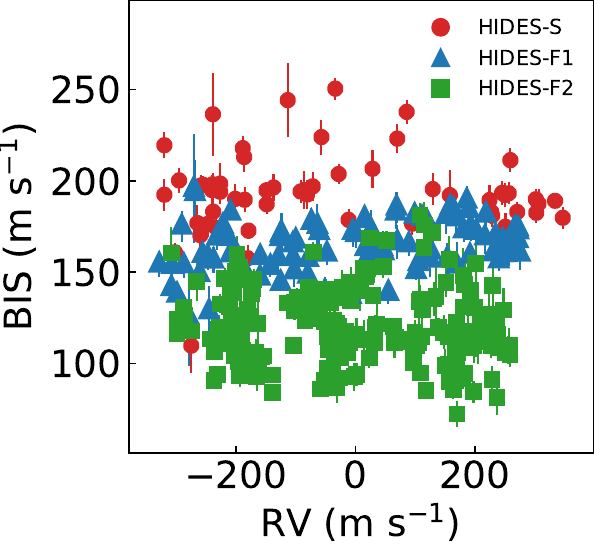}
\caption{
BIS against RVs of 11 Com. The data are taken by HIDES-S, -F1, and -F2 using 1.88m Telescope at OAO. The symbols are the same as those in Figure \ref{fig:HD107383_1pl_phase}. }\label{fig:HD107383_bis_rv}
\end{figure}

We also analyzed the line shape deformation by calculating BIS, BIS$^{\prime}$, and the S-index of Ca \emissiontype{II} lines, and their periodograms. 
As a result, no significant periodicity was found. 
And as seen in Figure \ref{fig:HD107383_bis_rv} the BIS of different instruments are averagely on different levels (HIDES-S: $\overline{\rm{BIS}}= 193.48\ \rm{m}\ \rm{s}^{-1}$, HIDES-F1: $\overline{\rm{BIS}}= 164.67\ \rm{m}\ \rm{s}^{-1}$, HIDES-F2: $\overline{\rm{BIS}}= 120.84\ \rm{m}\ \rm{s}^{-1}$), 
indicating that the observed line deformation was primarily driven by the instrumental profile rather than the star's intrinsic activity.
The star's projected rotational velocity is $v\sin i = 1.75\rm{km}\ \rm{s}^{-1}$ \citep{Takeda2008}. Adopting a stellar radius of $R_{\star} = 13.76 R_{\odot}$, we derived a maximum rotational period of $P_{\rm{rot}} = 397.9 \rm{d}$, which is slightly shorter than the RV variation of 434 d.

More scenarios can be responsible for regular RV variation and mimic planetary signals. Long secondary periods (LSP: \cite{Saio2015}) in red giant stars, probably caused by oscillatory convective mode, can be monitored through the RV method. 
They usually show photometric variations up to 1 mag and RV variations of a few $\rm{m}\ \rm{s}^{-1}$ in luminous ($L > 300 L_{\odot}$) giant stars \citep{Saio2015}. 
A new form of stellar variability, whose mechanism is not fully understood, was detected in K-giants $\gamma$ Dra (K5 III, $B-V=1.53$, \cite{Hatzes2018}) and Aldebaran (K5 III, $B-V=1.54$, \cite{Reichert2019}). 
They have RV semi-amplitude over 100 $\rm{m}\ \rm{s}^{-1}$, and they are likely to occur stars having larger $B-V$ values. 
However, 11 Com has photometric stability down to $\sigma=0.006$ mag and $B-V=0.99$ \citep{ESA1997}, which is distinct from the observed properties of LSP. 
\citet{Heeren2021} released the RV variations of $\epsilon$ Cyg A, which pretended to be a planetary signal but is likely an extreme example of a heartbeat binary system. 
The system has a close star separation of 15 au.
Although 11 Com A has a stellar companion, the heartbeat can be neglected due to the large separation of 999.0 au \citep{Roell2012}.

An asteroseismology analysis was provided by \citet{Ando2010}. From intensive RV monitoring, we estimated the RV amplitude of $\sim 10\ \rm{m}\ \rm{s}^{-1}$ caused by solar-like p-mode oscillation. Using the scaling relation in \citet{Kjeldsen1995}, we estimated the p-mode oscillation jitter amplitude of about $10.7\ \rm{m}\ \rm{s}^{-1}$, which has a good agreement with the observation. Although the oscillation should be responsible for the large RV scatter of the residuals, it is not the cause of long-period (i.e., a few hundred days) variations since the oscillation time scale is only a few hours.

Therefore, the complicated RV variations in the RV time series remain mysterious, and so far, we cannot conclude a self-consistent explanation for them.
Follow-up works may lift the veil of the truth, and they will appear in the forthcoming paper.

\newpage
\subsection{24 Boo: Two planet candidates}
24 Boo is a G3 IV star and reported to harbor a close-in giant planet in \citet{Takarada2018}. 
We started observation of 24 Boo at OAO in 2003 April. 
By 13-year monitoring with HIDES-S and -F1, we confirmed and released the planet in 2018. 
Since a 150-d variation remained in the residual of the single Keplerian model, we continued intensive observation of this star with HIDES-F1 and -F2. 
Until 2021 June, we have totally obtained 282 spectra with RVs. 
A single Keplerian model fitting yielded
$P = 30.33_{-0.01}^{+0.00} \ \rm{d}$, $K = 55.67_{-2.66}^{+2.31}\ \rm{m}\ \rm{s}^{-1}$, and $e = 0.032_{-0.023}^{+0.039}$. 
Adopting a stellar mass of $M_{\star}=1.05\ M_{\odot}$, we can derive the companion mass of $M_{\rm{p}}\sin i=0.883_{-0.043}^{+0.036}\ M_{\rm{J}}$ and semimajor-axis of $a = 0.194\ \rm{au}$. 
The rms scatter of the single Keplerian fit was $27.81\ \rm{m}\ \rm{s}^{-1}$. 

\begin{figure*}
\begin{center}
\includegraphics[scale=0.5]{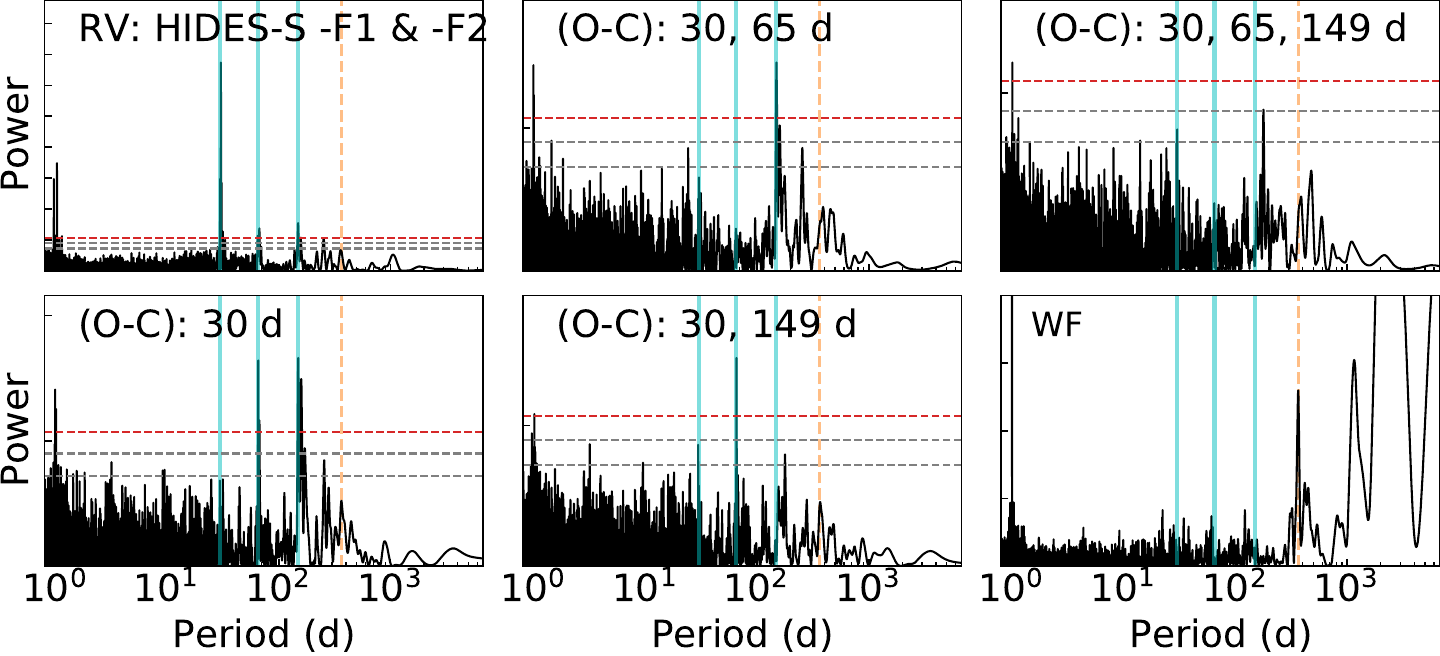}
\end{center}
\caption{
GLS periodograms for 24 Boo. 
Top left: The observed RVs. 
Bottom left: The RV residuals to the single Keplerian fitting of the known planet. 
Top middle: The RV residuals with the removal of the known planet and regular 65 d variation.
Bottom middle: The RV residuals with the removal of the known planet and regular 149 d variation.
Top right: The RV residuals to the 3-Keplerian fitting to the known planet, 65 d, and 149 d variations. 
Bottom right: The window function of the observed RVs. 
The horizontal lines represent 10\%, 1\%, and 0.1\% FAP levels from bottom to top. The vertical cyan solid lines indicate the focused periodic signals including a planetary signal at 30 d and two significant signals in the residuals respectively at 65 d and 149 d, and the vertical orange dashed line indicates 1 year. 
}\label{fig:HD127243_resid_ls}
\end{figure*}

\begin{figure}
\begin{center}
\includegraphics[scale=0.25]{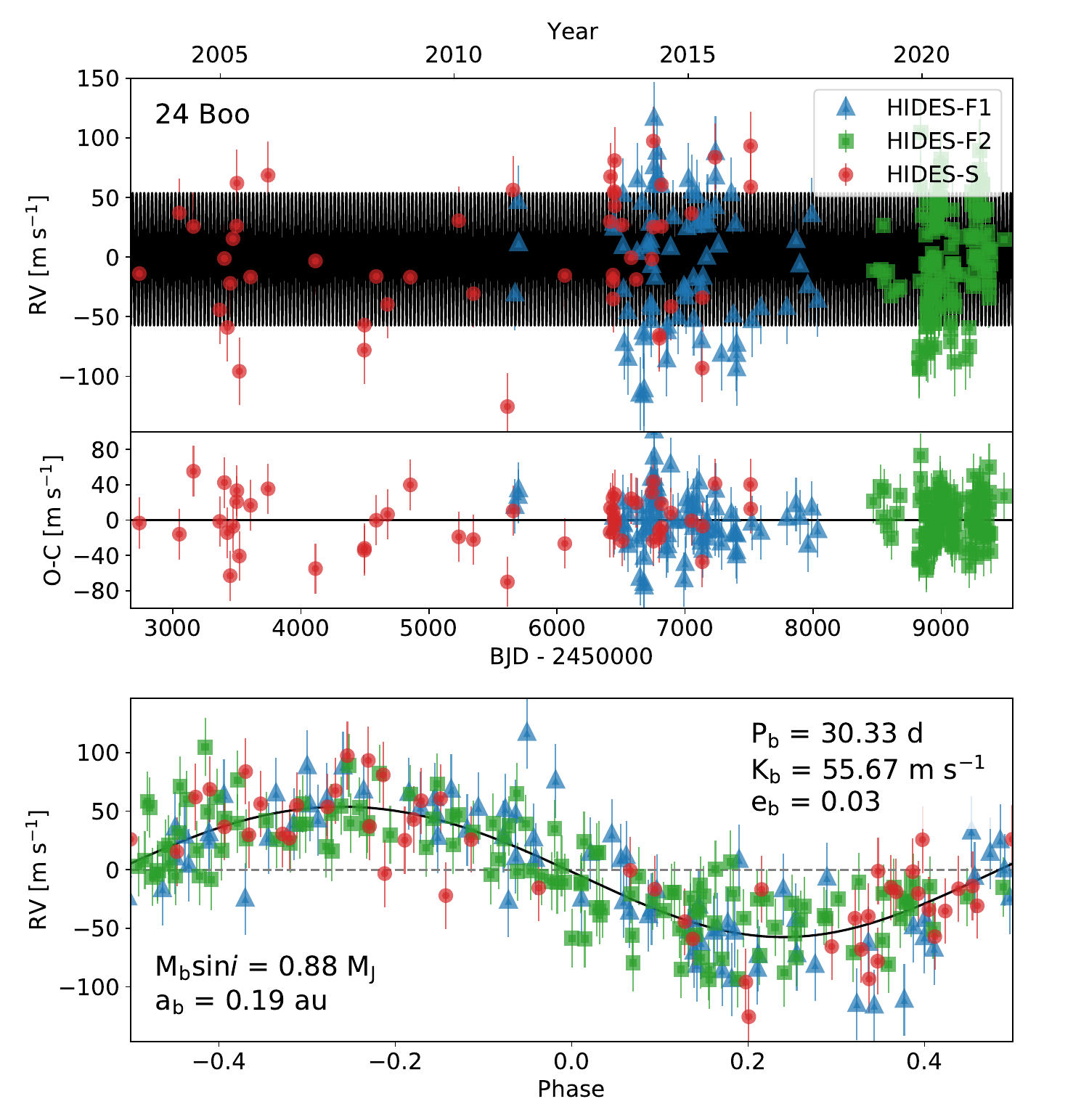}
\end{center}
\caption{
Single Keplerian solution to 24 Boo.
Top: Best fit single Keplerian curve in the full observation span, including fitted RV offsets between instruments and jitters that are included in the error bars. Mid: Residuals of the RVs with respect to the best-fit model. Bottom: the phase-folded orbit of the planet. Data taken by HIDES-S, -F1, and -F2 using 1.88m Telescope at OAO are shown by red circles, blue triangles, and green squares, respectively.
}\label{fig:HD127243_1pl_phase}
\end{figure}

\begin{figure*}
\begin{center}
\includegraphics[scale=0.45]{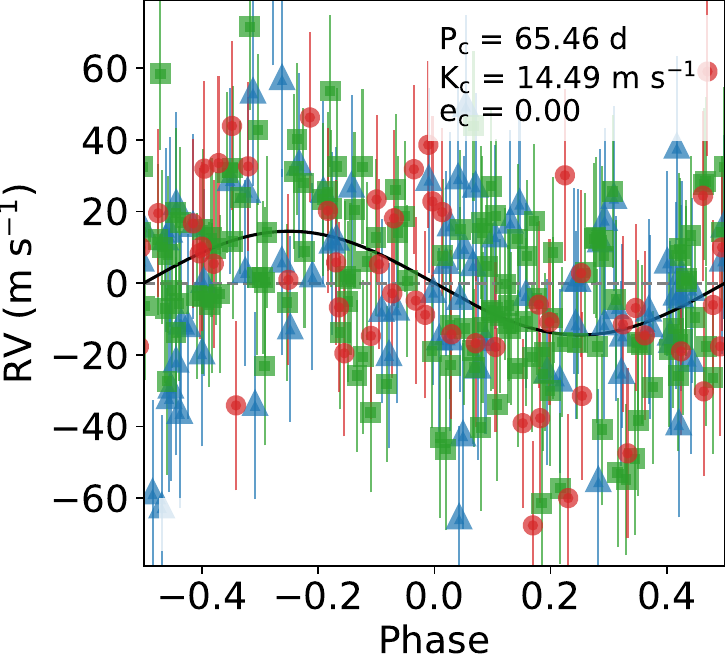}
\includegraphics[scale=0.45]{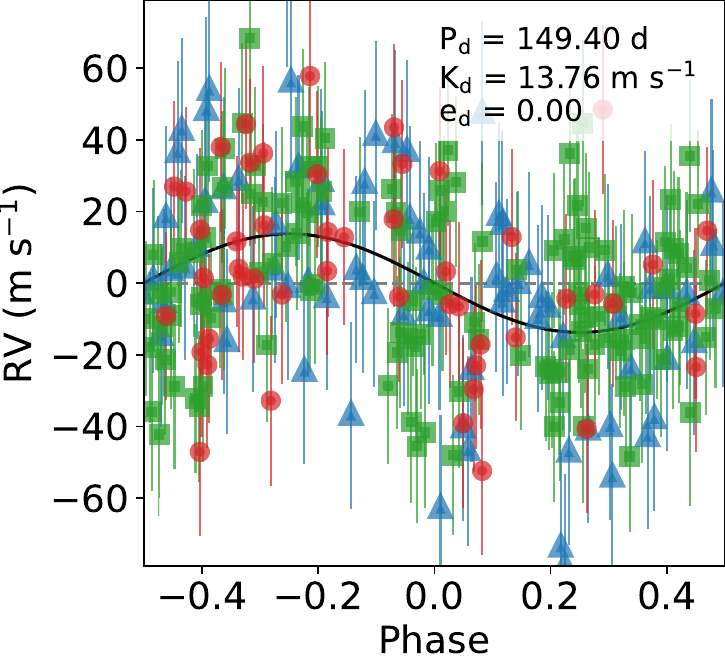}
\end{center}
\caption{
The phase-folded curves of the regular 65 d (left) and 149 d (right) variation in 24 Boo RVs. The symbols are the same as those in Figure \ref{fig:HD127243_1pl_phase}
}\label{fig:HD127243_resid_phase}
\end{figure*}

As seen in Figure \ref{fig:HD127243_resid_ls}, the RV residuals to the single Keplerian fitting exhibit regular variations at 65 d and 149 d. 
Considering large rms scatter in residuals, these variations also suggested possible outer companion(s). 
We then separately subtracted 65 d and 149 d signals by sinusoidal curve fitting, and the periodogram of residuals showed the other signal. 
We thus performed a triple Keplerian fitting with the known planet fixed and the eccentricities of prospective planets fixed to 0. 
Consequently, the rms scatter was {suppressed to $23.59\ \rm{m}\ \rm{s}^{-1}$ (decreased by $4.22\ \rm{m}\ \rm{s}^{-1}$)}, and we obtained $P_{\rm{c}} = 65.46\ \rm{d}$, $K_{\rm{c}} = 14.49\ \rm{m}\ \rm{s}^{-1}$, and $P_{\rm{d}} = 149.40\ \rm{d}$, $K_{\rm{c}} = 13.76\ \rm{m}\ \rm{s}^{-1}$. 
The semi-amplitudes of these variations are lower than rms scatter. 
If these variations are truly introduced by planets, it could be one planet with $M_{\rm{p}}\sin i=0.30\ M_{\rm{J}}$ and $a=0.32\ \rm{au}$, and another one planet with $M_{\rm{p}}\sin i=0.37\ M_{\rm{J}}$ and $a=0.56\ \rm{au}$ (Figure \ref{fig:HD127243_resid_phase}). 

We also calculated the line profiles and Ca \emissiontype{II} index, yet we did not find any periodicity related to 30 d, 65 d, and 149 d. 
As the star is slowly rotating with $v\sin i \lesssim 3.36 \rm{km}\ \rm{s}^{-1}$, we derived its maximum rotational period with a value of about 184 d, which is longer than either 65 d and 149 d. 
In addition, since these two signals are weak compared to the large jitter and scatter, we cannot validate the existence of the second and the third planet at this stage.

Furthermore, we note that the 24 Boo system is almost a twin of the HD 167768 system \citep{Teng2022c} in a few aspects. 
The central stars are both metal-poor ($\rm{Fe/H} \lesssim -0.7$) and deeply-evolved RGB stars ($\log g < 2.5$ cgs) in the galactic thick disk \citep{Takeda2008}, and they both harbor a close-in giant planet with two possible planet candidates at wider orbits.

\newpage
\subsection{41 Lyn and 14 And: Synchronous BIS variations}
41 Lyn (HD 81688, HR 2743, HIP 46471) is classified as a K0 III-IV star and reported to have a planet in \citet{Sato2008a}. 
The first spectrum was collected in 2003 March with HIDES-S, and we continued monitoring the star with HIDES-S, F1, and -F2 after the announcement of the planet until 2021 April.
The best-fit Keplerian model yields a near circular orbit with $P=183.93_{-0.09}^{+0.09}\ \rm{d}$, $K = 56.42_{-1.88}^{+1.87}\ \rm{m}\ \rm{s}^{-1}$, and $e = 0.040_{-0.031}^{+0.022}$. The rms scatter of RV residuals was $20.17\ \rm{m}\ \rm{s}^{-1}$. 
Adopting a stellar mass of $M_{\star}=1.07\ M_{\odot}$, we obtained the companion mass of $M_{\rm{p}}\sin i=1.654\ M_{\rm{J}}$ and semimajor-axis of $a=0.648\ \rm{au}$.  

14 And (HD 221345, HR 8930, HIP 116076) is classified as a K0 III star and reported to host a planet in \citet{Sato2008b}. 
The first spectrum was collected in 2004 January with HIDES-S, and we continued monitoring the star with HIDES-S, F1, and -F2 after the announcement of the planet until 2021 May.
The RV variation can be well fitted by a circular orbital with $P = 186.76_{-0.12}^{+0.11}\ \rm{d}$ and $K = 86.08_{-2.95}^{+2.76}\ \rm{m}\ \rm{s}^{-1}$. 
The rms scatter of RV residuals was $21.28\ \rm{m}\ \rm{s}^{-1}$. Adopting a stellar mass of $M_{\star}=1.78\ M_{\odot}$, we obtained the companion mass of $M_{\rm{p}}\sin i = 3.559\ M_{\rm{J}}$ and semimajor-axis of $a = 0.775\ \rm{au}$.
\begin{figure*}
\begin{center}
\includegraphics[scale=0.25]{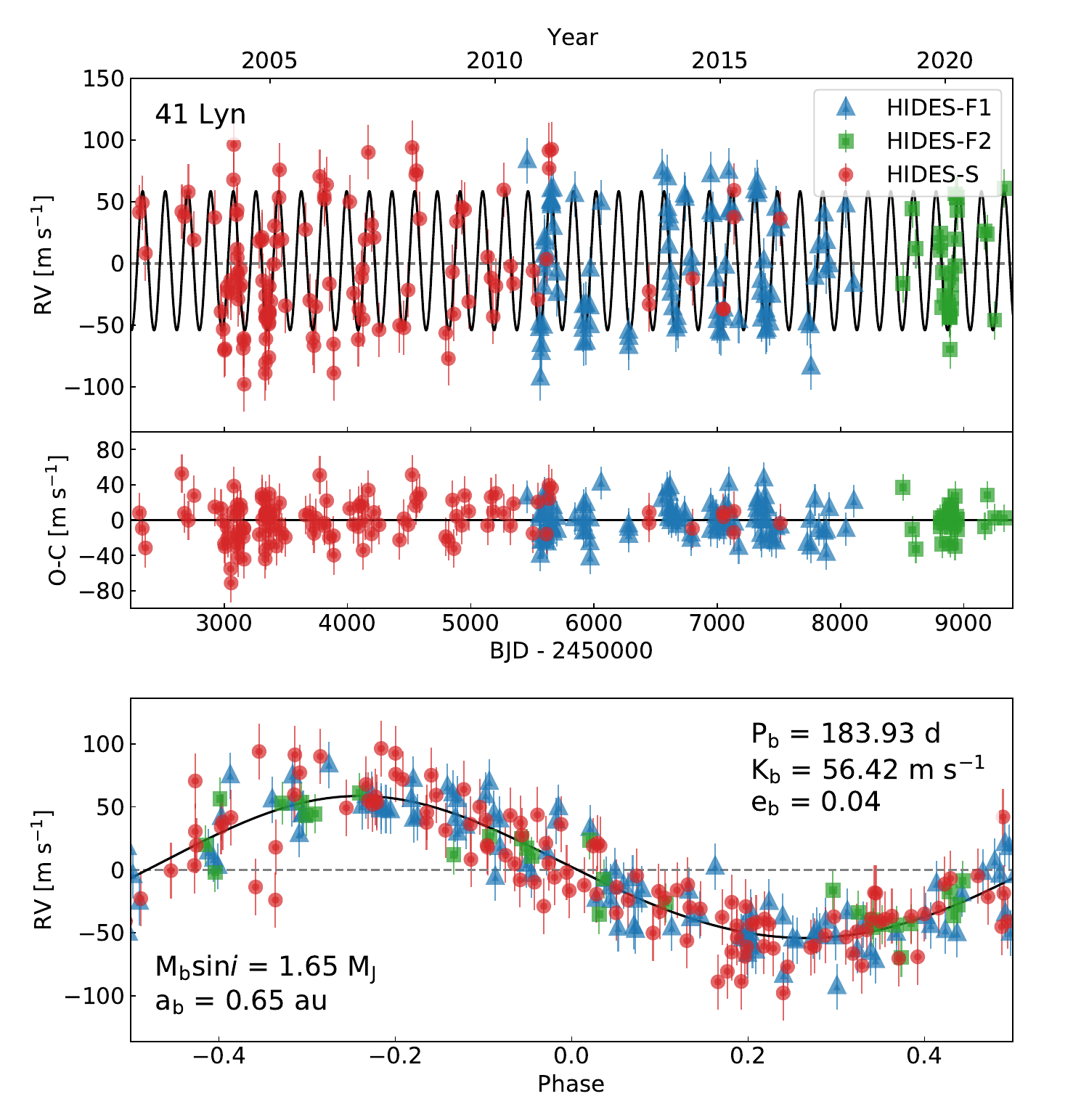}
\includegraphics[scale=0.25]{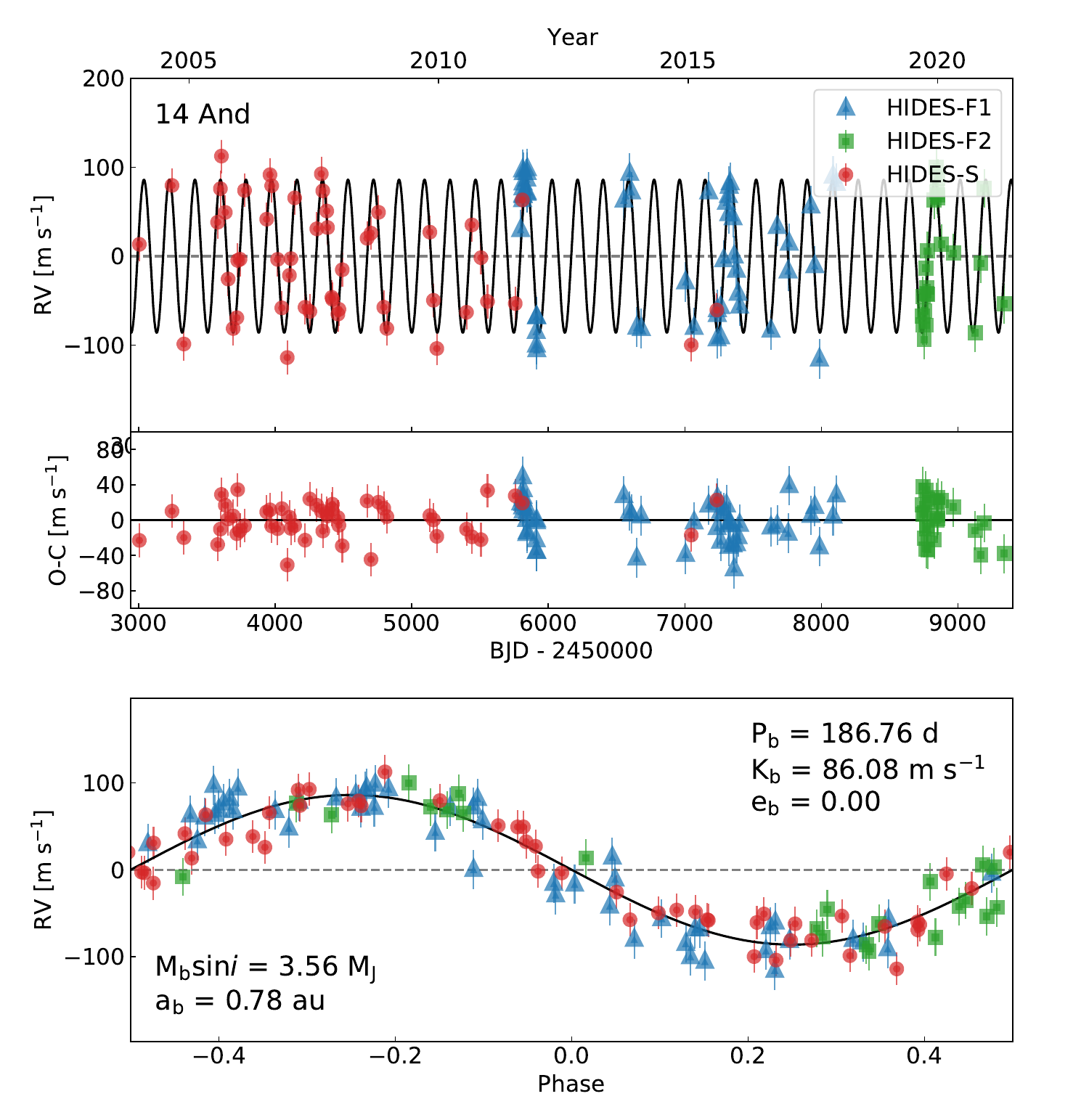}
\end{center}
\caption{
Single Keplerian solution to 41 Lyn (left) and 14 And (right).
Top: Best fit 1-Keplerian curve in the full observation span, including fitted RV offsets between instruments and jitters that are included in the error bars. Mid: Residuals of the RVs with respect to the best-fit model. Bottom: the phase-folded orbit of the planet. Data taken by HIDES-S, -F1, and -F2 using 1.88m Telescope at OAO are shown by red circles, blue triangles, and green squares, respectively.
}\label{fig:HD81688_1pl_phase}
\end{figure*}

\begin{figure*}
\begin{center}
\includegraphics[scale=0.5]{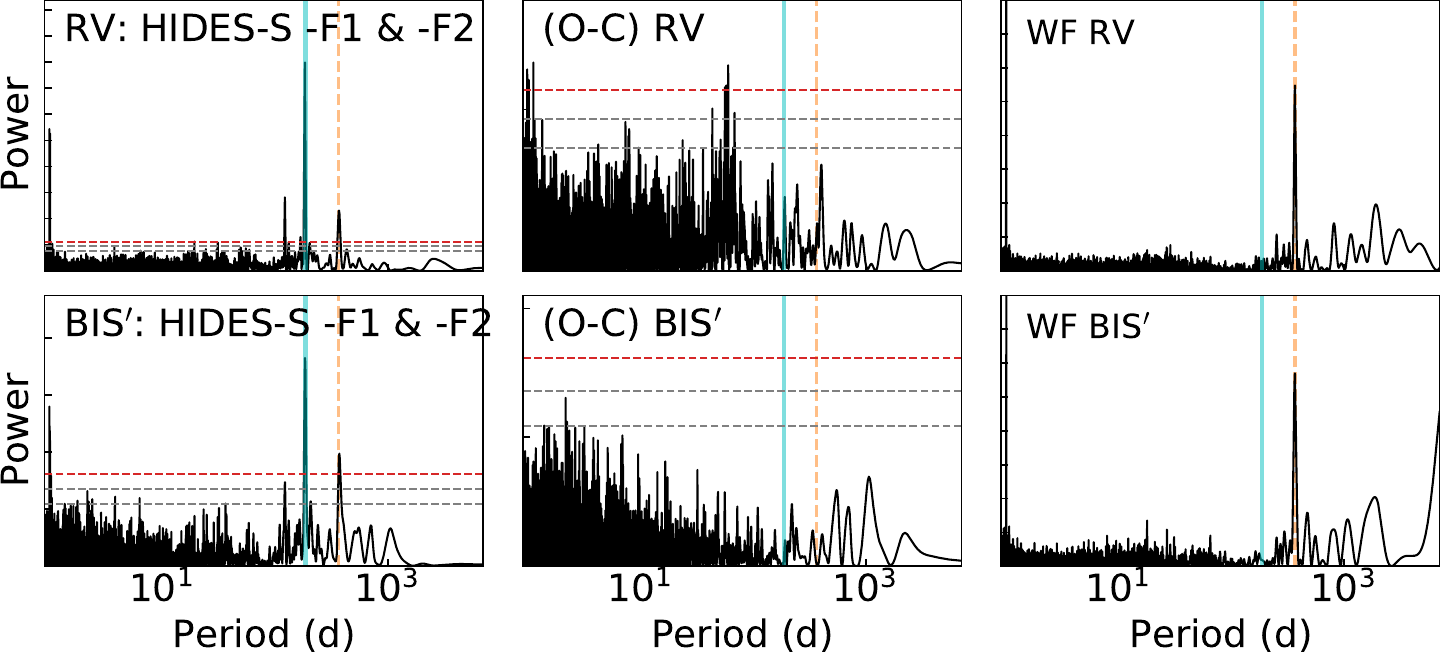}
\end{center}
\caption{
GLS periodograms for 41 Lyn. 
Top left: The observed RVs. 
Bottom left: The BIS$^{\prime}$. 
Top middle: The RV residuals to the 1-Keplerian fitting of the known planet.
Bottom middle: The BIS$^{\prime}$ residuals with removal of regular variation.
Top right: The window function of the observed RVs 
Bottom right: The window function of the BIS$^{\prime}$. 
The horizontal lines represent 10\%, 1\%, and 0.1\% FAP level from bottom to top. The vertical cyan solid lines indicate planetary signal, and the vertical orange dashed line indicates 1 year. 
}\label{fig:HD81688_RV_bis_GLS}
\end{figure*}

\begin{figure*}
\begin{center}
\includegraphics[scale=0.5]{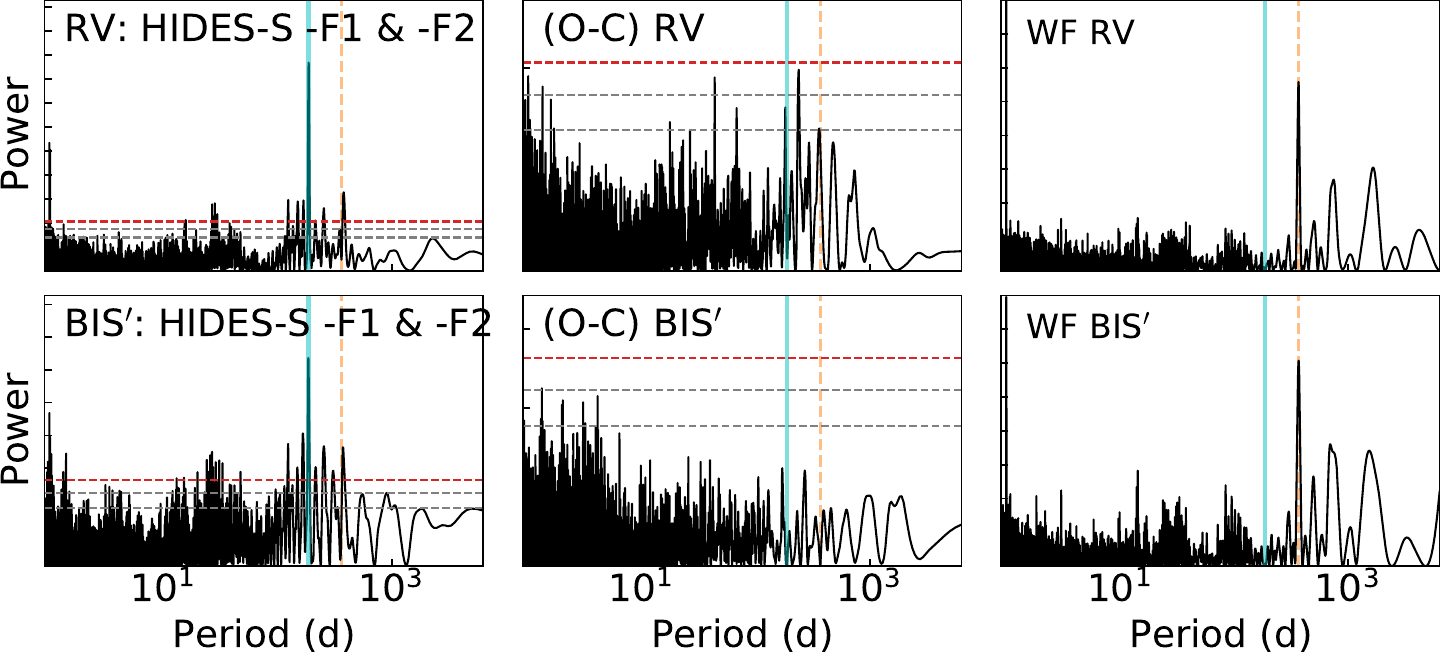}    
\end{center}
\caption{
GLS periodograms for 14 And. 
Top left: The observed RVs. 
Bottom left: The BIS$^{\prime}$. 
Top middle: The RV residuals to the 1-Keplerian fitting of the known planet.
Bottom middle: The BIS$^{\prime}$ residuals with removal of regular variation.
Top right: The window function of the observed RVs 
Bottom right: The window function of the BIS$^{\prime}$. 
The horizontal lines represent 10\%, 1\%, and 0.1\% FAP level from bottom to top. The vertical cyan solid lines indicate planetary signal, and the vertical orange dashed line indicates 1 year.
}\label{fig:HD221345_RV_bis_GLS}
\end{figure*}

\begin{figure*}
\begin{center}
\includegraphics[scale=0.35]{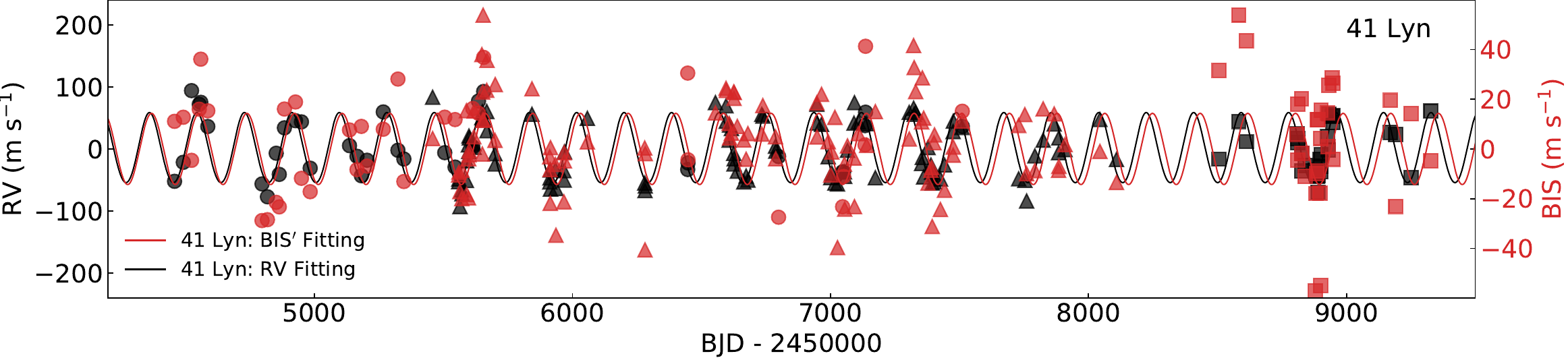}
\includegraphics[scale=0.35]{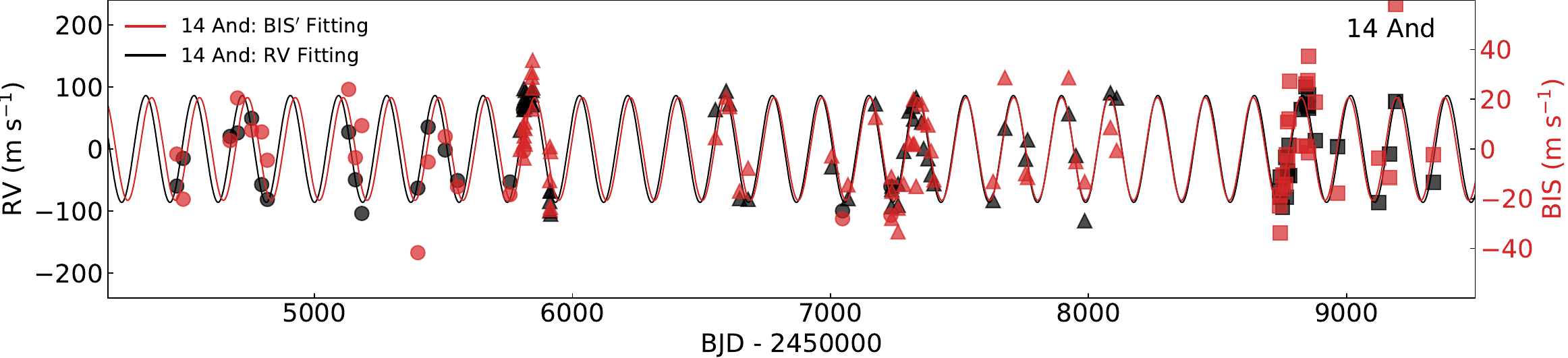} 
\end{center}
\caption{
Comparison between RVs and BIS of 41 Lyn (upper) and 14 And (lower): 
The RVs with offset removed and BIS with offset removed are shown in black and red markers, respectively. The markers with different shapes indicate the different instruments used: circle, triangle, and square represent HIDES-S, -F1, and -F2, respectively. The best fit 1-Keplerian orbit and BIS variation are given by black and red curves, respectively.
}\label{fig:HD81688_RV_bis_timeseries}
\end{figure*}

\begin{figure*}
\begin{center}
\includegraphics[scale=0.5]{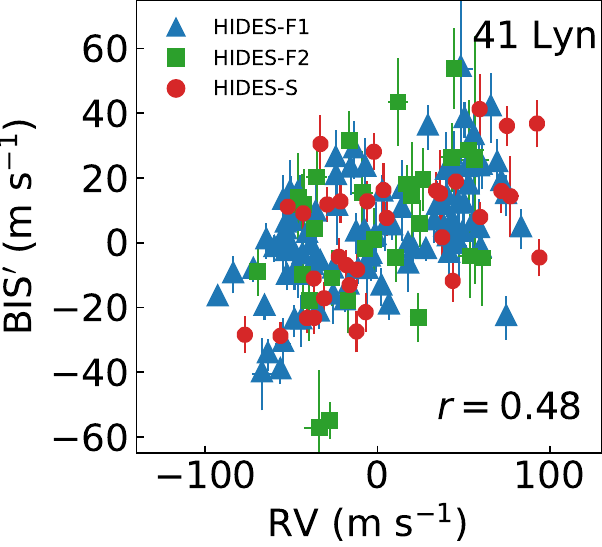}
\includegraphics[scale=0.5]{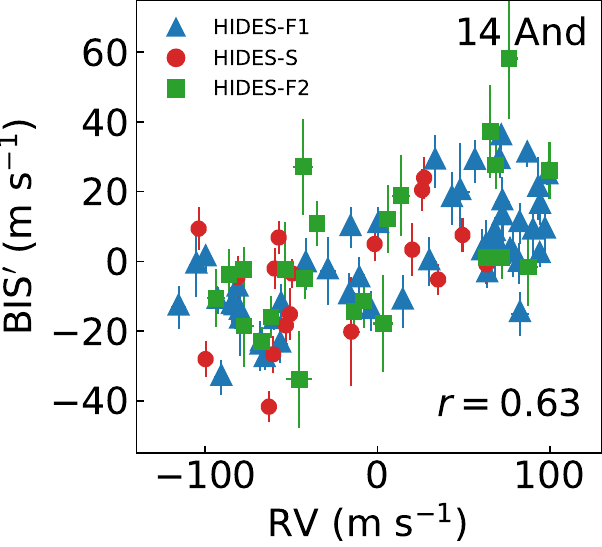}
\end{center}
\caption{
BIS$^{\prime}$ (mean removed BIS) against RVs of 41 Lyn (left) and 14 And (right). 
The data are taken by HIDES-S, -F1, and -F2 using 1.88m Telescope at OAO. The symbols are the same as those in Figure \ref{fig:HD81688_1pl_phase}.
}\label{fig:HD81688_RV_bis_rv}
\end{figure*}

The latest RV Keplerian fitting results (Figure \ref{fig:HD81688_1pl_phase}) of 41 Lyn and 14 And accord with previous reports \citep{Sato2008a, Sato2008b}. 
However, with long-baseline monitoring, we notice the line profiles of both of the stars regularly vary by time at periods of their fitted orbital periods.

Here, BIS$^{\prime}$ is applied to quantify the line profile variation, and it is defined as the mean removed bisector inverse span \citep{Teng2022a}. 
As illustrated by Figure \ref{fig:HD81688_RV_bis_GLS} and \ref{fig:HD221345_RV_bis_GLS}, GLS periodograms showed that RV data and BIS$^{\prime}$ data share almost the same periodicity for 41 Lyn and 14 And, respectively.
We then performed a sinusoidal curve fitting to BIS$^{\prime}$ for both of the stars. 
Similar to the circular orbital fitting to RVs (e.g., the aforementioned fitting of 14 And), the BIS$^{\prime}$ fitting included free parameters of semi-amplitude, period, phase of the curve, and offset to BIS$^{\prime}$ zero point of each instrument. 
The determination of best-fit parameters followed the same way as Keplerian fitting in this study (Section \ref{sec:observations}).

For 41 Lyn, we obtained $P_{\rm{BIS}^{\prime}} = 184.70\ \rm{d}$, which is almost equal to the best-fit orbital period of $P_{\rm{orb}} = 183.93\ \rm{d}$. 
Once again, for 14 And, it is $P_{\rm{BIS}^{\prime}} = 185.60\ \rm{d}$ to $P_{\rm{orb}} = 186.76\ \rm{d}$.
Moreover, BIS variations of both stars seemed synchronous to their RV variations, as the BIS phases coincided well with the RV phases. 
As shown in Figure \ref{fig:HD81688_RV_bis_timeseries}, 
we scale the amplitude of BIS$^{\prime}$ curve and find it almost overlaps with the RV curve for both of the stars. 
We further calculated the correlation coefficient with Pearson's $r$ between BIS$^{\prime}$ and RVs with the removal of their instrumental offsets.
Consequently, we obtained $r=0.48$ for 41 Lyn and $r=0.63$ for 14 And, which both indicate relatively strong correlation (Figure \ref{fig:HD81688_RV_bis_rv}).

The maximum rotational period can be estimated by projected rotational velocity $v\sin i$ and stellar radius $R_{\star}$. 
Adopting $v\sin i = 1.2\ \rm{km}\ \rm{s}^{-1}$\citep{Takeda2008} and $R_{\star} = 11.13\ R_{\odot}$ for 41 Lyn, and 
$v\sin i = 2.6\ \rm{km}\ \rm{s}^{-1}$\citep{Takeda2008} and $R_{\star} = 11.55\ R_{\odot}$ for 14 And, we obtain the maximum rotational period of 487 d and 225 d for 41 Lyn and 14 And, respectively. 
Combined with the fact that line profile correlates with RV, it could be inferred that 41 Lyn b (HD 81688 b) and 14 And b are questionable, and they are probably mimicked by line shape deformation due to the rotational modulation of stars.

Nonetheless, we are not yet able to confirm the rotational period of 41 Lyn and 14 And.
These two stars are considered to be chromospherically inactive stars for the inexistence of emission in the core of Ca II H lines \citep{Sato2008a, Sato2008b}. 
Hipparcos photometry revealed photometric stability of $\sigma =0.006\ \rm{mag}$ for both 41 Lyn and 14 And. 
We also analyzed the periodicity of the index of Ca \emissiontype{II} H lines $S_{\rm{H}}$ and Hipparcos photometry. Yet, no significant periodic signal was indicated by the GLS periodogram.

In addition, we noted another $\sim$58 d signal with FAP lower than 0.1\% in the periodogram of RV residuals of 41 Lyn. 
This RV variation could be interpreted by a sinusoidal curve with a period of $57.4\ \rm{d}$ and semi-amplitude of $5.9\ \rm{m}\ \rm{s}^{-1}$. 
At the current stage, we could not deduce the source of this RV variation with current data sets.  

\newpage
\subsection{HD 32518 and $\omega$ Ser: Unstable RV amplitudes from different observations}
HD 32518 (HR 1636, HIP 24003) is a K1 III star \citep{Dollinger2009} with one previously known planet. 
\citet{Dollinger2009} observed the star by using coud\'e \'echelle spectrograph of the 2m Alfred Jensch telescope of the Th\"uringer Landessternwarte and obtained 58 spectra. 
We observed this star with HIDES-F1 and -F2 and obtained 24 and 15 spectra, respectively, and our first HIDES observation was eight years after the end of \citet{Dollinger2009}'s observation. 
The best-fit Keplerian model with all data yielded the orbital parameters for the companion of $P = 157.35_{-0.08}^{+0.10}\ \rm{d}$, $K = 98.88_{-5.88}^{+5.60}\ \rm{m}\ \rm{s}^{-1}$, and $e=0.028_{-0.019}^{+0.034}$.
Adopting its stellar mass of $M_{\star}=1.13\ M_{\odot}$, we obtained the companion mass of $M_{\rm{p}}\sin i=2.849\ M_{\rm{J}}$ and semimajor-axis of $a=0.594\ \rm{au}$. 

$\omega$ Ser (HD 141680, HR 5888, HIP 77578) is a G8 III star, which was first reported to have RV variation in \citet{Sato2005}. 
The planet was announced in \citet{Sato2013b} with data collected from HIDES-S and HIDES-F1 since 2001 Feb. 
We continued to observe the star using HIDES-F1 and -F2 after the \citet{Sato2013b} and collected 179 RVs until April 2021. 
The best-fit Keplerian model with all RVs yielded the orbital parameters for the companion of $P = 278.59_{-0.61}^{+0.80}\ \rm{d}$, $K = 23.11_{-3.38}^{+1.94}\ \rm{m}\ \rm{s}^{-1}$, and $e = 0.180_{-0.127}^{+0.066}$. The rms was $19.42\ \rm{m}\ \rm{s}^{-1}$. 
Adopting its stellar mass of $M_{\star}=1.01\ M_{\odot}$, we obtained the companion mass of $M_{\rm{p}}\sin i=0.736\ M_{\rm{J}}$ and semimajor-axis of $a=0.838\ \rm{au}$. 

\begin{figure*}
\begin{center}
\includegraphics[scale=0.5]{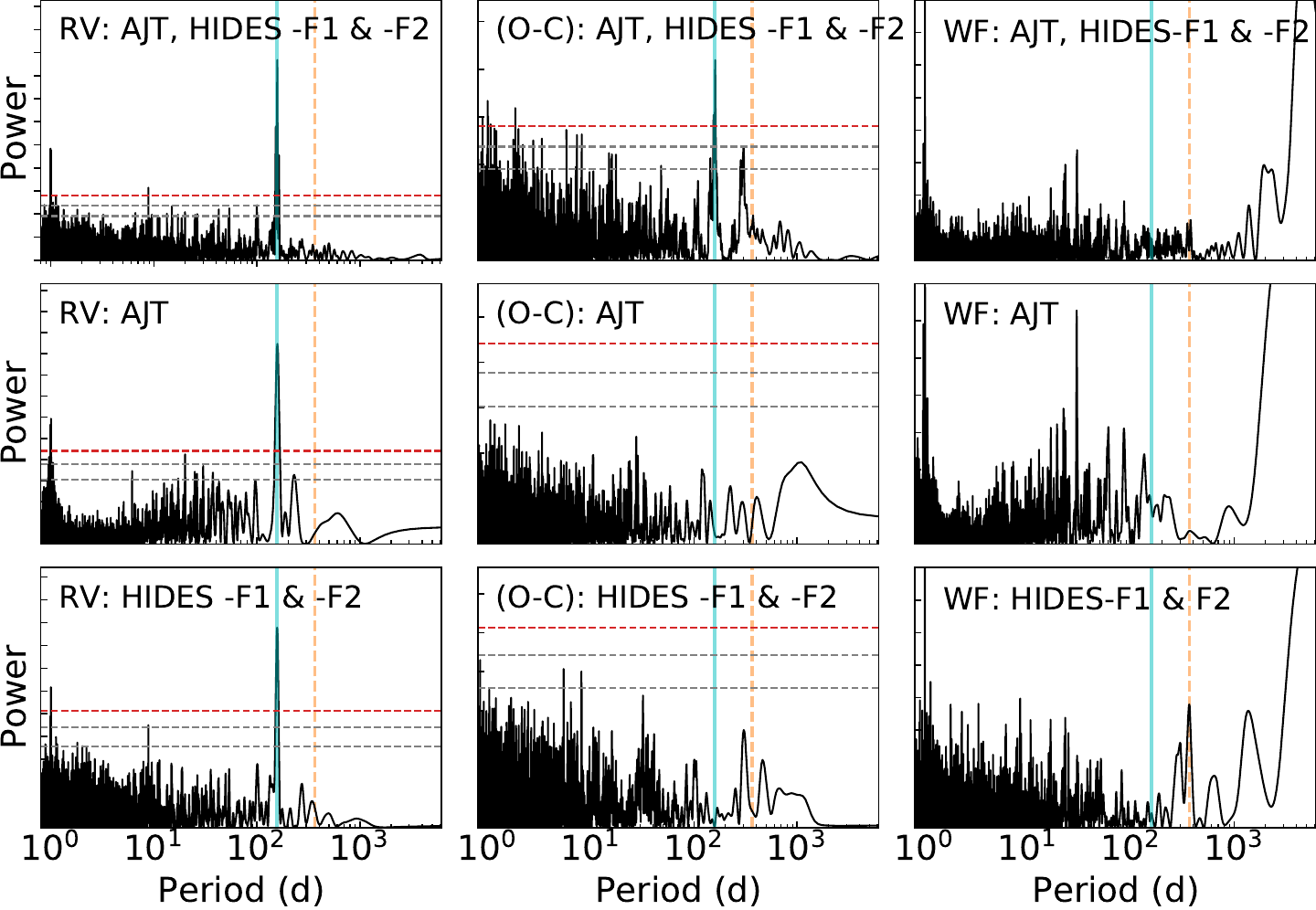} 
\end{center}
\caption{
GLS periodograms for HD 32518. 
From left to right column: The observed RVs, the RV residuals to the single Keplerian fitting of the known planet, and the window function of the observed RVs.
From top to bottom row: all observed data, data taken by the coud\'{e} echelle spectrograph on 2m Alfred Jensch Telescope (AJT) at Thuringia State Observatory, and data taken by HIDES-F1 and -F2 on 1.88m Telescope at OAO. 
The horizontal lines represent 10\%, 1\%, and 0.1\% FAP levels from bottom to top. The vertical cyan solid line indicates planetary signal, and the vertical orange dashed line indicates 1 year.
}\label{fig:HD32518_RV_Amp_GLS}
\end{figure*}

\begin{figure*}
\begin{center}
\includegraphics[scale=0.5]{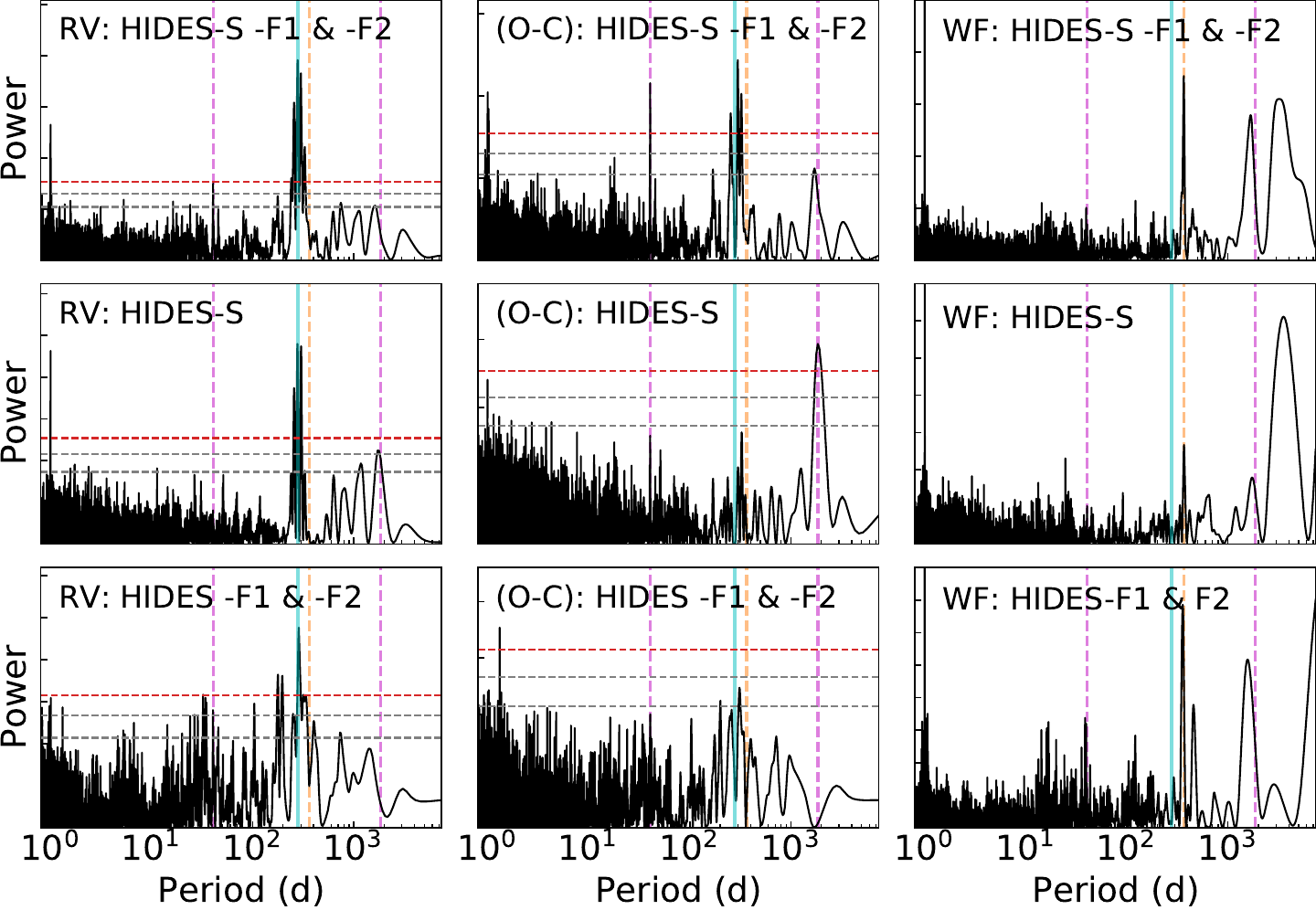}
\end{center}
\caption{
GLS periodograms for $\omega$ Ser. 
From left to right column: The observed RVs, the RV residuals to the single Keplerian fitting of the known planet, and the window function of the observed RVs.
From top to bottom row: all data taken by HIDES on 1.88m Telescope at OAO, data taken by HIDES slit mode (HIDES-S), and data taken by HIDES fiber mode (HIDES-F1 and -F2). 
The horizontal lines represent 10\%, 1\%, and 0.1\% FAP levels from bottom to top. The vertical cyan solid line indicates planetary signal, the vertical magenta dashed lines indicate extra periodic signals appeared in the residuals, and the vertical orange dashed line indicates 1 year.
}\label{fig:HD141680_RV_Amp_GLS}
\end{figure*}

\begin{figure*}
\begin{center}
\includegraphics[scale=0.25]{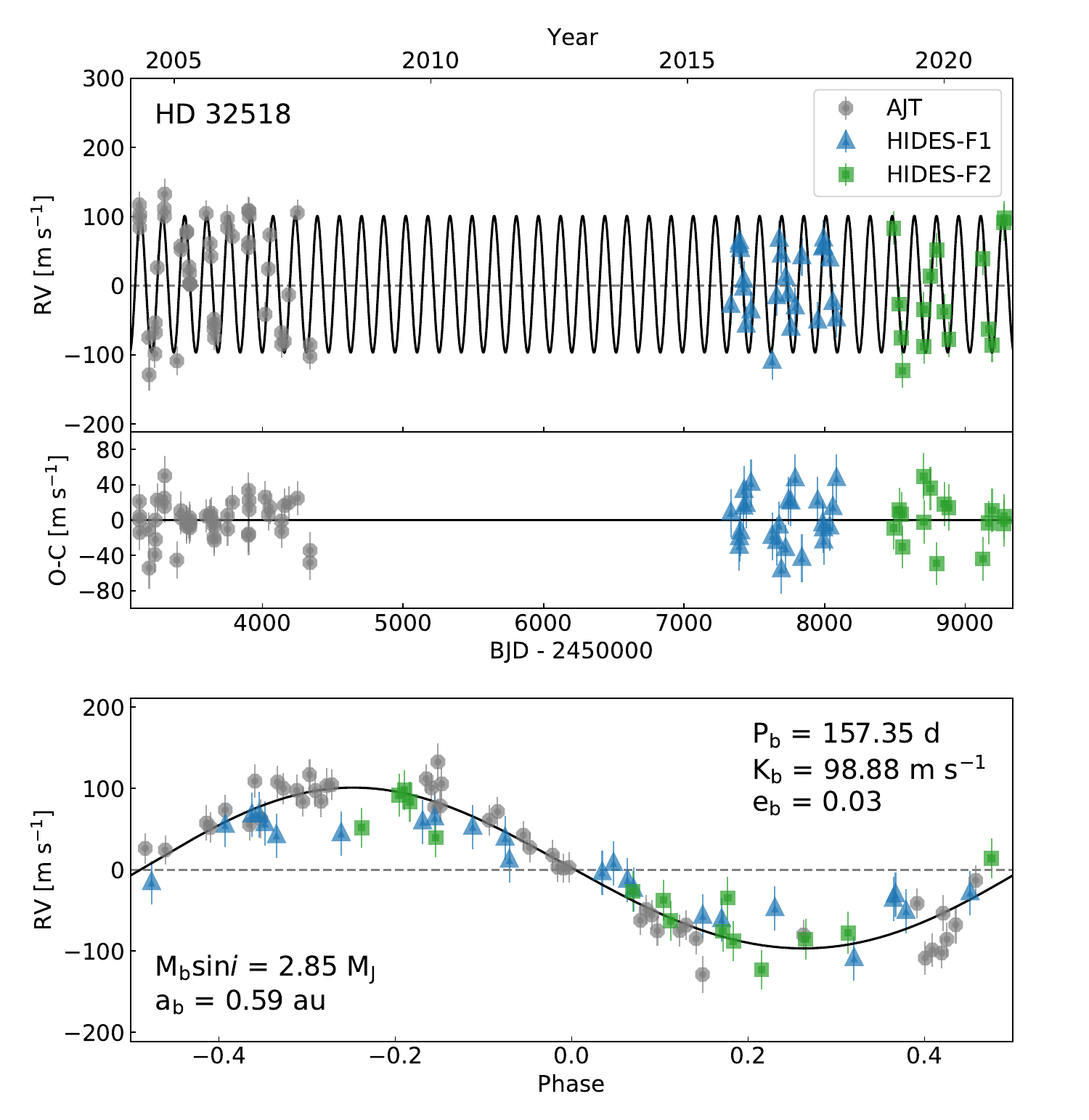}
\includegraphics[scale=0.25]{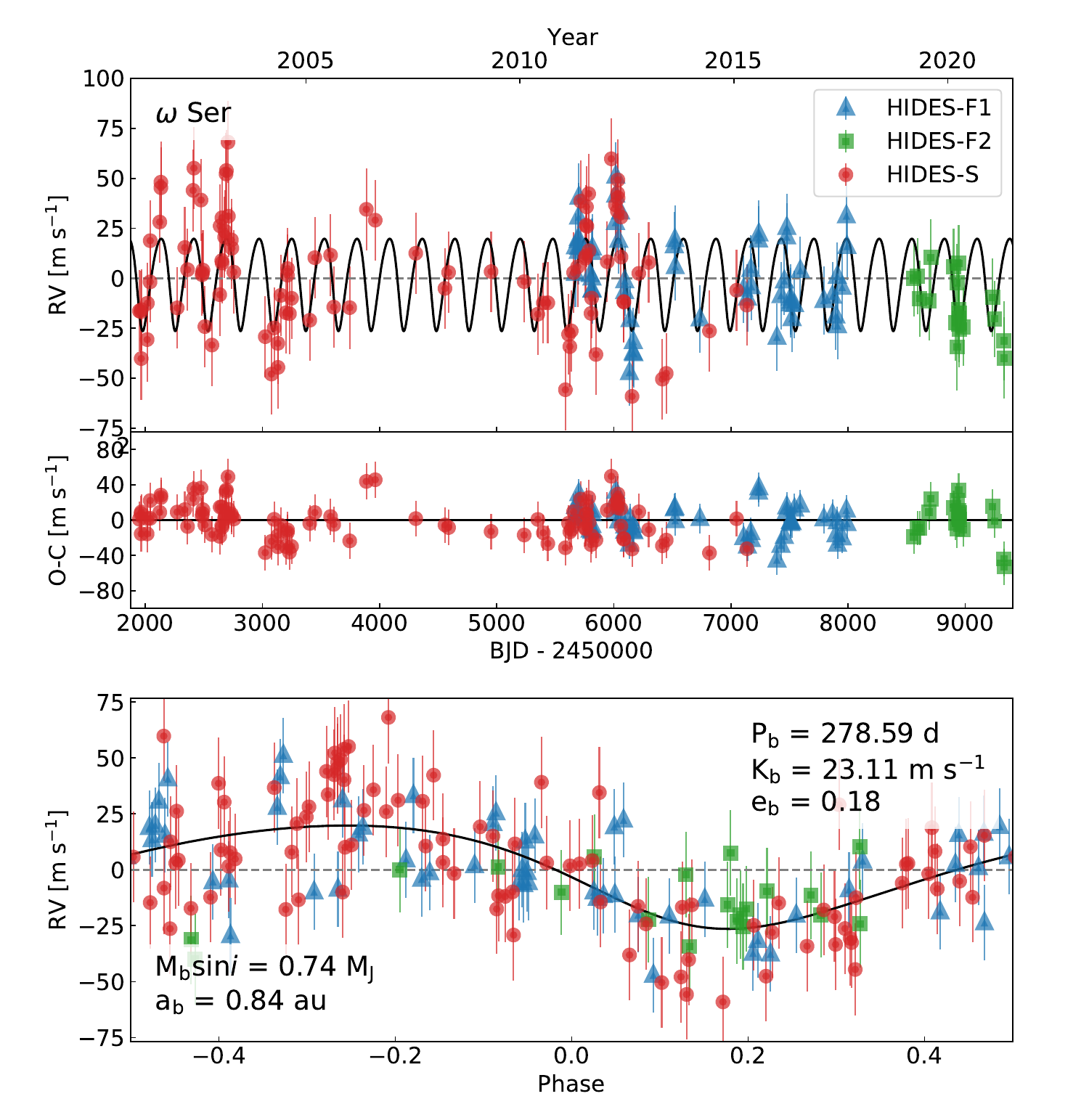}
\end{center}
\caption{
Single Keplerian solution to HD 32518 (left) and $\omega$ Ser (right).
Top: Best fit single Keplerian curve in the full observation span, including fitted RV offsets between instruments and jitters that are included in the error bars. Mid: Residuals of the RVs with respect to the best-fit model. Bottom: the phase-folded orbit of the planet. 
Data taken by coud\'{e} echelle spectrograph on 2m Alfred Jensch Telescope (AJT) at Thuringia State Observatory are shown by gray octagons.
Data taken by HIDES-S, -F1, and -F2 using 1.88m Telescope at OAO are shown by red circles, blue triangles, and green squares, respectively.
}\label{fig:HD32518_1pl_phase}
\end{figure*}

\begin{figure*}
\begin{center}
\includegraphics[scale=0.35]{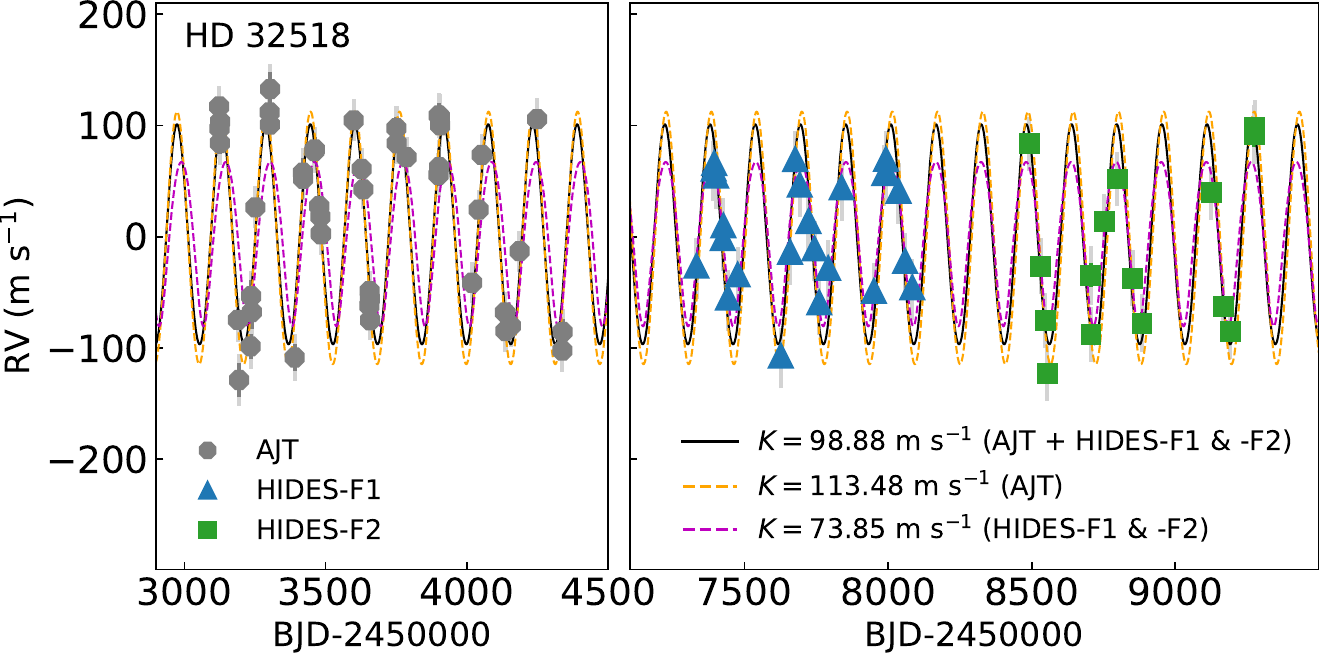}
\includegraphics[scale=0.35]{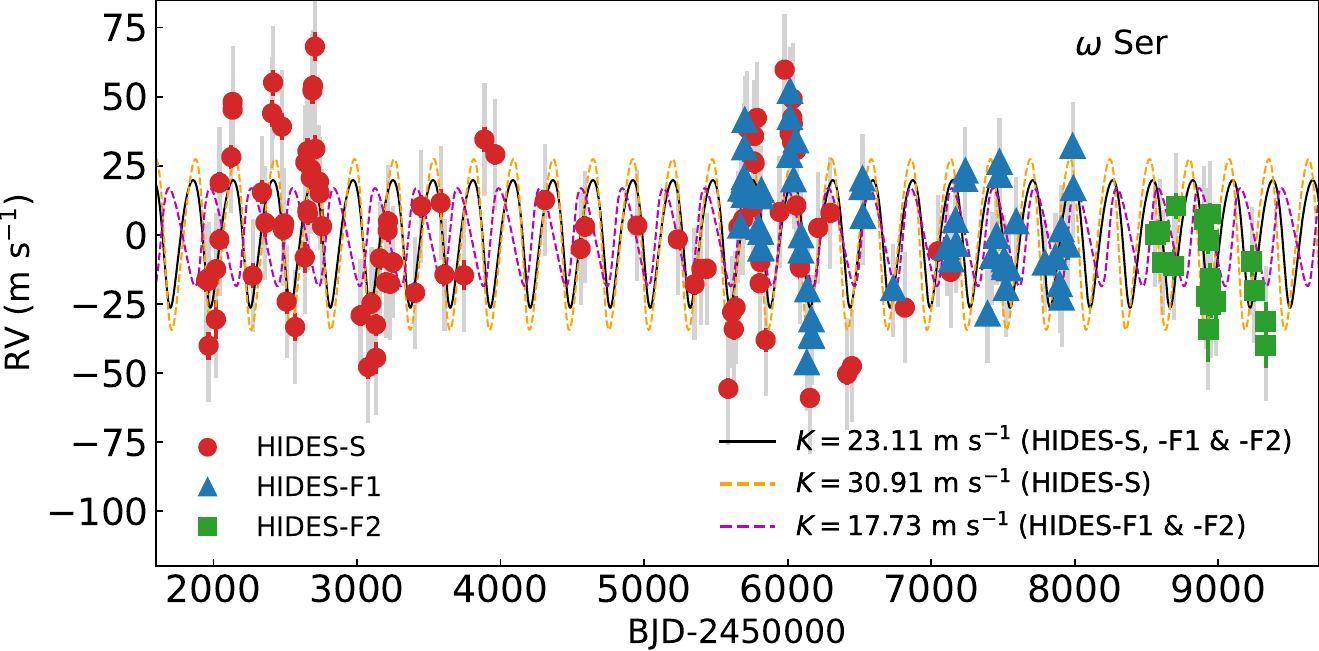}
\end{center}
\caption{
Comparison between different single Keplerian fitting results for 
HD 32518 (left) $\omega$ Ser (right) from different data sets.
(1) Black solid curve: best fits using all observed data.
(2) Yellow dashed curve: best fits using only data taken by the coud\'{e} echelle spectrograph on 2m Alfred Jensch Telescope (AJT) at Thuringia State Observatory for HD 32518, and data taken by HIDES slit mode (HIDES-S) for $\omega$ Ser.
(3) Magenta dashed curve: best fits using data taken by HIDES-F1 and -F2 on 1.88m Telescope at OAO.
RV markers are the same as those in Figure \ref{fig:HD32518_1pl_phase}.
}\label{fig:HD32518_RV_Amp}
\end{figure*}

According to our latest Keplerian fits, we noticed that the semi-amplitudes do not conform well with the previous results. 
For HD 32518, \citet{Dollinger2009} yields $K=115.83\pm 4.67 \rm{m}\ \rm{s}^{-1}$, which apparently differs from $K = 98.88_{-5.88}^{+5.60}\ \rm{m}\ \rm{s}^{-1}$ by this work, and for $\omega$ Ser, \citet{Sato2013b} yields $K=31.8_{-2.3}^{+2.3}\ \rm{m}\ \rm{s}^{-1}$, which is distinctive to $K = 23.11_{-3.38}^{+1.94}\ \rm{m}\ \rm{s}^{-1}$ by this work. 

Meanwhile, we also noted that there were strong signals that remained at the value of orbital periods in the GLS periodogram of the residuals for both of the stars (Figure \ref{fig:HD32518_RV_Amp_GLS} and \ref{fig:HD141680_RV_Amp_GLS}).
It hence suggested that the RV variations were not fully subtracted from the time series. 
The previous results were fully (HD 32518) or dominantly ($\omega$ Ser) constrained by data from one instrument. 
We, therefore, performed Keplerian orbital fits separately by data sets from different instruments. 
Here, we still considered HIDES-F1 and -F2 as two independent instruments, but we did not separate them into different fittings. 
As shown in Figure \ref{fig:HD32518_RV_Amp}, 
the fittings depending on different instruments were distinctive from each other. 

For HD 32518, the semi-amplitude from HIDES data,  {$K=73.85_{-6.56}^{+3.84}\ \rm{m}\ \rm{s}^{-1}$, was apparently lower than the one from AJT data, $K=113.48_{-4.17}^{+4.66}\ \rm{m}\ \rm{s}^{-1}$ over an 8$\sigma$ level}. 
As shown in the GLS periodogram of residuals (Figure \ref{fig:HD32518_RV_Amp_GLS}), the RV variation was fully subtracted in each separate data set. 
Furthermore, we confirmed that $\rm{BIS}$ of both HIDES-F1 and HIDES-F2 did not correlate with RV by Pearson's correlation {with their $r$-value equal to 0.03 and 0.21, respectively}, and they did not show strong periodicity around 157d.
The long time lag of HIDES after AJT observation made it impracticable to determine if the RV amplitude varied by time.
The rms scatter obtained from AJT data and HIDES data were both about $18\ \rm{m}\ \rm{s}^{-1}$.
We hence could not confirm the true reason for the large difference in RV semi-amplitude at the current stage. 

For $\omega$ Ser, the semi-amplitude from HIDES-F1 and F2 data, {$K=17.73_{-3.74}^{+4.24}\ \rm{m}\ \rm{s}^{-1}$, was apparently lower than the one from HIDES-S data, $K=30.91_{-4.21}^{+2.49}\ \rm{m}\ \rm{s}^{-1}$ over a 3$\sigma$ level}. 
As seen in the GLS periodogram of residuals (middle column in Figure \ref{fig:HD141680_RV_Amp_GLS}), the 278 d RV variation was clearly subtracted in separate data sets, while it is not fully subtracted in the full data. Moreover, full data residuals exhibited an additional periodicity of 40.54 d, and HIDES-S residuals exhibited another additional of 1852 d. 

Assuming a planetary situation, the 40.54 d signal in the full data set could be subtracted by a Keplerian orbit of $P=40.54\ \rm{d}$, $K=12.7\ \rm{m} \rm{s}^{-1}$, $e=0.28$, and with a planetary mass of $M_{\rm{p}}\sin i=0.2 M_{\rm{J}}$, and the rms scatter {suppressed to $\sim 17.75 \ \rm{m}\ \rm{s}^{-1}$ (decreased by $\sim 1.67 \ \rm{m}\ \rm{s}^{-1}$)}. 
The 1852 d signal in HIDES-S data set could be interpreted by another Keplerian orbit of $P=1840\ \rm{d}$, $K=14.5\ \rm{s}^{-1}$, $e=0.02$ and with a planetary mass of $M_{\rm{p}}\sin i=0.82 M_{\rm{J}}$, and the rms scatter {suppressed to $\sim 16.69 \ \rm{m}\ \rm{s}^{-1}$ (decreased by $\sim 2.73 \ \rm{m}\ \rm{s}^{-1}$)}. 

Yet, we could hardly treat them as planet candidates. 
According to the window function to $\omega$ Ser observations (right column in Figure \ref{fig:HD141680_RV_Amp_GLS}), peaks appeared close to 1852 d for all three sets, and another peak emerged at 40 d for HIDES-F1 and F2. 
These strongly suggested that the periodicity might be caused by an observational window, although they could be well-fitted by Keplerian orbits.  
In addition, the number of HIDES-F2 observations (21) is less than the number of HIDES-S (99) and HIDES-F1 (59). Since the RVs from HIDES-F1 around BJD 2456000 by-eye conformed well with RVs from HIDES-S, more observations from HIDES-F2 are necessary to determine a convincing RV semi-amplitude, as well as long-term periodicity.

\newpage
\section{Joint fits with astrometry}\label{sec:astrometry}
Among the revisited stars in this study, four stars: HD 5608, HD 14067, HD 120084, and HD 175679, showed significant astrometric acceleration from HGCA of $\chi^{2} > 14$ \citep{Brandt2018, Brandt2021}, 
which allows us to introduce astrometric data and perform joint constraints on the orbital parameters for the four systems. 

\begin{figure*}
\begin{center}
\includegraphics[scale=0.25]{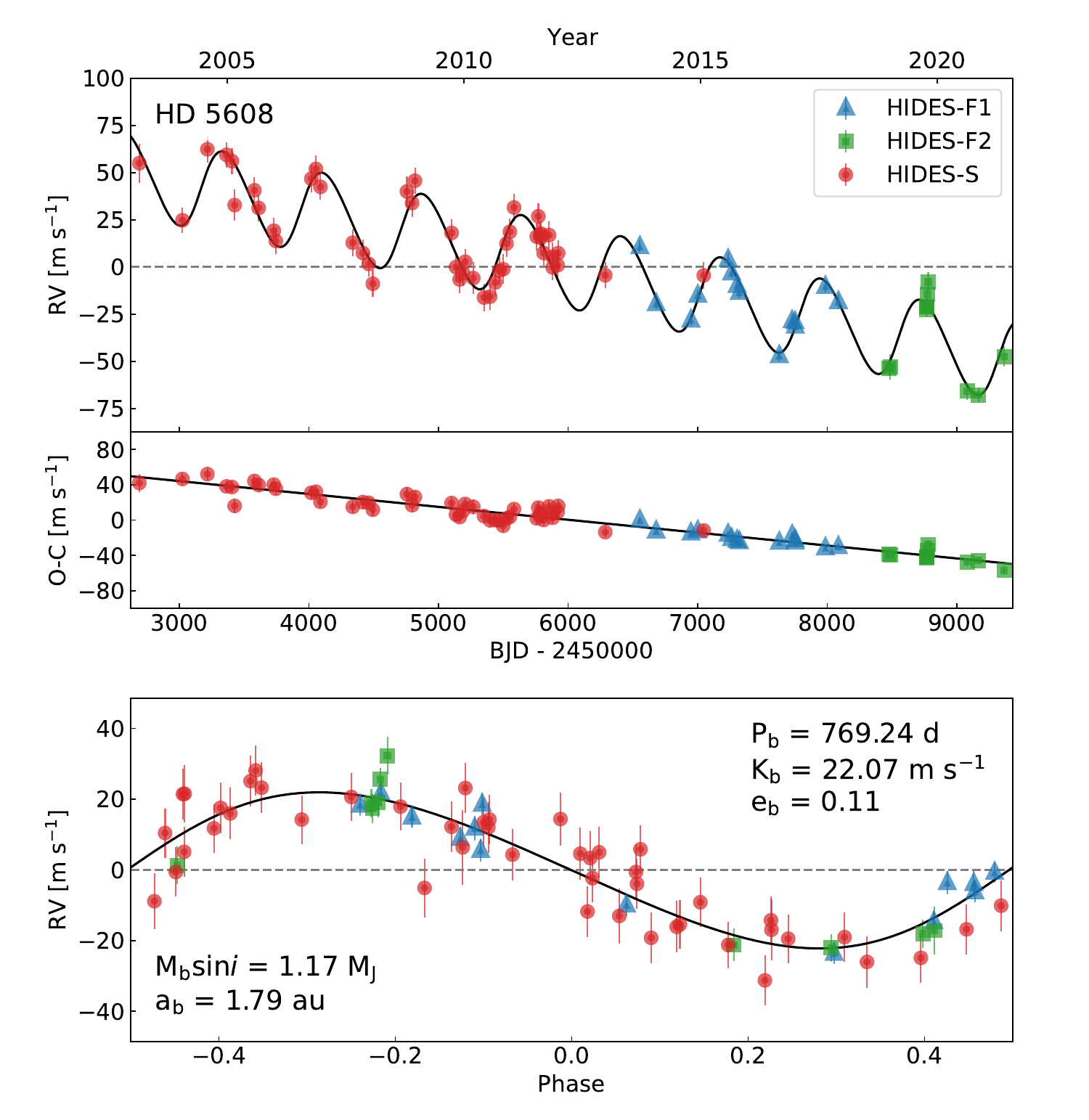}
\includegraphics[scale=0.25]{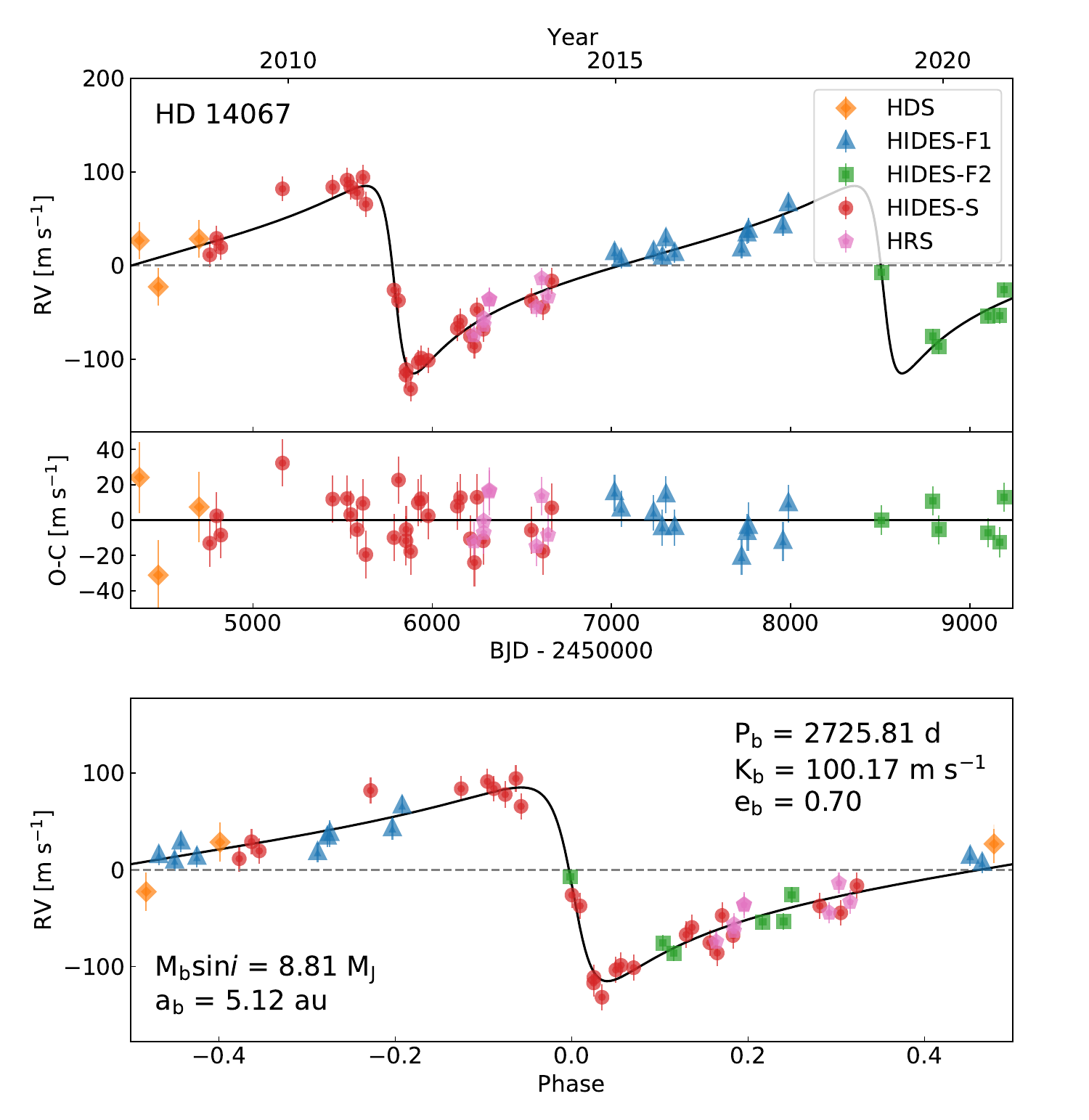}
\includegraphics[scale=0.25]{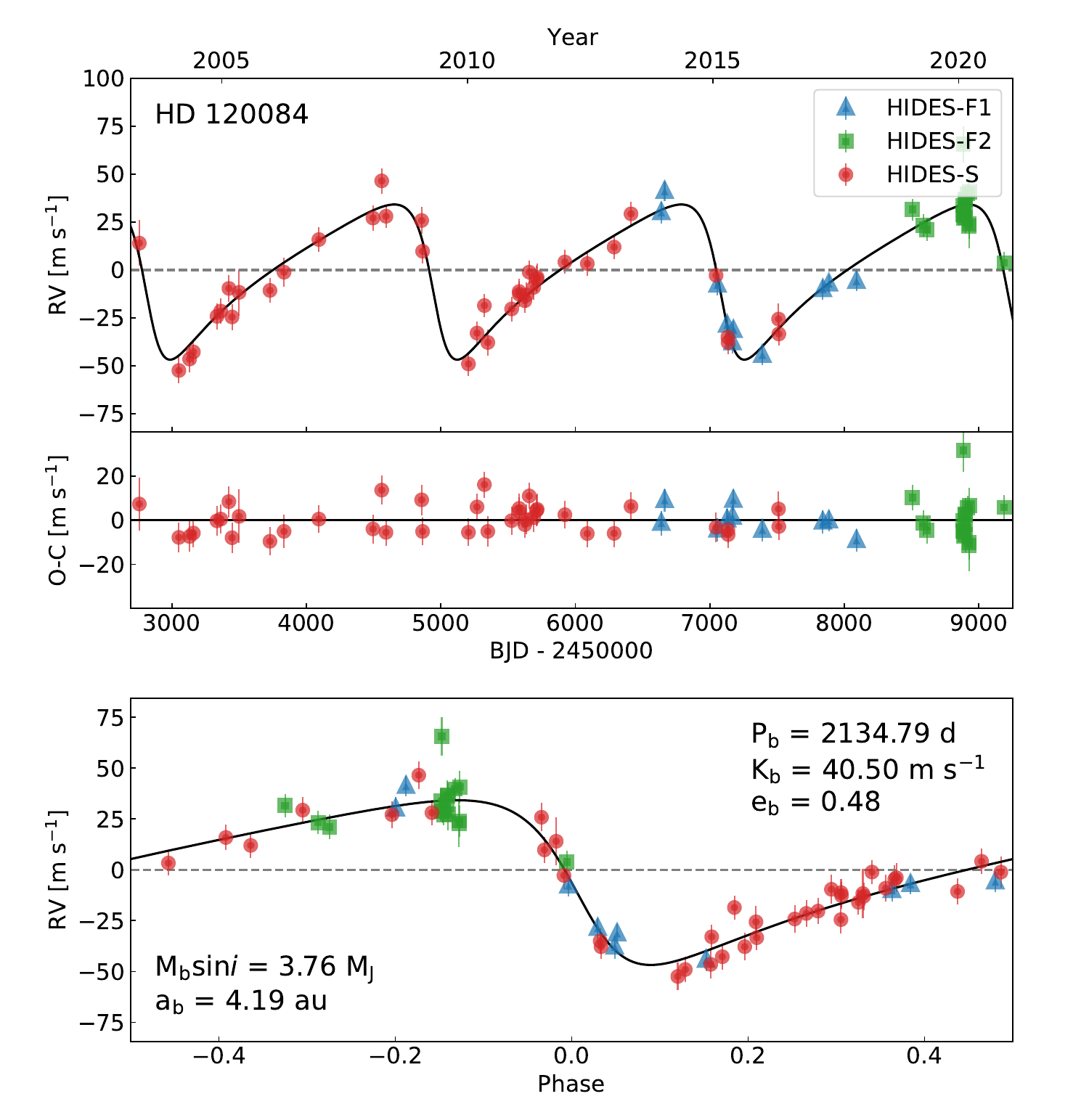}
\includegraphics[scale=0.25]{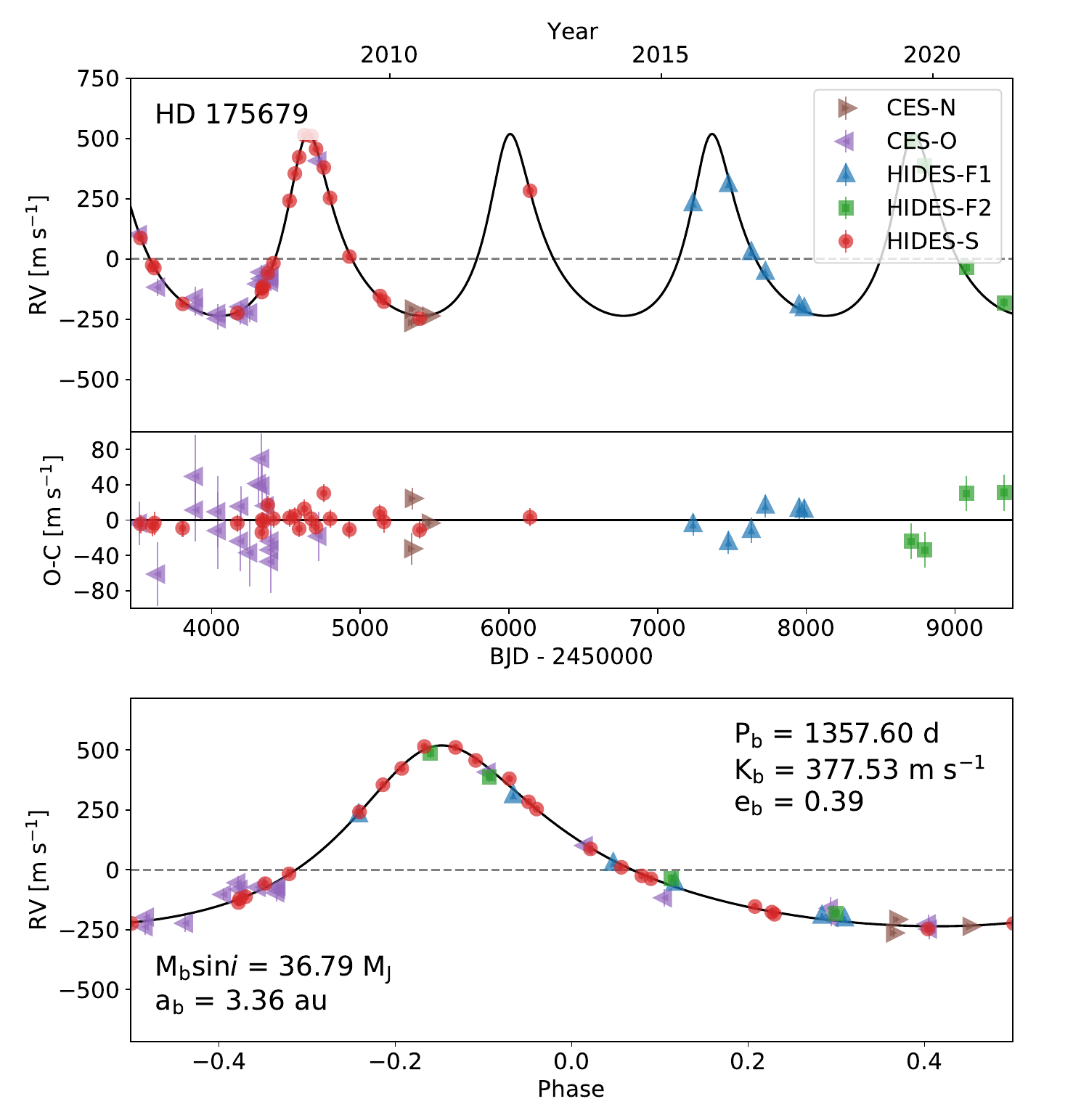}
\end{center}
\caption{
Single Keplerian solution to HD 5608, HD 14067, HD 120084, and HD175679.
Top: Best fit Single Keplerian curve in the full observation span, including fitted RV offsets between instruments and jitters that are included in the error bars. 
Residuals of the RVs with respect to the best-fit model. Bottom: phase-folded orbit of the planet. 
Data taken by HIDES-S, -F1, and -F2 using 1.88m Telescope at OAO are shown by red circles, blue triangles, and green squares, respectively.
Data taken by the High Dispersion Spectrograph (HDS) on Subaru Telescope at Mauna Kea are shown by orange spades. 
Data taken by the CES pre-upgrade (CES-O), CES post-upgrade (CES-N), and High-Resolution Spectrograph (HRS) on 2.16m telescope at Xinglong Observatory are shown by purple triangles and brown triangles, pink pentagons, respectively.
}\label{fig:HD5608_1pl_phase}
\end{figure*}

\begin{table*}[p]
\tbl{Joint fitting results of HD 5608, HD 14067, HD 120084 and HD 175679.}{%
\begin{tabular}{lcccc}\hline\hline
Object  & HD 5608 B & HD 14067 b & HD 120084 b & HD 175679 b\\
\hline
Primary mass $M_{\rm{pri}}\ (M_{\odot})$  & ${1.28}_{-0.23}^{+0.23}$ & ${2.40}_{-0.20}^{+0.20}$ & ${2.16}_{-0.27}^{+0.28}$ & ${2.70}_{-0.30}^{+0.30}$ \\
Secondary mass $M_{\rm{sec}}\ (M_{\rm{J}})$  & ${121.2}_{-7.0}^{+8.3}$ & ${15.3}_{-4.5}^{+6.2}$ &  ${6.4}_{-1.9}^{+2.9}$ & ${61.8}_{-9.4}^{+15}$\\
Semimajor-axis $a$ (au) & ${30.7}_{-4.9}^{+8.8}$ & ${5.15}_{-0.15}^{+0.14}$ & ${4.21}_{-0.19}^{+0.17}$ & ${3.36}_{-0.13}^{+0.12}$\\
$\sqrt{e} \sin \omega$ & ${-0.63}_{-0.16}^{+0.29}$ & ${0.773}_{-0.022}^{+0.022}$ & ${0.673}_{-0.028}^{+0.026}$ & ${-0.122}_{-0.013}^{+0.014}$ \\
$\sqrt{e} \cos \omega$  & ${-0.13}_{-0.33}^{+0.36}$ & ${-0.220}_{-0.051}^{+0.052}$ & ${-0.167}_{-0.058}^{+0.061}$ & ${0.5969}_{-0.0073}^{+0.0072}$\\
Inclination $i$ ($^{\circ}$)  & ${147.5}_{-13}^{+6.6}$ & ${118}_{-84}^{+31}$ & ${49}_{-21}^{+87}$ & ${41.3}_{-6.4}^{+109}$\\
Inclination $i$ ($i<90^{\circ}$)($^{\circ}$)$^{*}$  & - & ${39}_{-13}^{+28}$ & ${38}_{-12}^{+22}$ & ${39.2}_{-5.1}^{+6.2}$\\
Inclination $i$ ($i>90^{\circ}$)($^{\circ}$)$^{*}$ & ${147.5}_{-13}^{+6.6}$ & ${141}_{-23}^{+12}$ & ${139}_{-29}^{+15}$ & ${152.4}_{-4.4}^{+3.3}$\\
Ascending node $\Omega$ ($^{\circ}$) & ${141}_{-48}^{+56}$ & ${165}_{-49}^{+37}$ & ${156}_{-37}^{+129}$ & ${228}_{-150}^{+17}$\\
Mean longitude at $\lambda_{\rm{ref}}$ ($^{\circ}$)  & ${112}_{-76}^{+84}$ & ${27.3}_{-3.0}^{+3.0}$ & ${146.9}_{-2.8}^{+2.7}$ & ${137.93}_{-0.50}^{+0.51}$ \\
\hline
Period (yr) & ${146}_{-37}^{+74}$ & ${7.513}_{-0.038}^{+0.044}$ & ${5.864}_{-0.026}^{+0.024}$ & ${3.7109}_{-0.0036}^{+0.0035}$\\
Argument of periastron $\omega$ ($^{\circ}$)  & ${257}_{-35}^{+32}$ &  ${105.9}_{-3.8}^{+3.8}$ & ${103.9}_{-5.2}^{+5.1}$ & ${348.4}_{-1.3}^{+1.3}$\\
Eccentricity $e$ & ${0.53}_{-0.26}^{+0.18}$ & ${0.650}_{-0.028}^{+0.027}$ & ${0.483}_{-0.029}^{+0.027}$ & ${0.3714}_{-0.0079}^{+0.0080}$\\
Semimajor-axis $a$ (mas)  & ${524}_{-84}^{+151}$ & ${36.6}_{-1.0}^{+1.0}$ & ${40.5}_{-1.8}^{+1.7}$ & ${19.99}_{-0.76}^{+0.71}$\\
Periastron $T_{p}$ (JD-2450000)  & ${22895}_{-5098}^{+17464}$ & ${5797}_{-12}^{+11}$ & ${7083}_{-17}^{+17}$ & ${5990.1}_{-3.8}^{+3.7}$\\
Mass ratio  &   ${0.091}_{-0.015}^{+0.021}$ & ${0.0061}_{-0.0018}^{+0.0024}$ & ${0.00287}_{-0.00084}^{+0.0013}$ & ${0.0219}_{-0.0030}^{+0.0053}$\\
\hline
\end{tabular}}
\begin{tabnote}
\hangindent6pt\noindent
\hbox to6pt{\footnotemark[$*$]\hss}\unskip%
The inclination distributions are bimodal for companions without direct imaging measurements, so we separately report the values for prograde and retrograde orbits.
\end{tabnote}
\label{tab:astrometry_res}
\end{table*}

\begin{figure*}
\begin{center}
\includegraphics[scale=0.38]{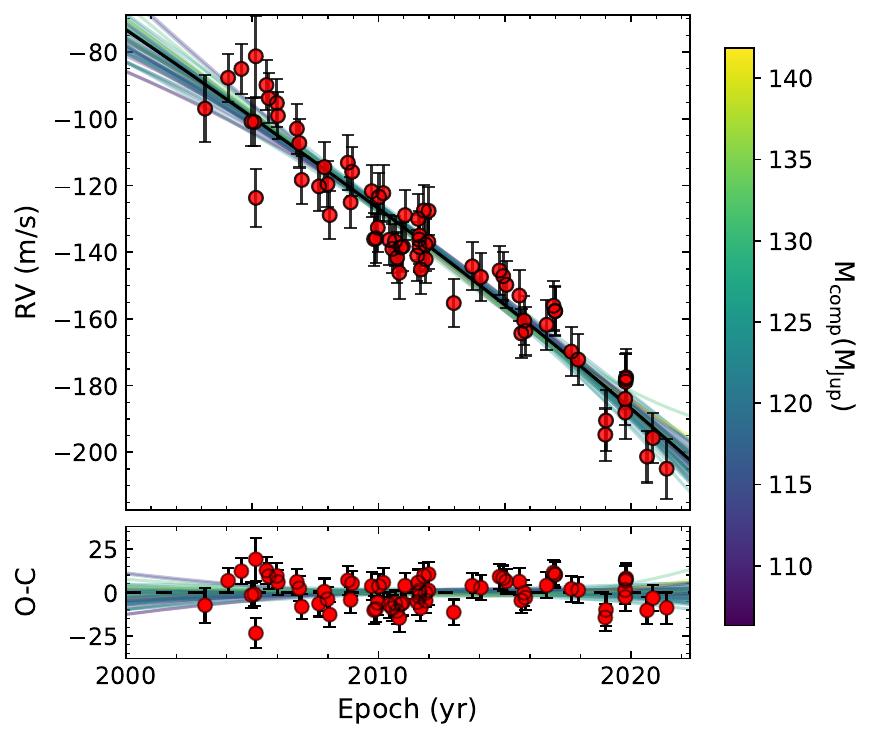} 
\includegraphics[scale=0.38]{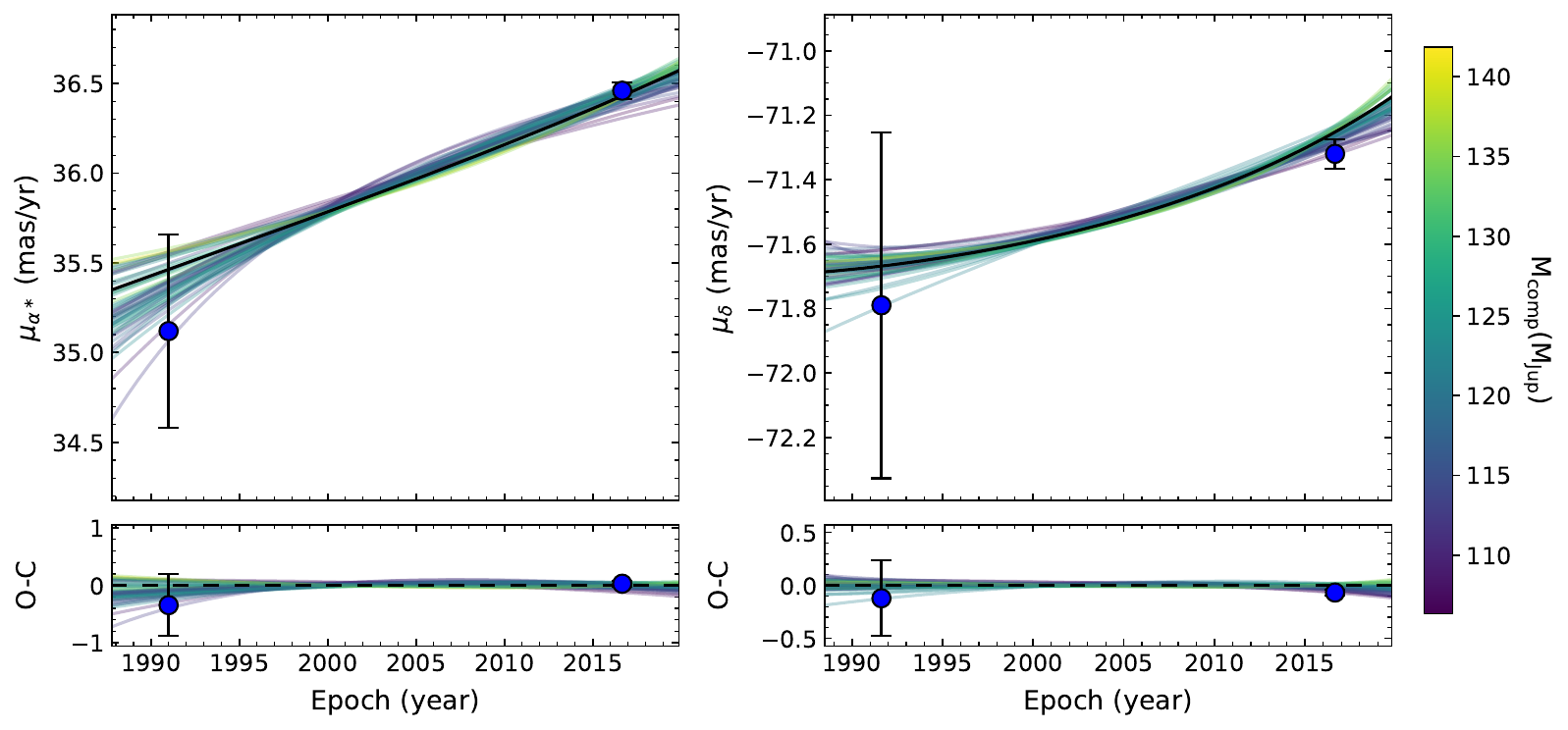}
\includegraphics[scale=0.38]{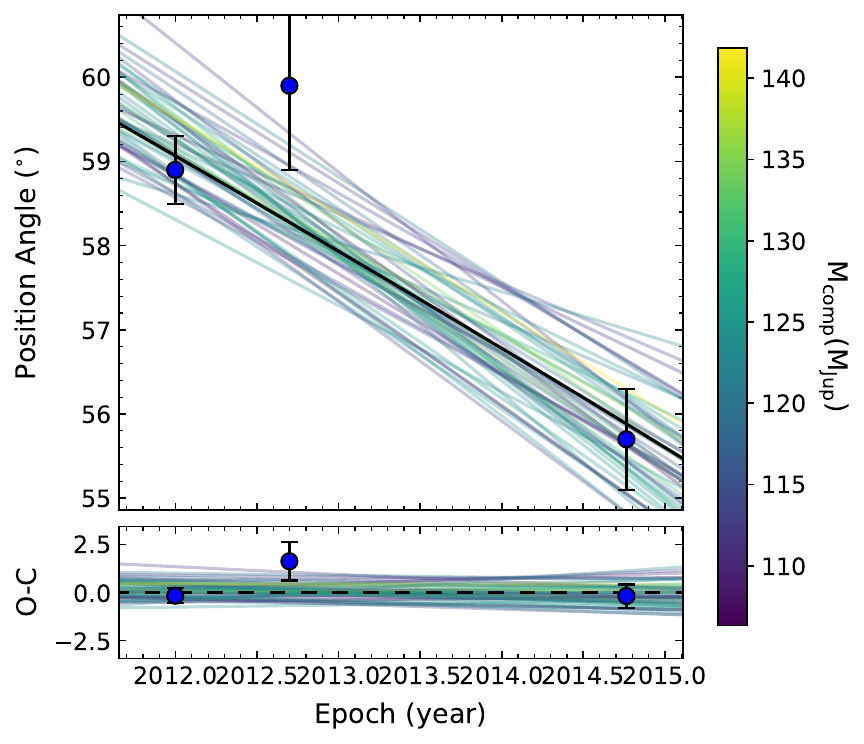}
\includegraphics[scale=0.38]{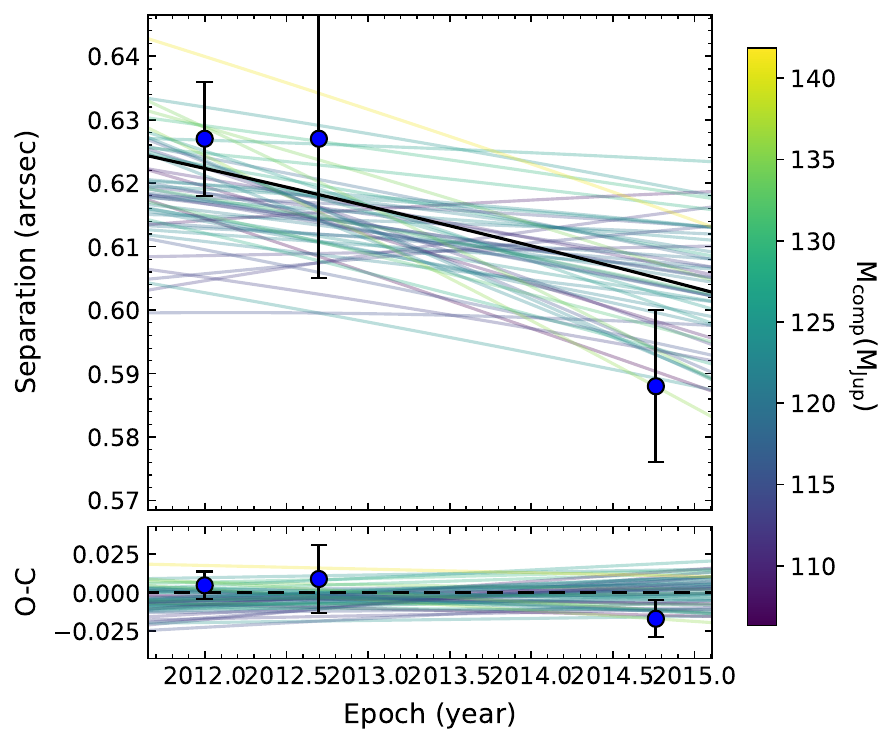}
\end{center}
\caption{
The joint fit result of the HD 5608 B using RV and astrometry with the removal of the known planet (HD 5608 b).
The best-fit orbit is indicated by black solid curves, while the randomly-selected 50 orbits are shown by lines with colors, corresponding to the estimated masses of HD 5608 B.
Upper left: The RV curve with offset-removed RVs in red circles.
Upper middle and upper right: The proper motion variations of HD 5608 along RA and Dec.
Bottom left and bottom right: the position angles (PAs) and the measured projected separations of HD 5608 B relative to HD 5608.
}\label{fig:HD5608_orvara}
\end{figure*}
\begin{figure*}
\begin{center}
\includegraphics[scale=0.38]{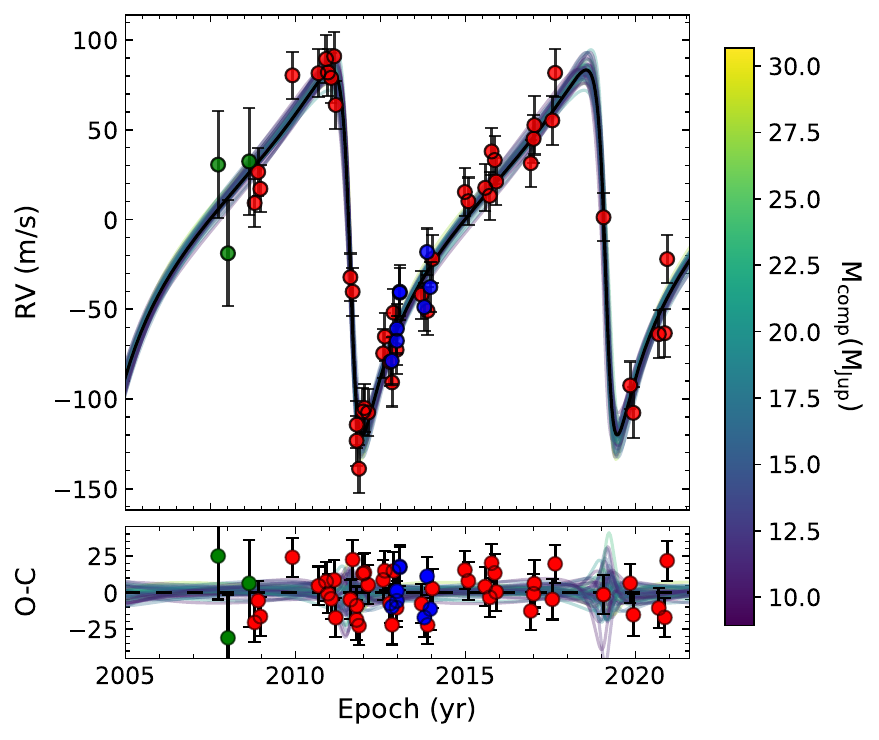}
\includegraphics[scale=0.38]{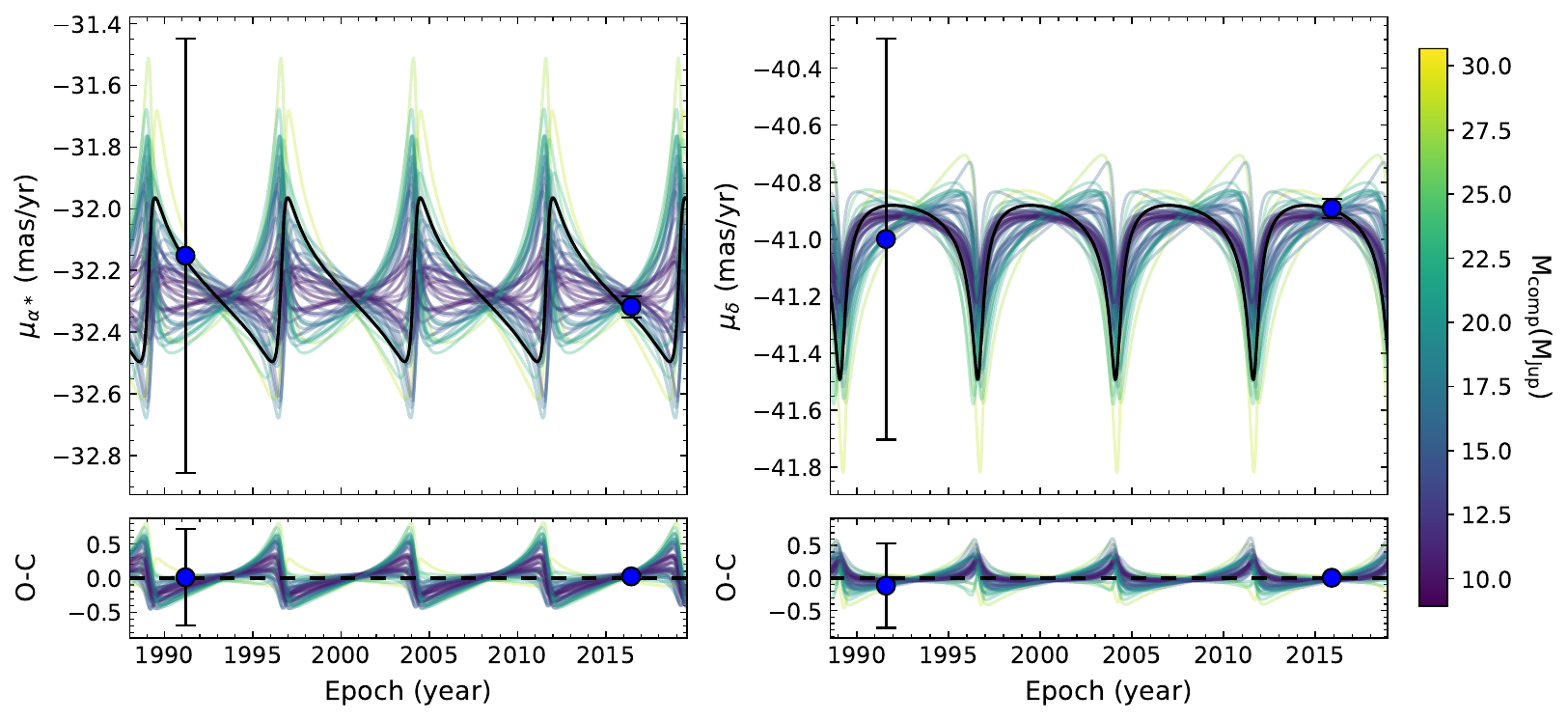}
\end{center}
\caption{
The joint fit of the HD 14067 b using RV and astrometry.
Left: The RV curve with RVs in red circles. 
Middle and right: The proper motion variations of HD 14067 along RA and Dec.
The best-fit orbital solution is shown by a black solid line, and another 50 orbits randomly taken from MCMC chains are shown by lines with colors, corresponding to the estimated masses of HD 14067 b. 
}\label{fig:HD14067_orvara}
\end{figure*}
\begin{figure*}
\begin{center}
\includegraphics[scale=0.38]{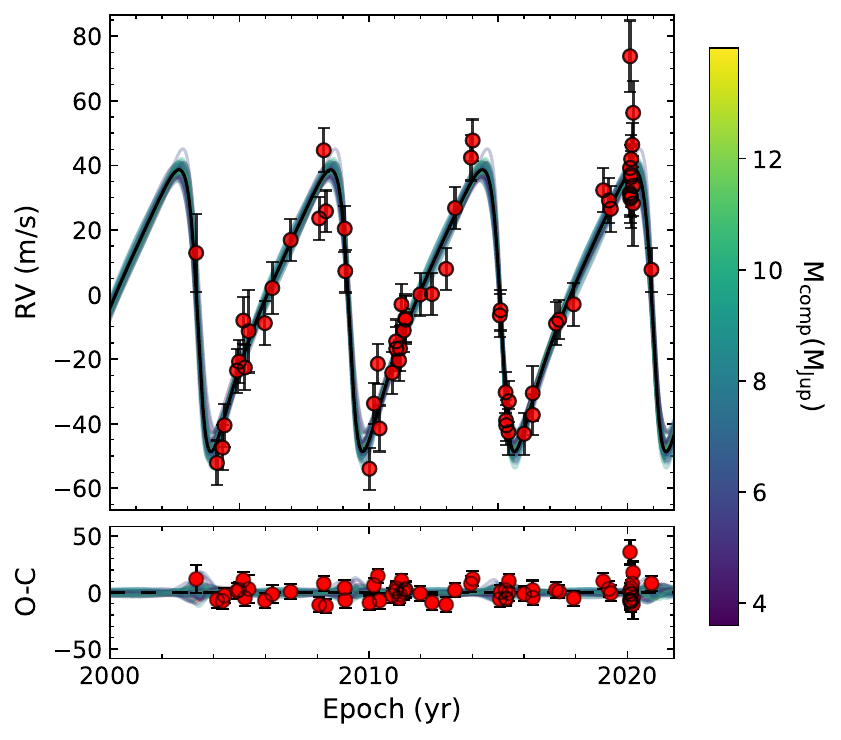}
\includegraphics[scale=0.38]{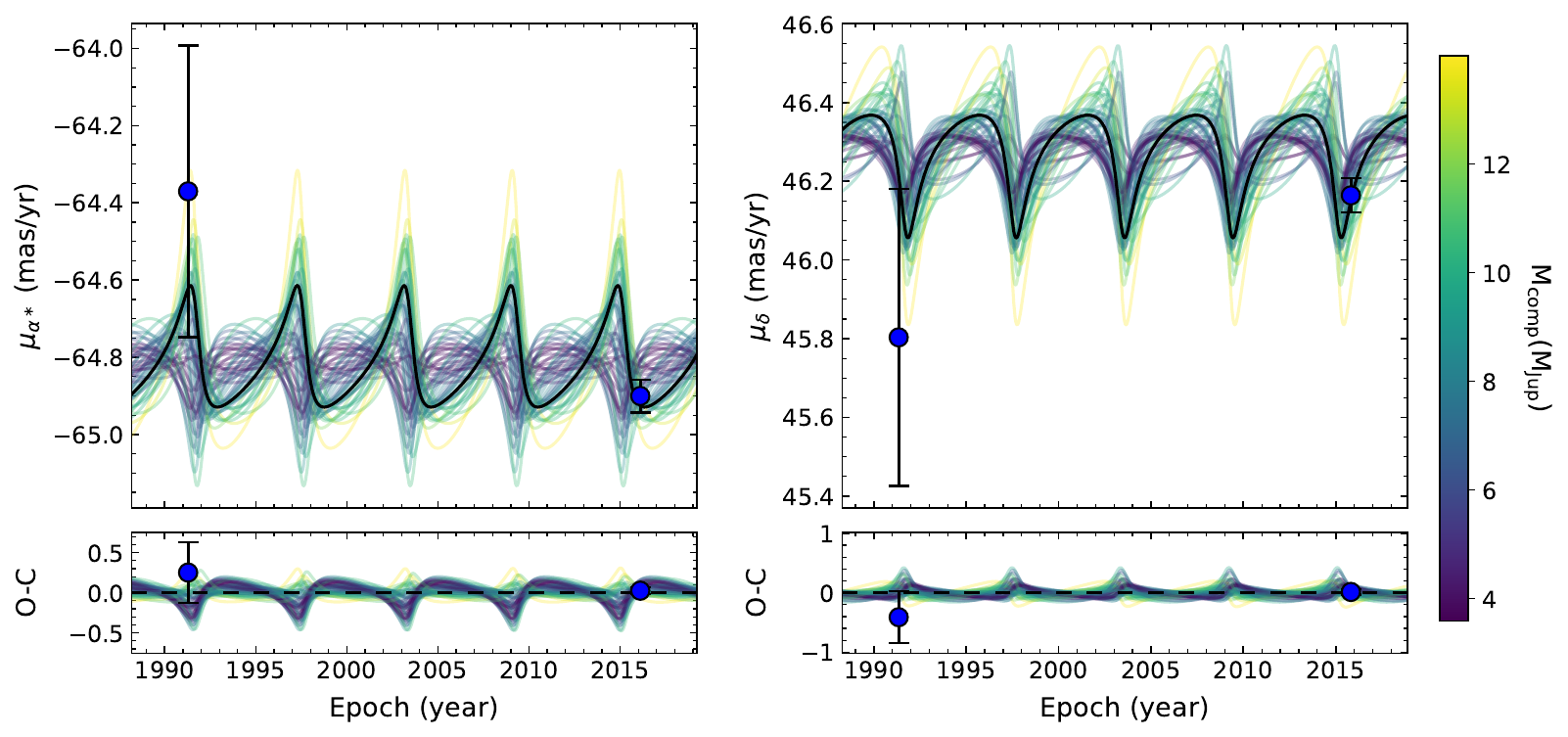}
\end{center}
\caption{
The joint fit of the HD 120084 b using RV and astrometry.
Left: The RV curve with RVs in red circles. 
Middle and right: The proper motion variations of HD 120084 along RA and Dec.
The best-fit orbital solution is shown by a black solid line, and another 50 orbits randomly taken from MCMC chains are shown by lines with colors, corresponding to the estimated masses of HD 120084 b. 
}\label{fig:HD120084_orvara}
\end{figure*}
\begin{figure*}
\begin{center}
\includegraphics[scale=0.38]{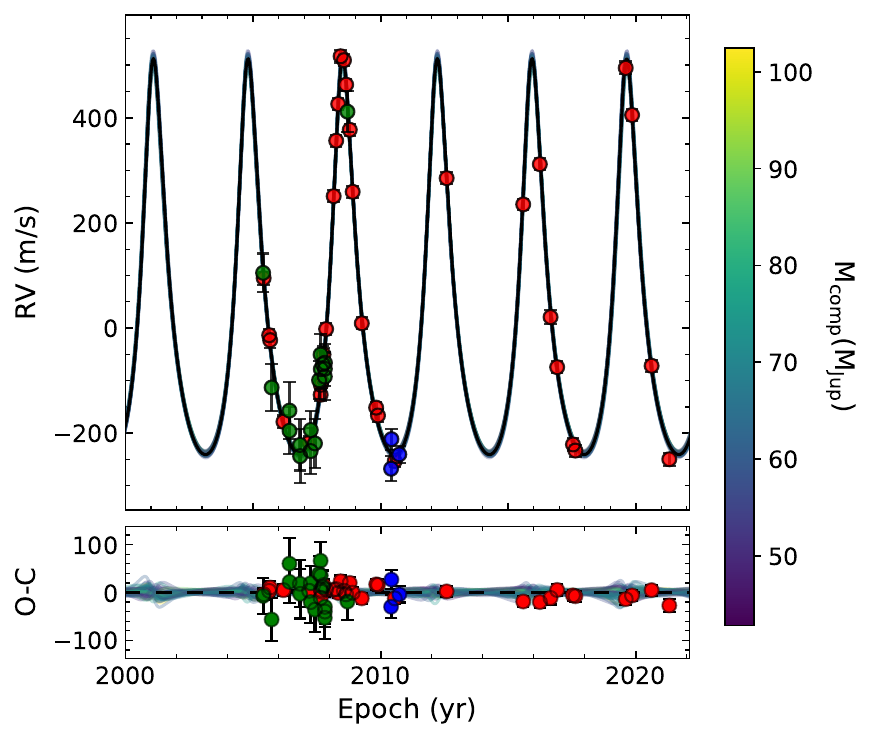}
\includegraphics[scale=0.38]{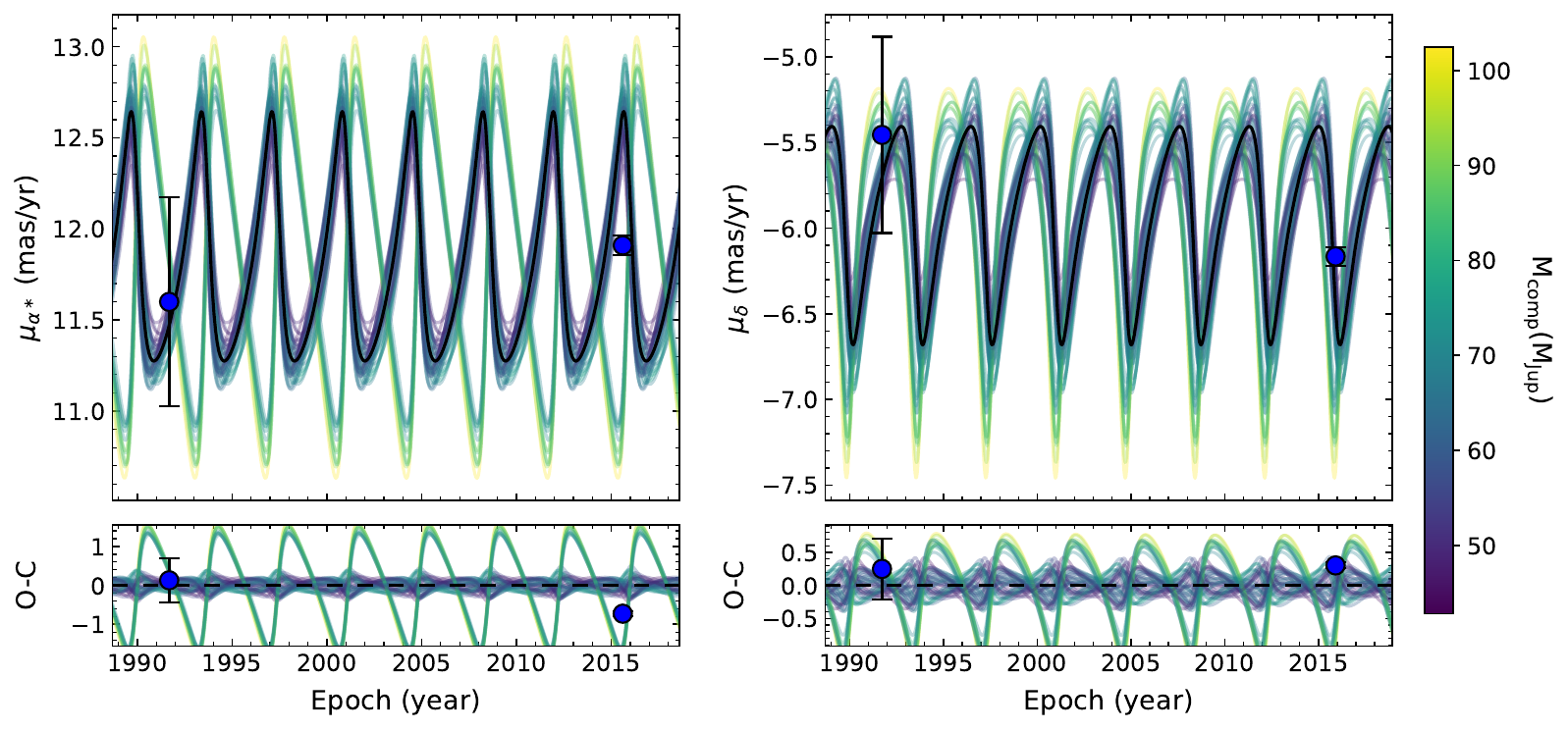}
\end{center}
\caption{
The joint fit of the HD 175679 b using RV and astrometry.
Left: The RV curve with RVs in red circles. 
Middle and right: The proper motion variations of HD 175679 along RA and Dec.
The best-fit orbital solution is shown by a black solid line, and another 50 orbits randomly taken from MCMC chains are shown by lines with colors, corresponding to the estimated masses of HD 175679 b. 
}\label{fig:HD175679_orvara}
\end{figure*}

As introduced in Section \ref{sec:res}, we refined the orbital parameters of the four systems by RV-only analyses (Figure \ref{fig:HD5608_1pl_phase}, and Table \ref{tab:orbpar_orbits}).
However, we did not directly adopt RVs in those analyses to the joint constraint, instead, we re-extracted the RVs from HIDES-S, -F1, and -F2 by using the same reference spectra.
This mainly aimed at two points:
\begin{itemize}
    \item To better express RV long-term trend in decade-long baseline;
    \item To avoid fake RV long-term trends or fake RV offsets introduced by over-fitting,
\end{itemize}
although this could include acceptable RV offsets between different observation modes (Appendix \ref{sec:inst_stability}) and lose some precision for HIDES-F1 and F2.
We then performed Keplerian orbital fits in the same way as described in Section \ref{sec:observations} and obtained new orbital parameters, RV long-term trend, RV offsets, and extra jitters. 
Specifically for HD 5608, the significant astrometric acceleration is triggered by the outer stellar companion and manifests as RV long-term acceleration, we thus subtracted the orbital motion due to the inner planetary companion with its best-fit model.
So far, we have finished the preparation of RV data.

The stars' absolute astrometry data were directly obtained from the HGCA EDR3 version. 
In addition, HD 5068 has relative astrometry for its distant companion from direct imaging \citep{Ryu2016}.

We performed joint constraints by using \texttt{orvara} software \citep{Brandt2021b}, which adopts astrometric and RV measurements and generates MCMC simulations to further estimate the orbital parameters.
In the joint constraints, we applied the default orbital parameters provided by \texttt{orvara}.
We set all parameters free with moderate priors. 
We adopted Gaussian priors on stellar mass according to the values derived in Section \ref{sec:opsp}, log-uniform prior on extra RV jitter between $1\ \rm{m}\ \rm{s}^{-1}$ and 5$\sigma$ of fitted extra jitters from \texttt{RadVel}.
and default priors provided by \texttt{orvara} on other parameters.

We generated MCMC simulations with \texttt{ptemcee} \citep{Foreman-Mackey2013, Vousden2016} to explore the parameter space. 
For each star, we generated 15 temperatures and 100 walkers with at least $4 \times 10^{5}$ steps per walk. 
Here we briefly address the definition of ``temperature'' from \citet{{Vousden2016}}. 
At higher temperatures, the posterior distribution approaches the prior, facilitating an efficient exploration of the entire prior volume by the chain without becoming trapped in areas of parameter space with high probability density. 
Conversely, at lower temperatures, a chain can more effectively sample from high-probability regions.
In each simulation, we recorded every 50 steps and set 20\% of the chain as the burn-in length.
The orbital parameters fitted and derived by joint constraints are given in Table \ref{tab:astrometry_res}. 

\subsection{HD 5608 B}
The significant acceleration in astrometry is triggered by its stellar companion which was first confirmed by direct imaging by \citet{Ryu2016}. 
Here, we comprehensively adopted the latest RVs and absolute and relative astrometry and confirmed the stellar companion. 
HD 5608 B has a best-fit mass of ${121.0}_{-7.0}^{+8.3}\ M_{\rm{J}}$ and it agrees with the mass estimation of $0.10 M{\odot}\ (105.8 M_{\rm{J}})$ in \citet{Ryu2016} within 2$\sigma$. 
In addition, HD 5608 B has a large estimated eccentricity of 0.53, resulting in a periastron distance of $a(1-e) \sim 246\ \rm{au}$. 
The best-fit orbit with RV and astrometry measurements is shown in Figure \ref{fig:HD5608_orvara}.

We note that the inner planetary companion has a slight eccentricity of about 0.1, thus it is reasonable to consider the Kozai mechanism as a trigger to the eccentricity. 
The oscillation timescale of the Kozai mechanism can be estimated according to \citet{Holman1997}:
\begin{equation}
\label{eq:kozai}
P_{\rm Kozai} \sim \frac{M_{\rm A}}{M_{\rm B}} 
\frac{P_{\rm B}^{2}}{P_{\rm pl,0}}(1-e_{\rm B}^{2})^{3/2}
\end{equation}
where $M_{\rm A}$ is the mass of the primary star, $M_{\rm B}$ is the mass of stellar companion, $P_{\rm B}$ is the period of the companion, $P_{\rm pl,0}$ is the initial orbital period of the planet, and $e_{\rm B}$ is the eccentricity of the stellar companion. 
Taking the best fit of HD 5608 B into the equation, we obtained the timescale $P_{\rm Kozai} \sim 68 \rm{kyr}$, which is sufficiently shorter than the system's age (over Gyr).

We expect future observations, including continuous RV monitoring and direct imaging, which can lead to tighter constraints on the orbital configuration of the HD 5608 system.
We also hope to deeply investigate the stability and Kozai mechanism of this system with new observations so as to recover the evolutionary progress of the system. 

\subsection{HD 14067 b}
From the latest RVs, we confirmed that the wide-orbit solution ($P \sim 2800 \rm{d}$) of HD 14067 b given by \citet{Wang2014} should be the true scenario.
By combining the RVs with astrometry data, we estimated a best-fit semimajor-axis ($a$) of ${5.14}_{-0.22}^{+0.21}$ au, a mass of ${15.3}_{-4.5}^{+6.2} M_{\rm{J}}$, and an inclination of ${39}_{-13}^{+28\circ}$ or ${147.5}_{-13}^{+6.6\circ}$.
Furthermore, the joint constraint also supports the highly eccentric orbit with the best fit of $e = {0.650}_{-0.028}^{+0.027}$.
The best-fit orbit with RV and astrometry measurements is shown in Figure \ref{fig:HD14067_orvara}.

\subsection{HD 120084 b}
HD 120084 b has a best-fit semimajor-axis ($a$) of ${4.21}_{-0.19}^{+0.17}$ au, a mass of ${6.4}_{-1.9}^{+2.9} M_{\rm{J}}$, and an inclination of ${38}_{-12}^{+22\circ}$ or ${139}_{-29}^{+15\circ}$. 
The best-fit orbit with RV and astrometry measurements is shown in Figure \ref{fig:HD120084_orvara}.

\subsection{HD 175679 b}
From the joint constraint, HD 175679 b has a best-fit semimajor-axis ($a$) of ${3.36}_{-0.13}^{+0.12}$ au, mass of ${62.2}_{-9.5}^{+15} M_{\rm{J}}$, and inclination of ${39.2}_{-5.1}^{+6.2\circ}$ or ${152.4}_{-4.4}^{+3.3\circ}$. 
Our fitting answers the remaining question in \citet{Wang2012}: the companion is more likely to be a massive brown dwarf rather than a low-mass star. 
The best-fit orbit with RV and astrometry measurements is shown in Figure \ref{fig:HD175679_orvara}.

\newpage
\section{Discussion}\label{sec:discussion}
Hunting for extra companions in the known systems around evolved stars in Okayama Planet Search Program (OPSP) was inspired by the increasing world of multiple-planet systems and their particular role in terms of the formation and evolution of planetary systems. 
The revisited systems were mostly published years ago, with only RV measurements covering several orbital phases. 
In addition, the masses of these central stars were estimated with coarse isochrones. 
Thus, we both refined the orbital parameter of the planets with new RVs, 
and the re-estimated properties of their central stars, 
so that stars and planets could be distributed in proper populations, and RV residuals could be examined for extra signal.

In this section, we focus on the multiplicity of planetary systems around evolved stars. 
We investigate its correlation with the metallicity of host stars and the total mass of planets in the systems.

\subsection{Sample selection}\label{sec:discussion-sample}
\begin{figure}
\begin{center}
\includegraphics[scale=0.45]{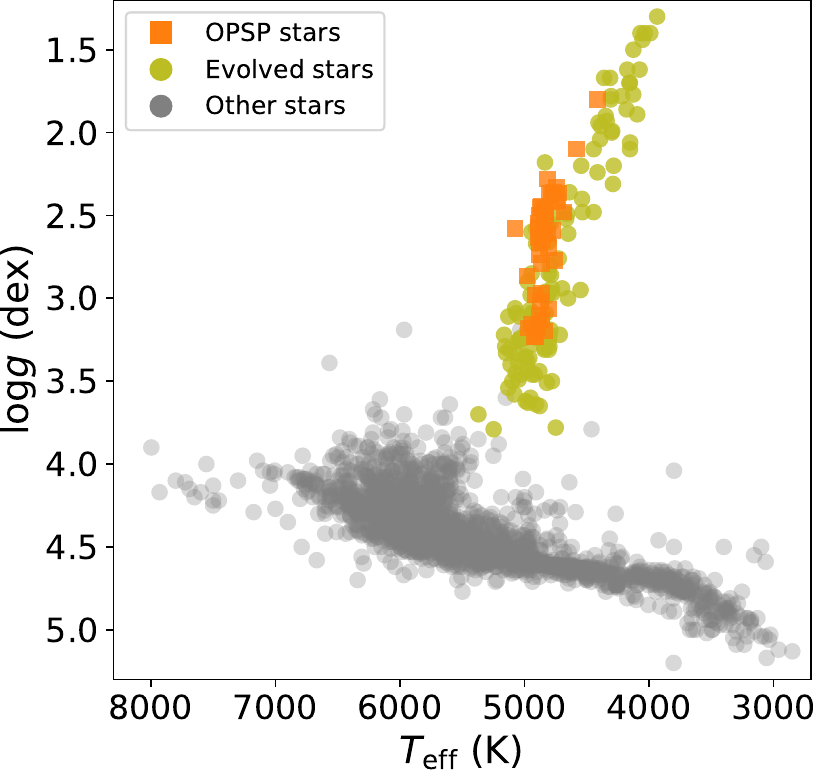}
\end{center}
\caption{
HR diagram of planet-harboring stars. 
The evolved stars selected for this study are marked with colors. Specifically, the ones observed by OPSP are marked with orange squares.
}\label{fig:f_HR_full_sample}
\end{figure}
\begin{figure}
\begin{center}
\includegraphics[scale=0.45]{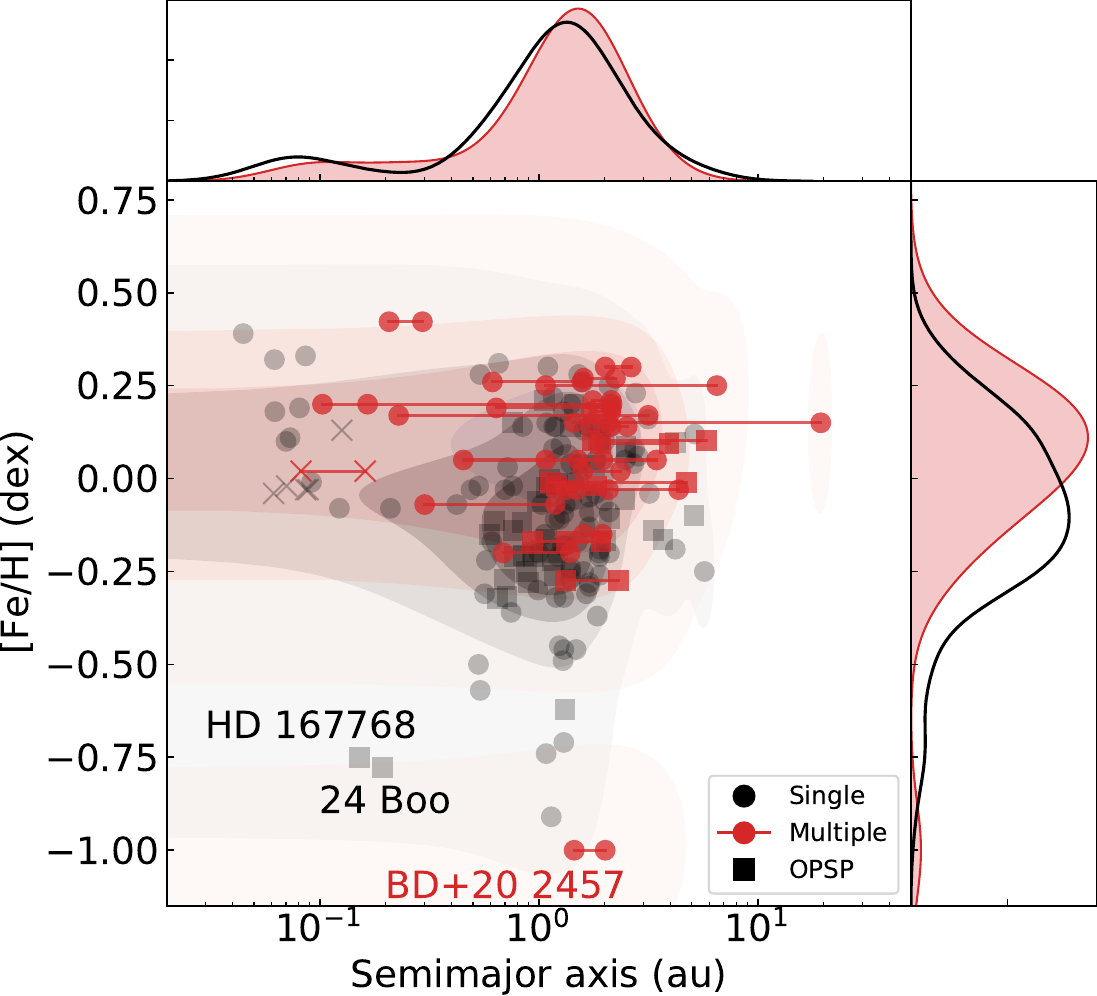}
\end{center}
\caption{
Metallicity as a function of semimajor axis for known planetary systems orbiting evolved stars. 
Each data point corresponds to a confirmed planet, with multiple-planet systems indicated by a red marker and connected by a straight line. 
Planets identified by the OPSP are marked in squares, and those without a measured mass are indicated by a cross. 
All contour plots are flattened for visualization purposes, with red contours indicating multiple-planet systems and non-colored contours indicating single-planet systems.
}\label{fig:f_sma_met_full_sample}
\end{figure}
Our sample selection was based on a data crossing from both \url{Exoplanet.eu} and NASA exoplanet archive (\url{https://exoplanetarchive.ipac.caltech.edu})\footnote{We acquired the data on 2022 December 17th.}.
We basically followed planetary systems and their parameters given by \url{Exoplanet.eu},
but we filled in the blanks in \url{Exoplanet.eu} if there were records provided by the NASA exoplanet archive or SIMBAD (\url{http://simbad.u-strasbg.fr/simbad}).
We also supplemented the confirmed planetary systems in the NASA exoplanet archive to our samples if they did not emerge in \url{Exoplanet.eu}.

We selected planetary systems around evolved stars. 
The evolved stars were defined as
$\log g < 3.8$ (less than one-fourth of the surface gravity of the Sun), $T_{\rm{eff}} < 5400\ \rm{K}$, and $R > 2 R_{\odot}$ 
to ensure that host stars are evolved, and they should be out of the main sequence.
Consequently, we obtained 172 planetary systems (Figure \ref{fig:f_HR_full_sample} and \ref{fig:f_sma_met_full_sample}), including 143 single-planet systems ($N_{\rm{pl}} = 1$) and 29 multiple-planet systems ($N_{\rm{pl}} \ge 2$), 156 systems detected by RVs, and 16 systems detected by transit.

For the selected samples, we believe that there would not be a large bias to cover planets having short or intermediate orbital periods of $P \lesssim 10^3\ \rm{d}$. 
It is because most of the systems were detected by RVs, whose observational time spans were normally longer than a few years, and those detected by transit in $Kepler$ were also observed at a time scale of four years. 
The current planet discoveries around evolved stars are all gas giant planets. The least massive one is \textit{Kepler}-56 b \citep{Huber2013}, with a minimum mass of $M_{\rm{p}} = 0.06 M_{\rm{J}}$.

In order to estimate the properties of planetary systems around evolved stars, we performed a bootstrap resampling, which is efficient for small samples that cannot be described by Gaussian distribution, or for samples whose distribution is unknown. 
We also expected that, via resampling, we could reduce the influence of extreme cases and recover less biased statistical properties, e.g., median value.
We randomly shuffled all the selected systems $1 \times 10^{5}$ times, and we picked up $\lceil 2\times\sqrt{N} \rceil=27$ samples in each shuffle.
In addition, we injected moderate Gaussian noises into the parameters of each sample.
Specifically, we set $\sigma = 0.1$ dex to metallicity (the iron abundance [Fe/H]) of each host star, and $\sigma = 0.2 M_{\rm{p,tot}}$ to the total planet mass in each system.

At first glance, we noted that the resampled metallicity is slightly higher than the one directly calculated from the true 172 samples.
The resampling revealed a mean and median of  
$-0.04 \pm 0.05$ and $-0.02 \pm 0.06$, respectively,
while the true samples were $-0.07$ and $-0.04$.
These are biased by planet-harboring stars with extremely low metallicity, e.g., HD 4760 \citep{Niedzielski2021}, BD+20 2457 \citep{Niedzielski2009}, etc.

\subsection{Color cut-off bias in OPSP}
Originally, a color cut-off of $0.6 \lesssim B-V \lesssim 1.0$ was adopted in OPSP star selection standards to achieve intrinsic RV stability of $\sigma \sim 20\ \rm{m}\ \rm{s}^{-1}$. 
We basically kept this standard during the operation of the program, including the extensions of EAPS-Net.
However, some studies revealed that a cut-off could result in the exclusion of the more metal-rich and low-gravity stars {from the evolved star samples in different programs} \citep{Mortier2013, Wittenmyer2017}.
{In addition, \citet{Mortier2013} calculated $B-V$ values with different given $\log g$ and $\rm{[Fe/H]}$ values for a given $T_{\rm{eff}}$ of 4850 K with a calibration of \citet{Sekiguchi2000} which was summarized from 537 Infrared Space Observatory standard stars. Here, 4850 K is a typical temperature of an evolved star in OPSP and other planet survey around evolved stars. From their calculation, a star having $\log g = 2.5$, a typical gravity in OPSP, will be probably ruled out with $B-V < 1.0$ cut-off if the star has higher $\rm{[Fe/H]}$ value.}
{Since the metal-rich stars are ruled out due to the strong $B-V$ selection effect, planetary systems around metal-rich stars were not surveyed as a result.}

{Similarly, we have found the OPSP planet-harboring stars tend to be metal-poor than the sample that we selected in the above subsection. We derived mean and median of metallicity ($\rm{[Fe/H]}$; we all use $\rm{[Fe/H]}$ hereafter for metallicity hereafter) from bootstrap resampling and obtained $-0.13 \pm 0.08$ and $-0.12\pm 0.08$, respectively. These values} were lower than the ones of the whole evolved planet-harboring star samples of $-0.04 \pm 0.05$ and $-0.02 \pm 0.06$ at $\lesssim 2\sigma$ level.
A Kolmogorov-Smirnov test (K-S test: \cite{Kolmogorov1933}, \cite{Smirnov1948}) revealed a $p$-value equal to zero, which indicated a distinction between OPSP planet-harboring stars and the whole sample. 
{Here, we note that the sample selected in the above subsection should be somewhat biased, because this sample was a most inclusive one among planetary systems around evolved stars, and it was obtained from a mixture of planetary systems around evolved stars from different surveys, including RV survey and transit survey with different star selection criteria.}
{But this whole sample is enough for the evidence of the selection effect of color cut-off of OPSP for its metal-poor contribution to the whole sample.}
{Additionally, the shuffle in bootstrap resampling was uniformly conducted, which would not introduce new selection effect when calculating the mean and median value. Therefore, we believe the our metallicity estimation can properly reflect the metal-poor selection effect due to $B-V$ color cut-off in OPSP.}
{Furthermore, we anticipate that a large unbiased sample with $\sim$50-100 planetary systems around evolved stars can be established in the future, and reliable statistics and selection effect can be diagnosed within this sample.}

The positive correlation between the occurrence of giant planets and metallicity could be concluded from surveys without a color cut-off, e.g., \citet{Reffert2015, Jones2016, Wittenmyer2016b}. 
\citet{Wolthoff2022} recently investigated the aforementioned three surveys and also yielded a positive result in evolved stars.
However, the color cut-off bias around evolved stars could lead to opposition, manifesting a null correlation. 
\citet{Wittenmyer2017} showed the positive correlation could be hardly summarized from surveys, e.g., OPSP \citep{Takeda2008} and \citet{Mortier2013}, which contained color cut-off to avoid redder stars. 
{As for our OPSP, we have not yet finished the survey, and giving an updated correlation of giant planet occurrence within OPSP is out of the scope of this paper. The relevant study will be presented in a forthcoming paper.}

The positive giant planet-metallicity correlation could be considered as evidence for the core-accretion model (e.g., \cite{Pollack1996}), 
in which planet formation requires solid materials to concrete a core before the accretion of gas in the protoplanetary disk, and thus higher stellar metallicity relates to the planet's higher formation rate.
Mechanism independent from metallicity, e.g., gravitational instability \citep{Boss1997}, 
in which a companion directly forms by the gravitational collapse of the gas without a solidified progenitor, can also contribute to planet formation.

Recently, some studies revealed that the gas giant companions of different masses might refer to different populations and form via different mechanisms. 
More specifically, planets as massive as Jupiters are likely to form via core-accretion, while brown dwarfs or planets more massive than several Jupiter masses may form via gravitational disk instability.
\citet{Schlaufman2018} showed that planets or substellar companions more massive than a boundary, (e.g., $10 M_{\rm{J}}$) may not orbit metal-rich stars, and \citet{Maldonado2017} and \citet{Maldonado2019} confirmed that BD hosts do not show the giant-planet metallicity correlation.
\citet{Santos2017, Adibekyan2019}, and \citet{Sousa2021} claimed different formation mechanisms of planets with masses above and below $4 M_{\rm{J}}$, and \citet{Adibekyan2019} also argued that planets of the same (high) mass could be formed through different channels.
Thus, the planet-metallicity correlation should be a key to the formation scenario of giant planets.

\subsection{Metallicity and Multiplicity}
\begin{figure*}
\begin{center}
\includegraphics[scale=0.45]{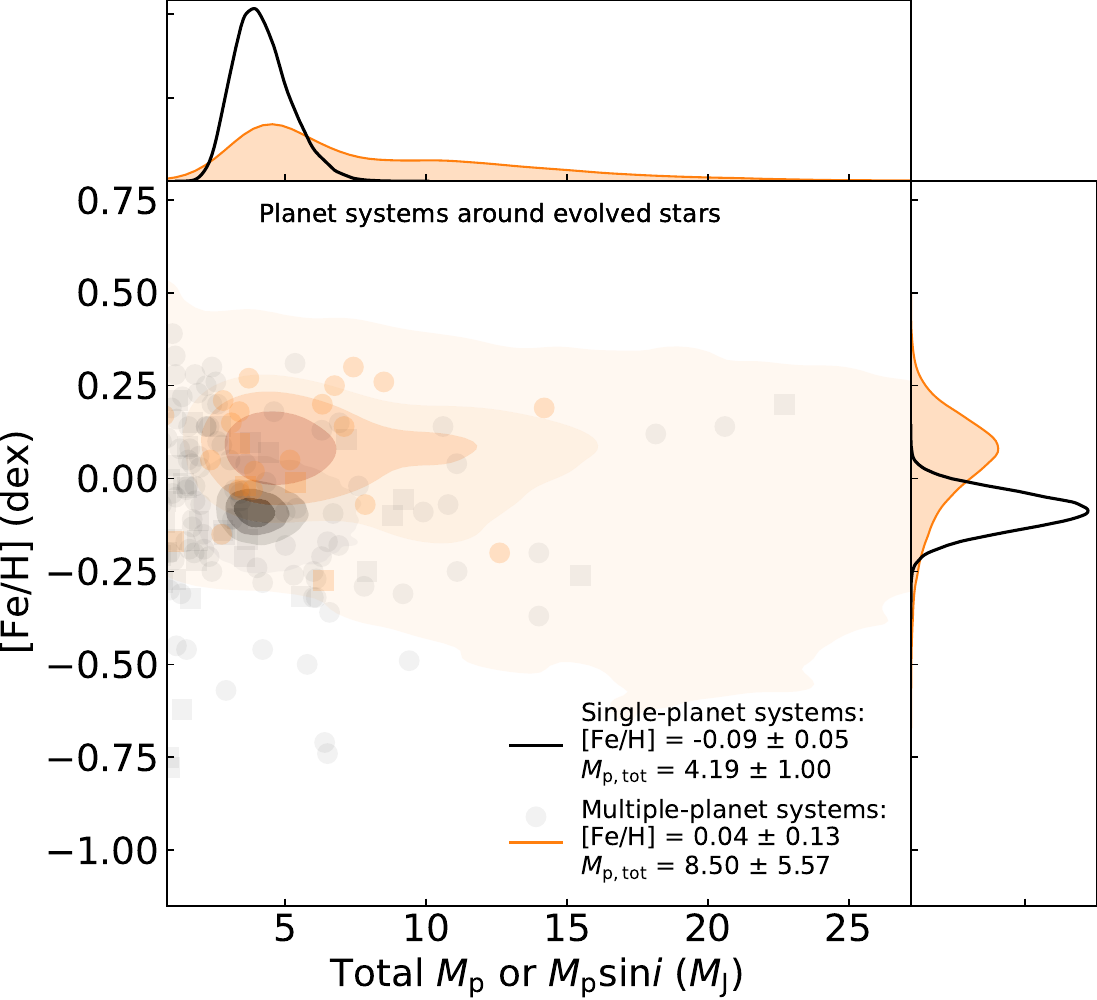}
\includegraphics[scale=0.45]{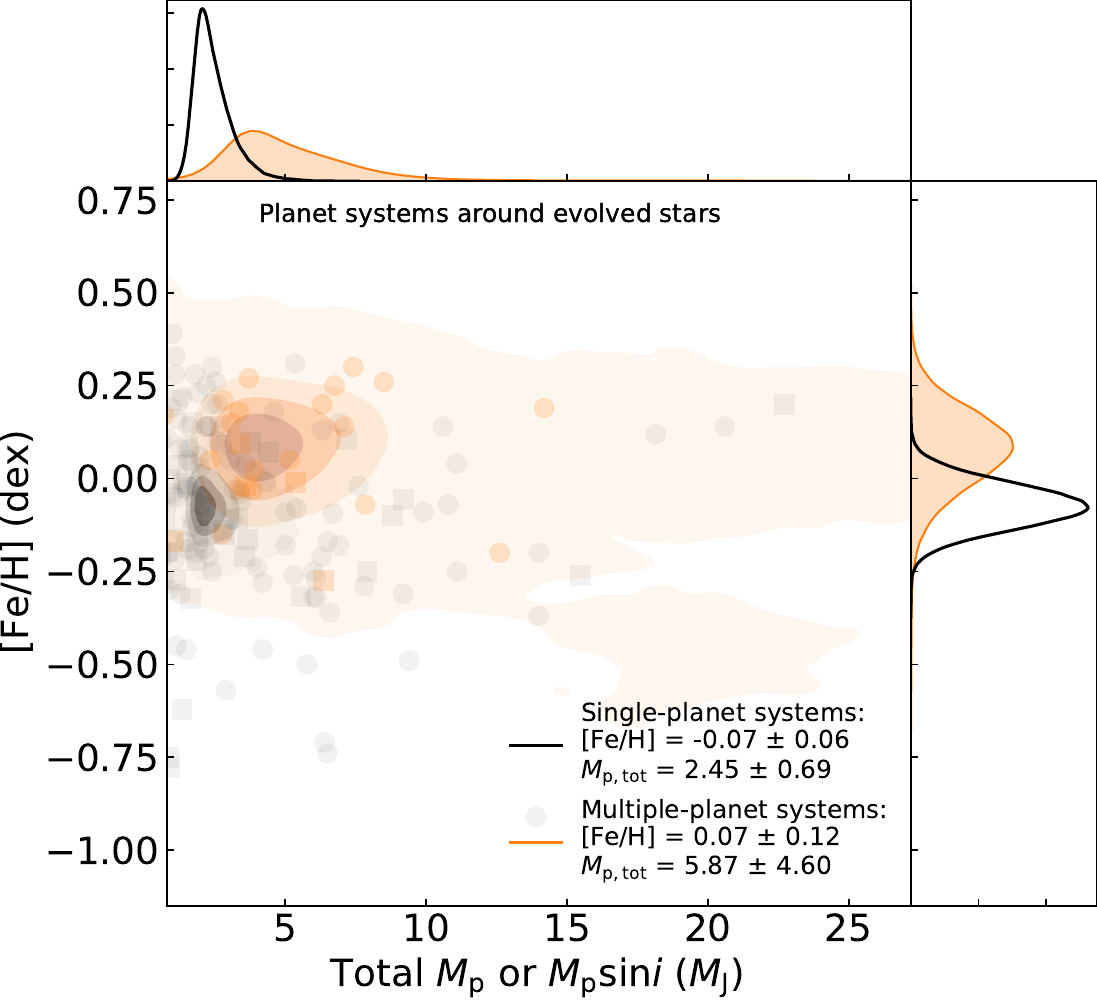}
\end{center}
\caption{
Distribution of resampled metallicity against total planet mass for all selected planetary systems around evolved stars.
(Left: distribution of mean values. Right: distribution of median values.)
The orange contours refer to multiple-planet systems while non-colored ones refer to single-planet systems. 
The scattered plots are the selected planetary systems around evolved stars to generate the resampling.
}\label{fig:f_bootstrap_full_sample}
\end{figure*}
\begin{figure*}
\begin{center}
\includegraphics[scale=0.45]{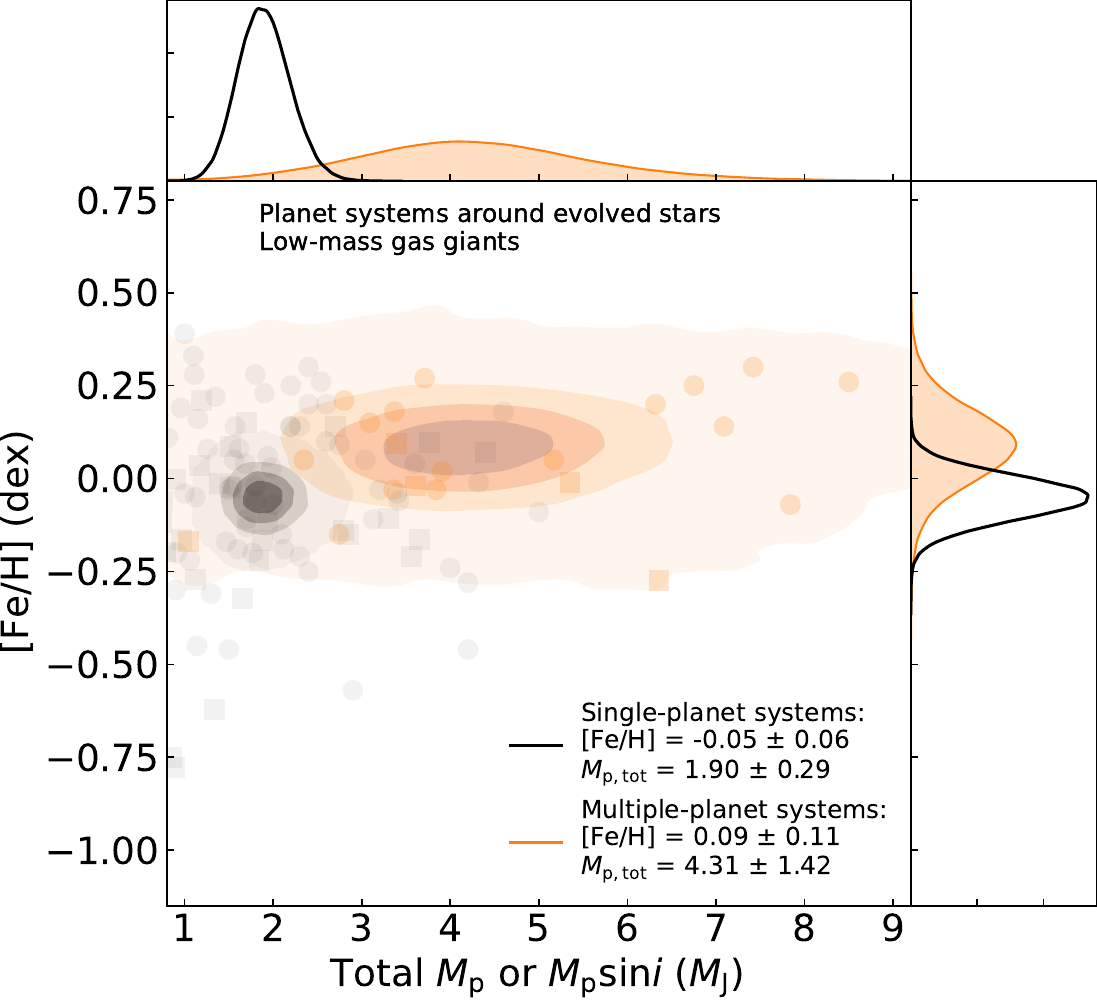}
\includegraphics[scale=0.45]{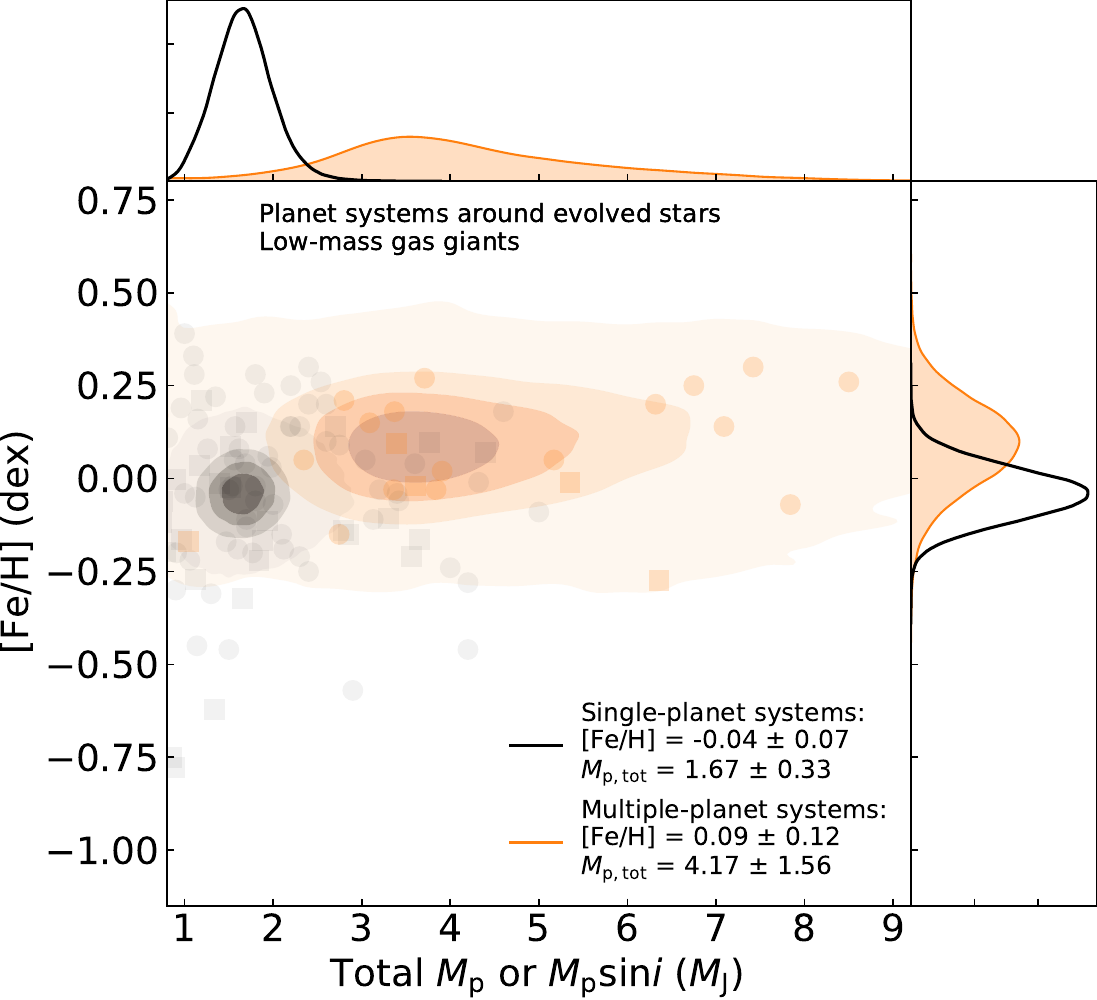}
\end{center}
\caption{
Distribution of resampled metallicity against total planet mass for planetary systems harboring Jupiter-mass planets around evolved stars.
(Left: distribution of mean values. Right: distribution of median values.)
The orange contours refer to multiple-planet systems while non-colored ones refer to single-planet systems. 
The scattered plots are the selected planetary systems around evolved stars to generate the resampling.
}\label{fig:f_bootstrap_lm}
\end{figure*}
\begin{figure*}
\begin{center}
\includegraphics[scale=0.45]{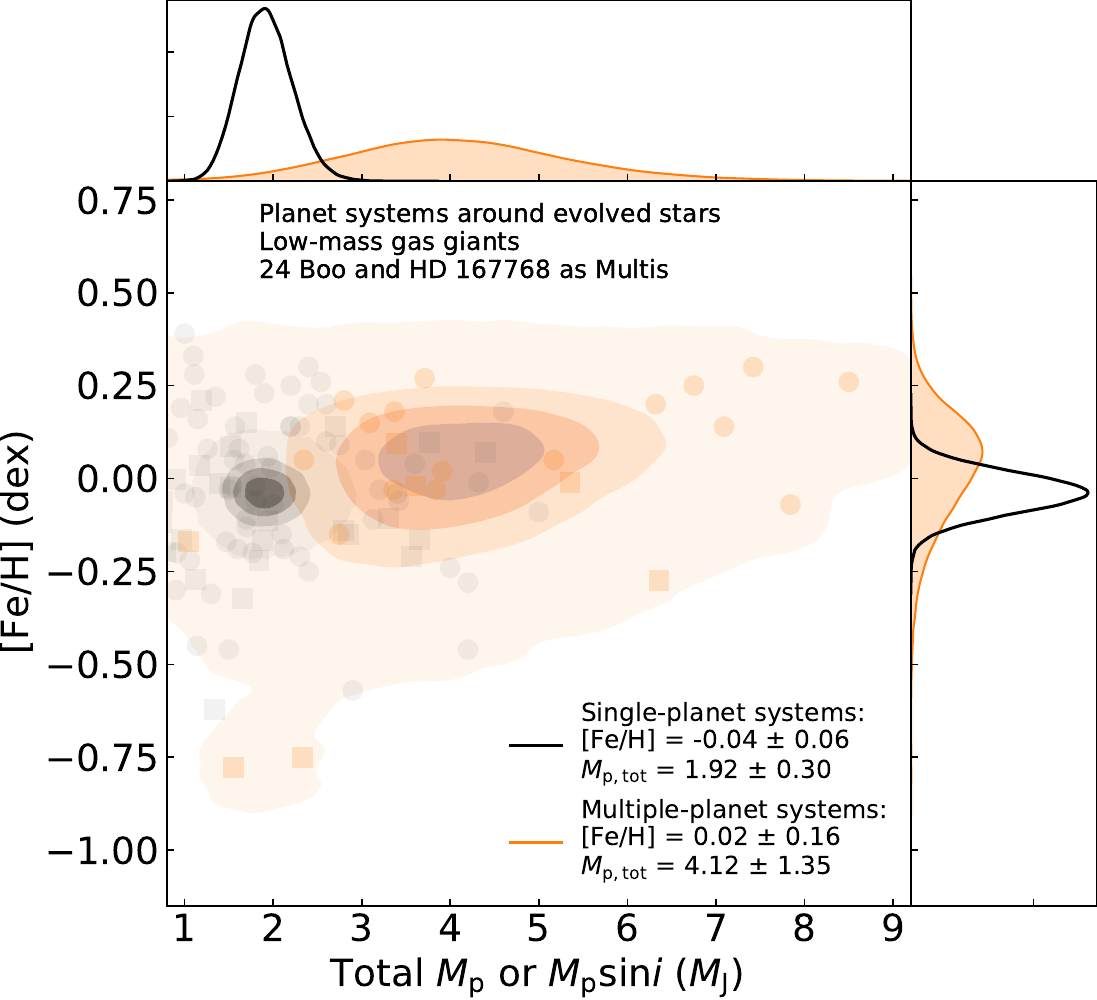}
\includegraphics[scale=0.45]{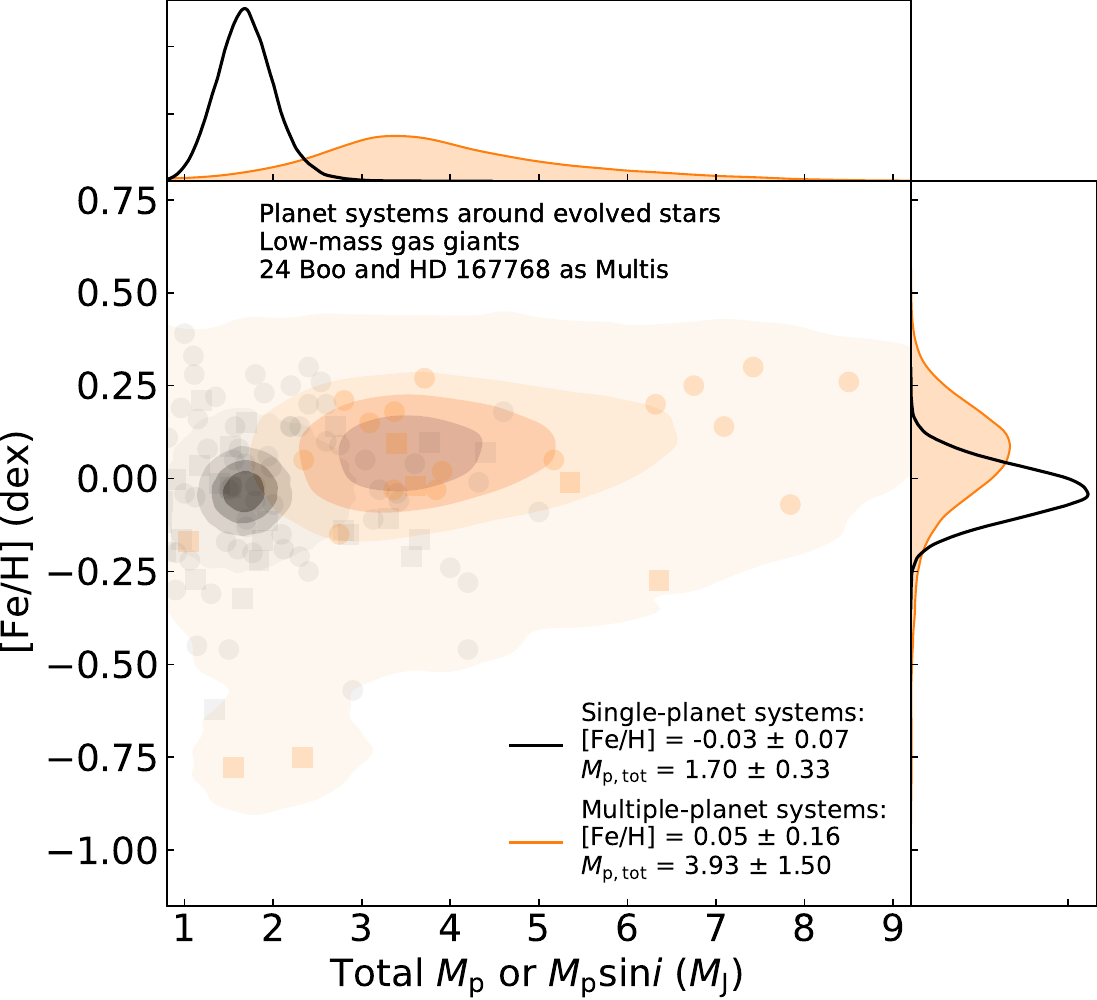}
\end{center}
\caption{
Distribution of resampled metallicity against total planet mass for planetary systems harboring Jupiter-mass planets around evolved stars and 24 Boo and HD 167768 system as multiple-planet systems
(Left: distribution of mean values. Right: distribution of median values.)
The orange contours refer to multiple-planet systems while non-colored ones refer to single-planet systems. 
The scattered plots are the selected planetary systems around evolved stars to generate the resampling.
}\label{fig:f_bootstrap_teng2022cd_lm}
\end{figure*}
\citet{Fischer2005} firstly pointed out that high stellar metallicity appeared to be correlated with the presence of multiple-planet systems. 
In this study, we tested this trend on evolved stars.

First, we divided the whole samples into single-planet systems (hereafter: singles) and multiple-planet systems (hereafter: multis).
The resampling revealed that the number ratio of multis to singles around evolved stars was $0.21 \pm 0.10$, which is lower than one of all detected planetary systems ($\sim 0.28$ by \url{exoplanet.eu}).
Qualitatively, as a planetary system and its host star evolve, the system could lose planet(s) due to tidal interactions between the planet(s) and its host stars, between planets, or between planet(s) and fly-by body(ies). 
Thus the ratio of multis to singles should be averagely lower at the higher evolutionary stage of the planetary systems. 

For metallicity, the resampling revealed the mean metallicities of multis and singles were, respectively, $0.04 \pm 0.12$ and $-0.09 \pm 0.06$, suggesting multis should be averagely more metal-rich (Figure \ref{fig:f_bootstrap_full_sample}).
A K-S test with $p$-value equal to zero revealed the distinction between the two samples.
The resampling also revealed that the average total planet mass of multis is slightly higher than that of singles. 
But it should be noted that the mean value might not reflect the actual case.
The masses of gas giant companions and the metallicity of host stars differ significantly. 
A brown dwarf can be $\sim10^{2}$ times more massive than a Saturn-mass planet,
and a planet-harboring star in the galactic thick disk is averagely more metal-poor than the one in the galactic thin disk.
Therefore, we checked the median values of metallicity and total planet mass from the resampling.
Similarly, multis appeared to be more metal-rich and more massive than singles. 
But the median total masses of both multis and singles were significantly lower than the mean values by a few Jupiter masses, which indicated the overestimation in total masses by the mean value.

As introduced in the former subsection, the occurrence of planets as massive as Jupiters tends to agree well with the giant-planet metallicity correlation, while the occurrence of massive gas companions does not.
Hence, we further focused on planetary systems containing Jupiter-mass planets.
We restricted each system's mean planet mass to a maximum of $5 M_{\rm{J}}$ (for simplicity, we call them Jupiter-mass planets), and we examined the metallicity, multiplicity, and total planet mass.
Consequently, the resampling revealed mean metallicities of $0.09 \pm 0.11$ and $-0.05 \pm 0.06$, 
and mean total planet mass of $4.31 \pm 1.42\ M_{\rm{J}}$ and $1.90 \pm 0.29\ M_{\rm{J}}$, respectively for multis and singles (Figure \ref{fig:f_bootstrap_lm}). 
The median metallicity was illustrated similarly by $0.09 \pm 0.11$ and $-0.04 \pm 0.06$ respectively for multis and singles. 
Meanwhile, the mass difference between the median and mean was reduced to $< 0.3\ M_{\rm{J}}$ with the restriction to the planet masses (Figure \ref{fig:f_bootstrap_lm}).
Both revealed the trend that metallicity positively correlates with multiplicity and total planet mass for planetary systems having Jupiter-mass planets.
As the formation of the core of a Jupiter-like planet depends on the metallicity of the star and its disk, we thus believe the positive correlation can be considered to be evidence for the core-accretion model (e.g., \cite{Ida2004}).

Moreover, we became interested in two planetary systems during our discussions on planetary systems having Jupiter-mass planets.
The systems, 24 Boo and HD 167768, have been identified as single stars but with potential planet candidates, as indicated by our current work and previous research by \citet{Teng2022c}.
Both host stars belong to the thick disc population metal-poor with $\rm{[Fe/H]} \lesssim -0.7$ but enhanced alpha element abundances.
To investigate the impact of these two systems, we conducted additional resampling by considering them as multis,
utilizing the masses of the planet candidates (two for 24 Boo and two for HD 167768) obtained from this study and \citet{Teng2022c}.
Consequently, we confirmed a marginal positive correlation between metallicity and multiplicity/total mass, 
but the mean and median metallicities of multis decreased respectively by 0.07 and 0.04 dex, respectively, compared to when these systems were assumed to be singles (Figure \ref{fig:f_bootstrap_teng2022cd_lm}).

Thus, according to the rare cases of 24 Boo and HD 167768, we hereby raise three questions in this paper:
\begin{itemize}
    \item Is [Fe/H] the best (or a proper) parameter to describe the metallicity of a planetary system?
    \item Does a giant planet in the galactic thick disk refer to the same planet population as the one in the thin disk?
    \item If a giant planet in thick disk forms via the core-accretion mechanism, is the core composition the same or different from the one of planets in think disk?
\end{itemize}
Answering these questions is out of the scope of this data-release paper.

\newpage
\section{Summary}\label{sec:summary}
In the present study, we conducted a revisitation of 32 evolved stars known to harbor planets, as part of the Okayama Planet Search Program (OPSP) and its extended collaborative network, the East Asian Planet Search Network (EAPS-Net), using the most recent precise radial velocity (RV) measurements obtained at the Okayama Astrophysical Observatory (OAO).

We have confirmed one new wide-orbit planet in the known 75 Cet system (75 Cet c: $P = 2051.62\ \rm{d}$, $M_{\rm{p}}\sin i = 0.91 M_{\rm{J}}$), and we have confirmed the RV long-term trends in five stars (HD 5608, $\kappa$ CrB, HD 167042, 18 Del, and HD 208897), which indicates the possible outer massive companions.
Additionally, we have further investigated the radial velocity (RV) variability of 11 stars for different reasons.
\begin{itemize}
\item $\epsilon$ Tau, 11 Com, and 24 Boo: We have detected additional periodic RV variations.
\item 41 Lyn and 14 And: We have found line profiles synchronously vary with RV by time. 
\item HD 32518 and $\omega$ Ser: We have measured that the amplitudes of RV variations are not stable from different data sets.
\item HD 5608, HD 14067, HD 120084, and HD 175679: We have jointly constrained the orbital parameters by RVs and astrometry data.
\end{itemize}

Furthermore, our analysis has provided confirmation that the strict $B-V$ color cut-off employed in the OPSP has resulted in a sample that is more metal-deficient. 
This conclusion was drawn from a comparison of planetary systems discovered by the OPSP to a complete sample of planetary systems that exist around evolved stars. 
Most significantly, our investigation of the complete sample reveals a tendency for metal-rich stars to possess a greater number of multiple-planet systems with a higher total planetary mass. 
This correlation is particularly pronounced in planetary systems that contain Jupiter-mass planets, {which can be strong evidence for the core-accretion planet formation model.}

\newpage
\begin{ack}

This research is primarily based on data collected at the Okayama Astrophysical Observatory (OAO), which is operated by the National Astronomical Observatory of Japan. 
We are grateful to all the staff members of OAO for their support during the observations.
We thank the observatory for allowing us to use the data obtained during the engineering times at OAO, including a portion of observations of the RV standard star $\tau$ Cet, $\iota$ Per, and $\epsilon$ Vir.
We thank the students of Tokyo Institute of Technology and Kobe University for their kind help with the observations at OAO. 
I.B., S.S., M.Y., are grateful to TUBITAK National Observatory, IKI, KFU and AST for their partial support in using RTT-150 (Russian - Turkish 1.5-m telescope in Antalya).
We express our special thanks to Yoichi Takeda for the support in stellar property analysis. 
We thank John A. Johnson, Brendan P. Bowler, Debra A. Fischer, and Andrew W. Howard for their kind help with the RV data observed at Lick observatory. We thank the anonymous referee for the valuable comments and suggestions to improve the manuscript. 

This research adopted data obtained from the Maunakea observatory.
The authors wish to recognize and acknowledge the very significant cultural role and reverence that the summit of Maunakea has always had within the indigenous Hawaiian community.  We are most fortunate to have the opportunity to conduct observations from this mountain.

B.S. was partially supported by MEXT's program ``Promotion of Environmental Improvement for Independence of Young Researchers" under the Special Coordination Funds for Promoting Science and Technology, and by Grant-in-Aid for Young Scientists (B) 17740106 and 20740101, Grant-in-Aid for Scientific Research (C) 23540263, Grant-in-Aid for Scientific Research on Innovative Areas 18H05442 from the Japan Society for the Promotion of Science (JSPS), and by Satellite Research in 2017-2020 from Astrobiology Center, NINS.
H.I. was supported by JSPS KAKENHI Grant Numbers JP16H02169, JP23244038.
M.Y. was supported by The Scientific and Technological Research Council of Turkey (TUBITAK), the project number of 114F099. 
I.B. was financed by subsidy FZSM-2023-0015 of the Ministry of Education and Science of the Russian Federation allocated to the Kazan Federal University for the State assignment in the sphere of scientific activities.

This research has made use of \textsc{IRAF}. It is distributed by the National Optical Astronomy Observatories, which is operated by the Association of Universities for Research in Astronomy, Inc. under a cooperative agreement with the National Science Foundation, USA.
This research has made use of the following \texttt{Python} packages for scientific calculation: 
\begin{itemize}
    \item[] Randomization, Likelihood maximization, MAP, K-S test: \texttt{NumPy} \citep{Harris2020}, \texttt{SciPy} \citep{Virtanen2020},
    \item[] GLS periodogram: \texttt{astropy} \citep{Astropy2013},
    \item[] Stellar properties: \texttt{isoclassify} \citep{Huber2017}, 
    \item[] Keplerian orbital fit: \texttt{RadVel} \citep{Fulton2017, Fulton2018}, \texttt{htof} \citep{Brandt2021}, \texttt{orvara} \citep{Brandt2021b},
    \item[] Gaussian Process: \texttt{celerite} \citep{Foreman-Mackey2017}
    \item[] MCMC sampling: \texttt{emcee} \citep{Foreman-Mackey2013}, \texttt{ptemcee} \citep{Foreman-Mackey2013, Vousden2016}.
\end{itemize}

This research has made use of the SIMBAD database, operated at CDS, Strasbourg, France.
This work has made use of data from the European Space Agency (ESA) mission
{\it Gaia} (\url{https://www.cosmos.esa.int/gaia}), processed by the {\it Gaia}
Data Processing and Analysis Consortium (DPAC,
\url{https://www.cosmos.esa.int/web/gaia/dpac/consortium}). Funding for the DPAC
has been provided by national institutions, in particular, the institutions
participating in the {\it Gaia} Multilateral Agreement.
This research has made use of the NASA Exoplanet Archive, which is operated by the California Institute of Technology, under contract with the National Aeronautics and Space Administration under the Exoplanet Exploration Program.
This research has made use of data obtained from tools provided by the portal \url{exoplanet.eu} of The Extrasolar Planets Encyclopedia.
\end{ack}

\newpage
\appendix
\section{Instrumental Stability of HIDES}\label{sec:inst_stability}
The instrumental stability of HIDES-S and -F1 was examined and discussed in \citet{Teng2022a}. 
With RV standard stars, we ensured that HIDES-S and -F1 could be stable with their individual modes at a level of $\lesssim 4 \rm{m\>s^{-1}}$.
Here, we examine the stability of HIDES-F2 independently and the 20-year long-term stability of HIDES with the combination of three individual observation modes.
We note that a slight order shift happened to the CCD in November 2019 due to an earthquake near the observatory, hence we also separately examine the stability of HIDES-F2 pre- and post-earthquake. 

\begin{figure*}
\begin{center}
\includegraphics[scale=0.5]{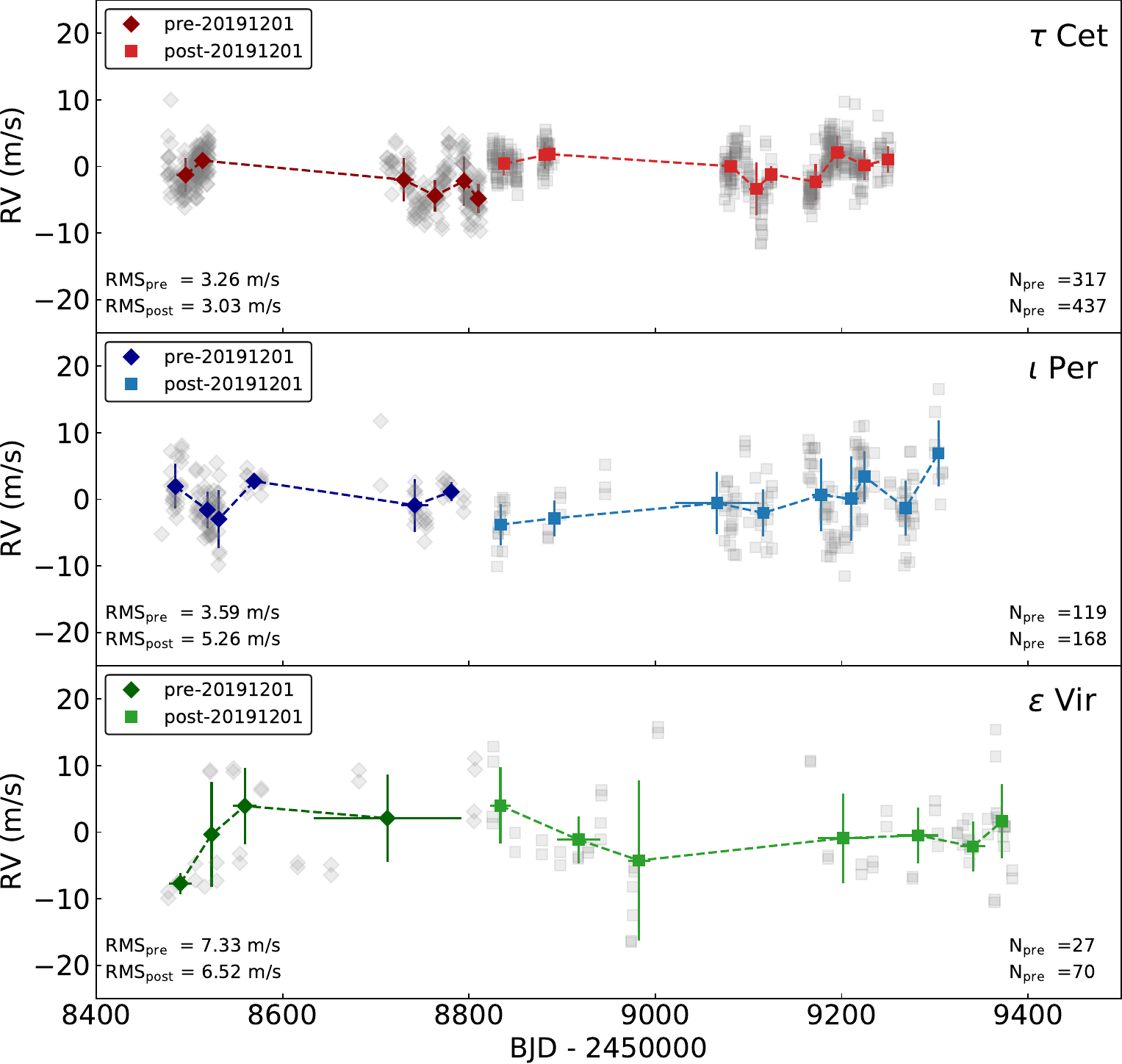}
\end{center}
\caption{
The RVs of $\tau$ Cet, $\iota$ Per, and $\epsilon$ Vir obtained by HIDES-F2. Single measurements are marked by non-colored markers, and binned measurements are marked by colored markers connected by dashed lines for indication of trends. 
We use 20191201 to indicate the earthquake, and HIDES was not scheduled between 2019 November 25th and 2019 December 6th.
}\label{fig:F2_rvs}
\end{figure*}

Three different RV standard stars were adopted to examine the instrumental stability of HIDES-F2 (Figure \ref{fig:F2_rvs}).
$\tau$ Cet (HD 10700, G8 V, $V=3.5$), known to have stable RV among solar-type stars, was set to be one of the RV standard stars in this study. 
The typical exposure time of $\tau$ Cet was set to be 5min, and the RVs were all measured from the same reference spectrum. 
Consequently, we obtained the root mean squares of pre- and post-earthquake of $3.26\ \rm{m\>s^{-1}}$ ($N_{\rm{obs}}=317$) and $3.03\ \rm{m\ s^{-1}}$ ($N_{\rm{obs}}=437$), and we did not observe significant RV offset between the occurrence of the earthquake.
It suggested the negligible influence of the order shift due to the earthquake for stars that emerged in OPSP. 
Similarly, we examined the stability of another two RV standard stars $\iota$ Per and (HD 19373, G0 V, $V=4.05$) $\epsilon$ Vir (HD 113226, G8 III, $V=2.79$) in OPSP \citep{Sato2005}.
Two stars showed typical errors of $3-7\ \rm{m\>s^{-1}}$ and negligible RV difference pre- and post-earthquake for stars that emerged in OPSP. 

\begin{figure*}
\begin{center}
\includegraphics[scale=0.5]{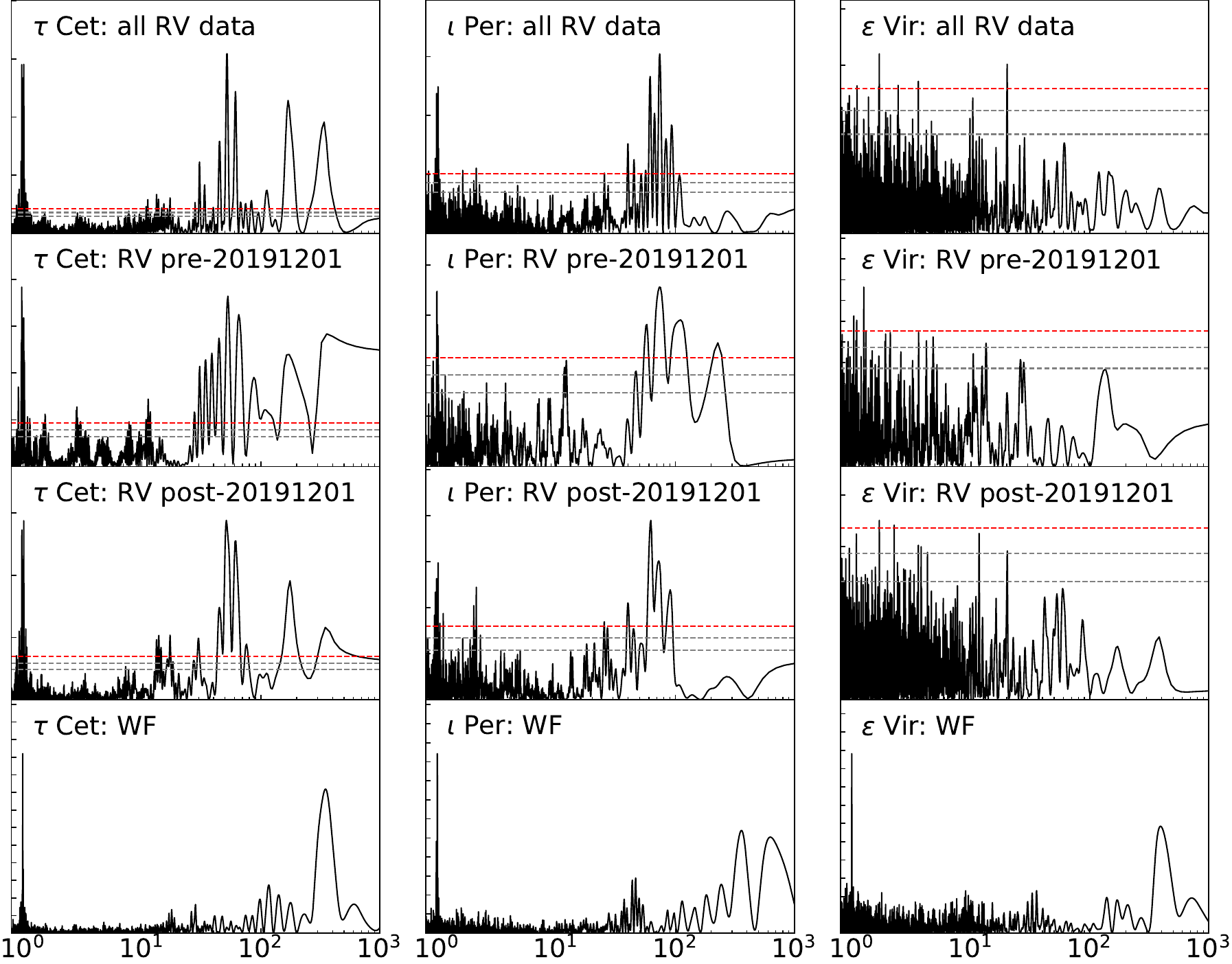}
\end{center}
\caption{
The GLS periodograms of RVs of $\tau$ Cet (left panel), $\iota$ Per (mid panel), and $\epsilon$ Vir 
(right panel) obtained by HIDES-F2.
In each panel, we present the GLS periodograms of each star derived from different data sets and the window function.
In each subplot, we indicate FAP levels of 10\%, 1\%, and 0.1\% with the horizontal lines from bottom to top.
}\label{fig:F2_gls}
\end{figure*}

To search for instrumental variations, we calculated GLS periodogram of the full RVs of the three RV standard stars. 
As shown in Figure \ref{fig:F2_gls}, we could discover a significant variation of approximately two months from RVs of both $\tau$ Cet and $\iota$ Per with estimated amplitudes of $\lesssim 3 \rm{m\ s^{-1}}$.
However, such a variation could not be derived from $\epsilon$ Vir RVs and some previous studies on $\tau$ Cet, e.g., the planet discovery of $\tau$ Cet with HARPS and Keck \citep{Feng2017}.
Therefore, we infer that the two-month variation with low amplitude might be introduced from the instrument itself.

\begin{figure*}
\begin{center}
\includegraphics[scale=0.55]{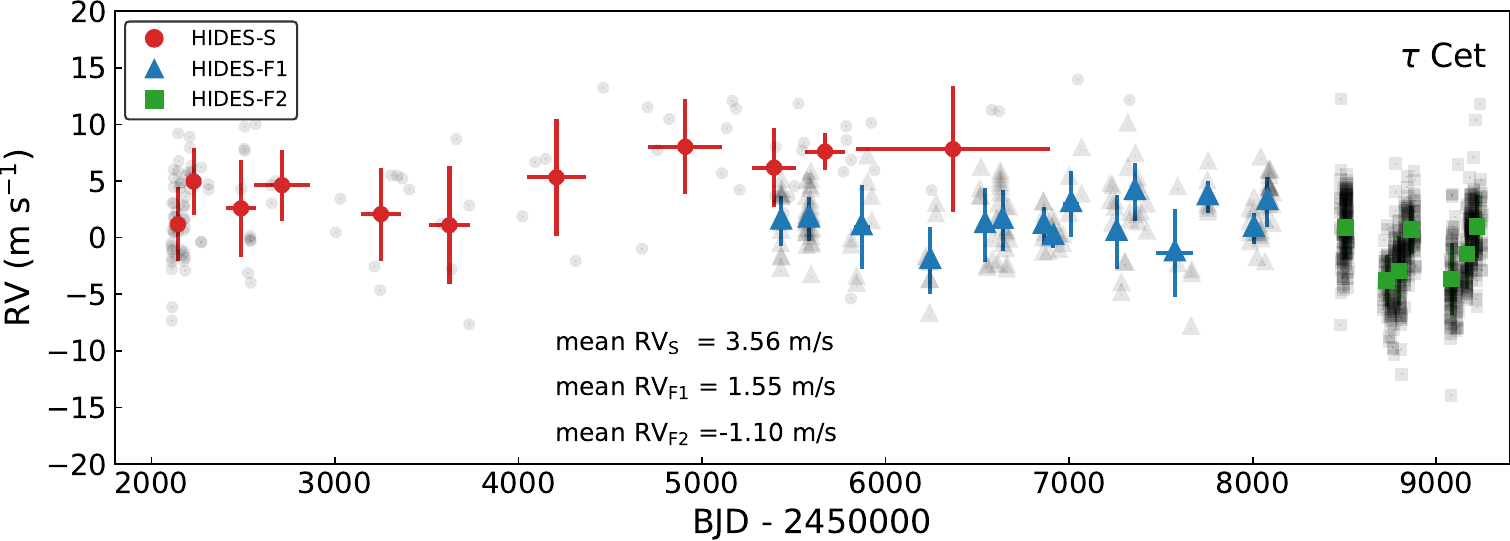}
\end{center}
\caption{
The RVs of $\tau$ Cet measured by a unified reference spectrum. 
Single measurements are marked by non-colored markers, and binned measurements are marked by colored markers.
}\label{fig:HD10700}
\end{figure*}

As for the stability of HIDES over 20 years, it was diagnosed by RVs of $\tau$ Cet for its mild RV variation and long-term stability.
We re-extracted the RVs obtained by HIDES-S, -F1, and -F2 with the same reference spectrum, and we empirically eliminated the RVs with observational errors over $6\ \rm{m\ s^{-1}}$.
As shown in Figure \ref{fig:HD10700}, we could apparently observe an offset between HIDES-S and (-F1 and -F2) at a maximum of $\sim 5\ \rm{m\ s^{-1}}$, and an offset between HIDES-F1 and -F2 of $\sim 2.5\ \rm{m\ s^{-1}}$.
In most cases, we could ignore the instrumental RV offsets by setting them as free parameters in the Keplerian orbital fit.
In this study, it was not suggested to simply ignore the RV offsets for stars showing astrometric accelerations, because we could trace astrometric accelerations with long-term consecutive RV monitoring based on a unified zero-point.

In the case of HIDES, however, the RV standard stars were sparsely observed by HIDES-S and -F1. 
It was thus never a good strategy to fit and remove the trends with lines or curvatures because of the overcorrection to the zero-point.
Instead, we could adopt a more moderate correction by ignoring the slight instrumental RV variations inside individual observation modes but shifting the RV offsets between individual modes based on the mean level.
It had two primary benefits in this study:
\begin{itemize}
    \item It did not influence the RV orbit fitting in Section \ref{sec:orbfit} since, as mentioned before, the RV offsets to an arbitrary zero-point in the Keplerian model were set to be free parameters.
    \item It moderately flattened the RV offsets between different observation modes so that the RV variations of the revisited stars can be well described in the longest time scale of 20 years. 
\end{itemize}
This moderate correction was applied to four systems (HD 5608, HD 14067, HD 120084, and HD 175679; see Section \ref{sec:astrometry}) when we performed the joint constraint with RV and astrometry.

\section{RV fitting results for normal stars}
\begin{figure*}
\begin{center}
\includegraphics[scale=0.25]{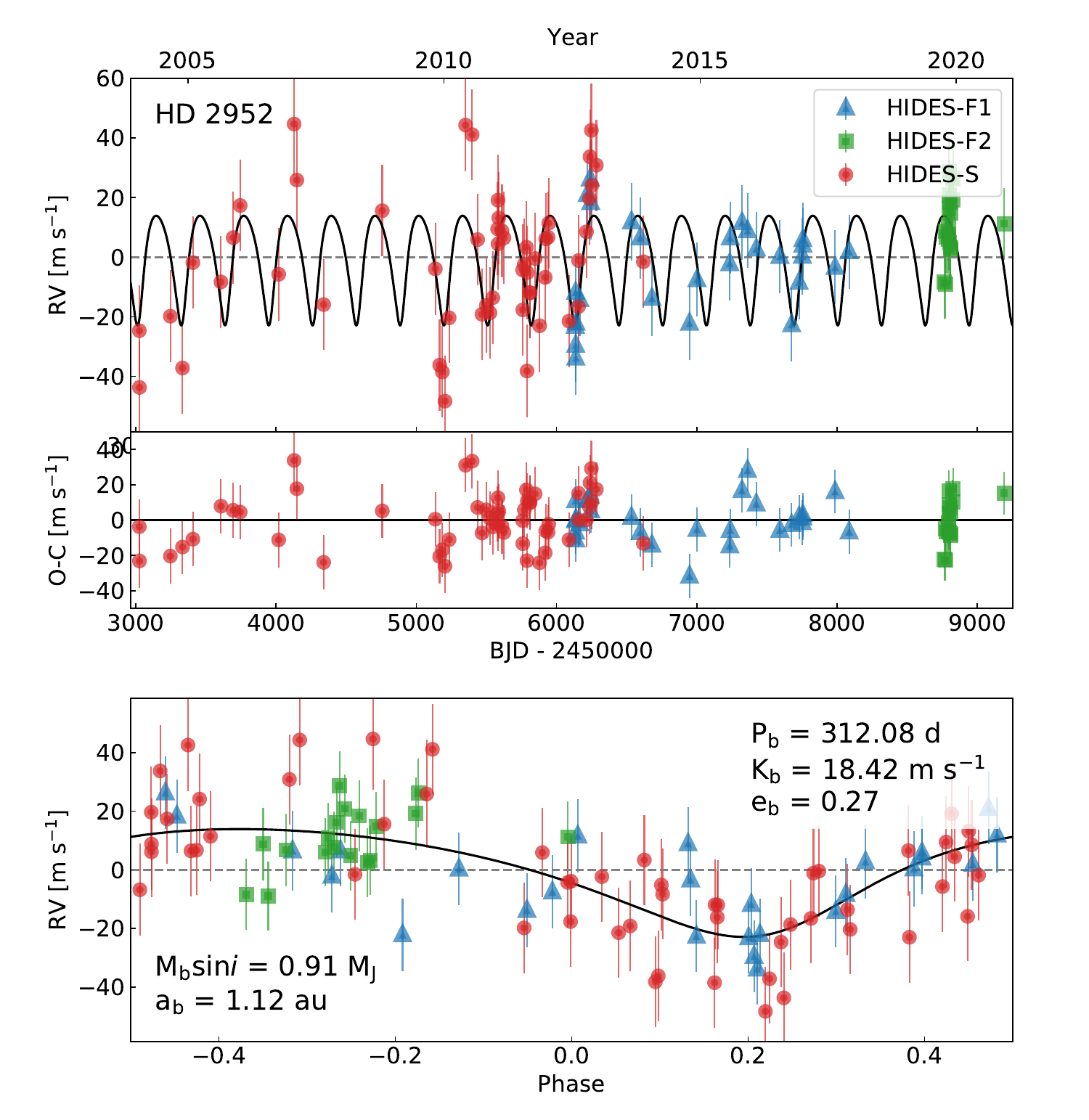}
\includegraphics[scale=0.25]{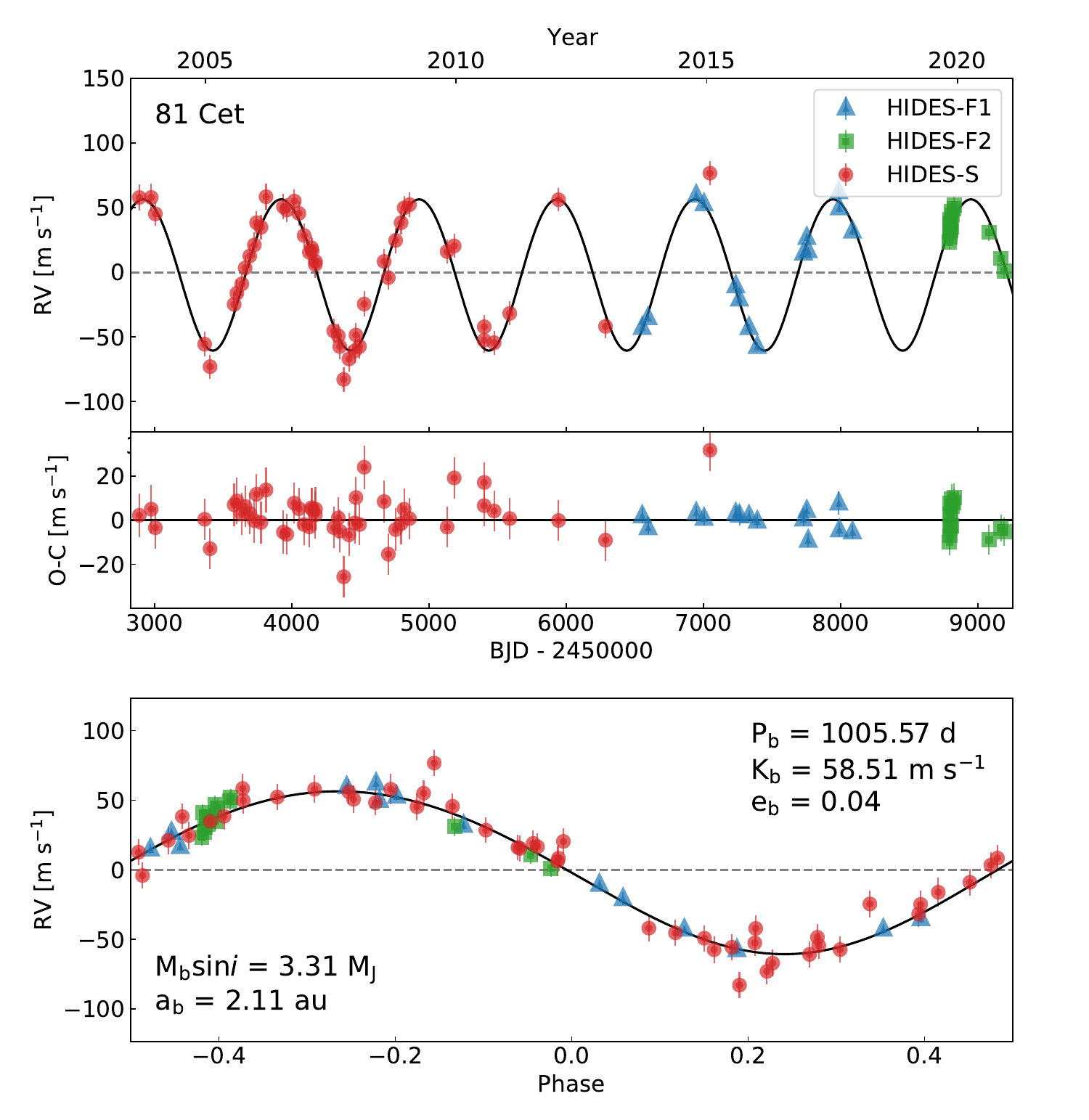}
\includegraphics[scale=0.25]{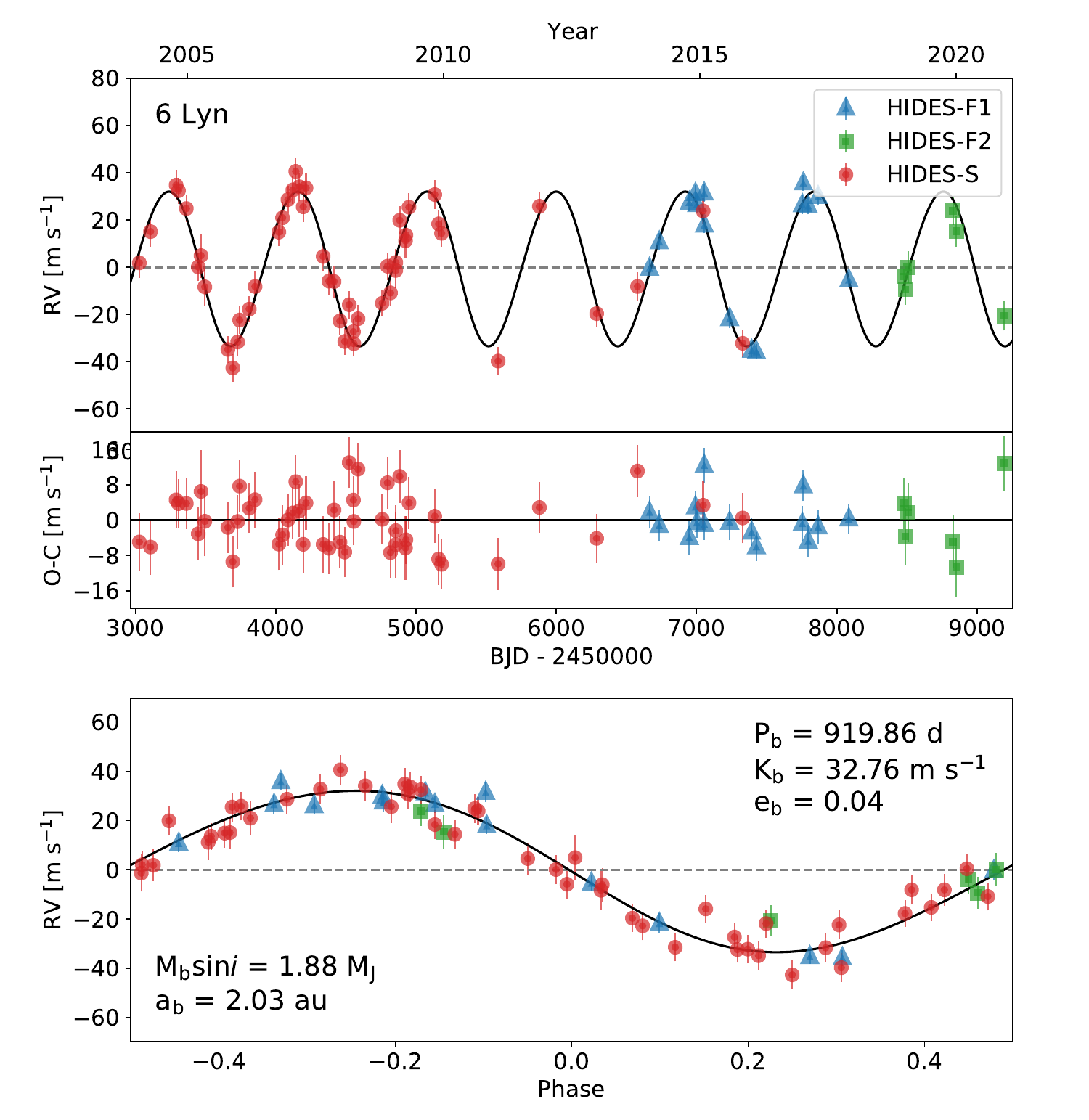}
\includegraphics[scale=0.25]{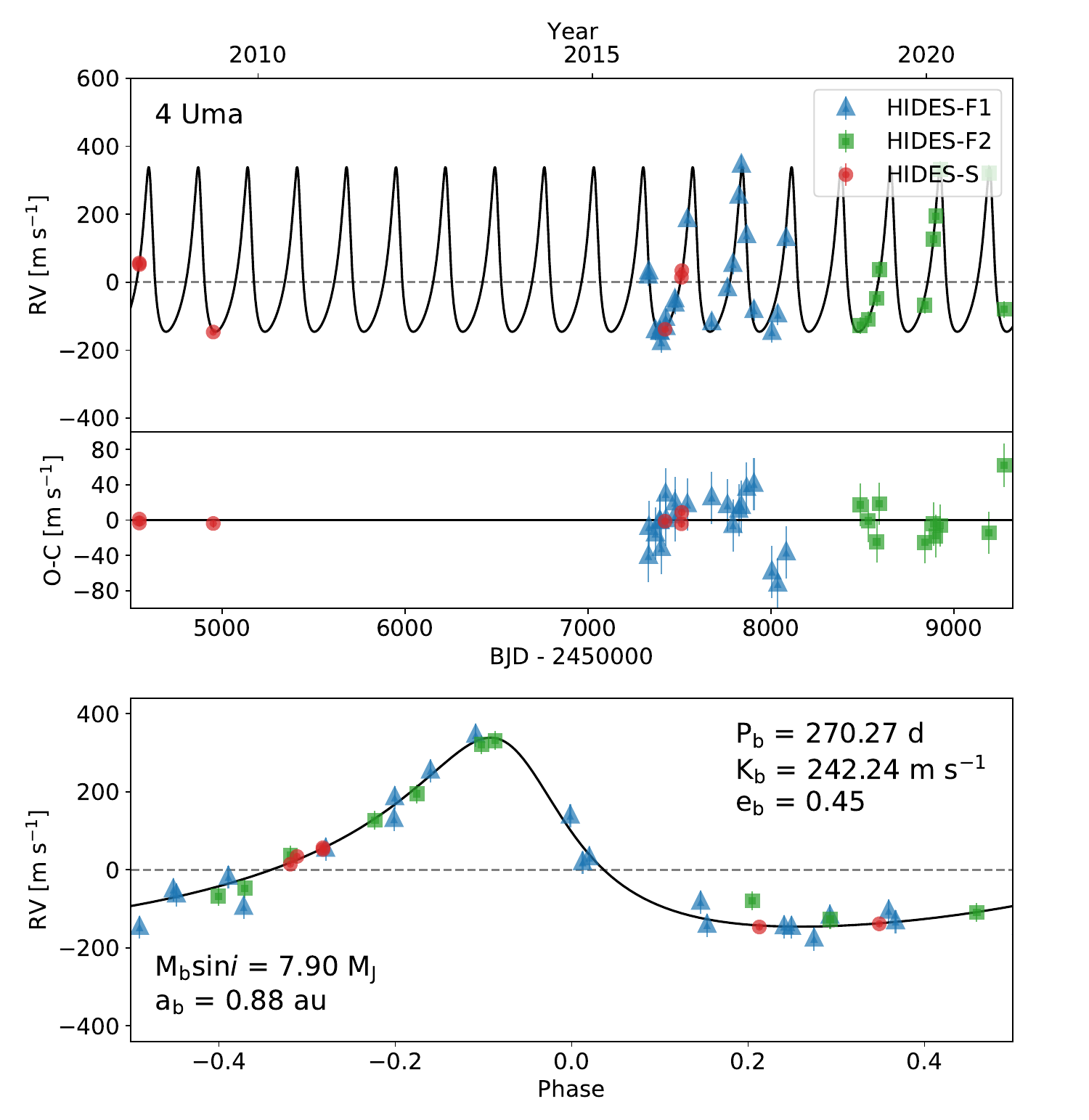}
\end{center}
\caption{
Keplerian orbital fit to HD 2952, HD 4732, HD 16400, 6 Lyn, and 4 Uma.
}\label{fig:HD2952_phase}
\end{figure*}

\begin{figure*}
\begin{center}
\includegraphics[scale=0.25]{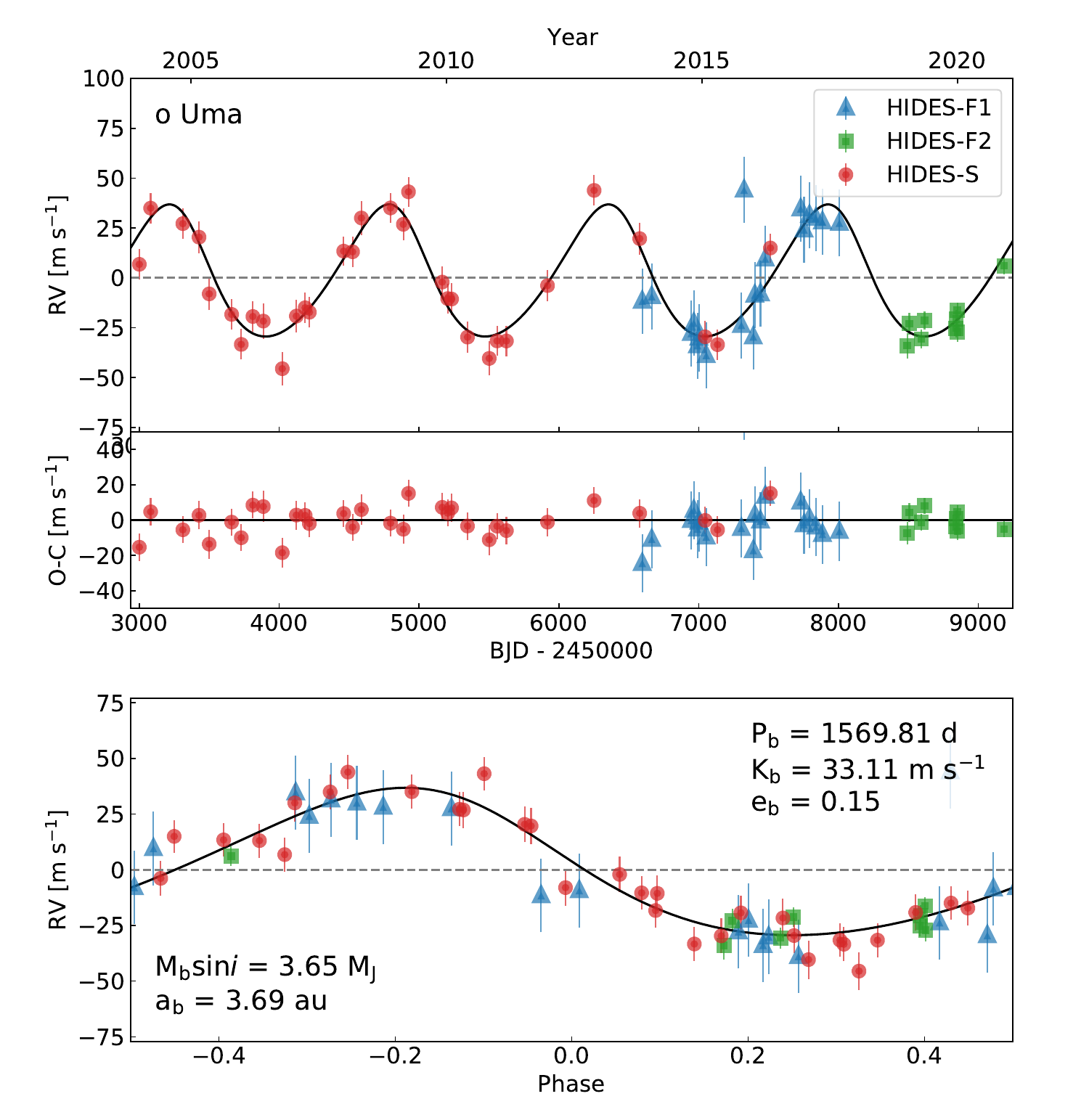}
\includegraphics[scale=0.25]{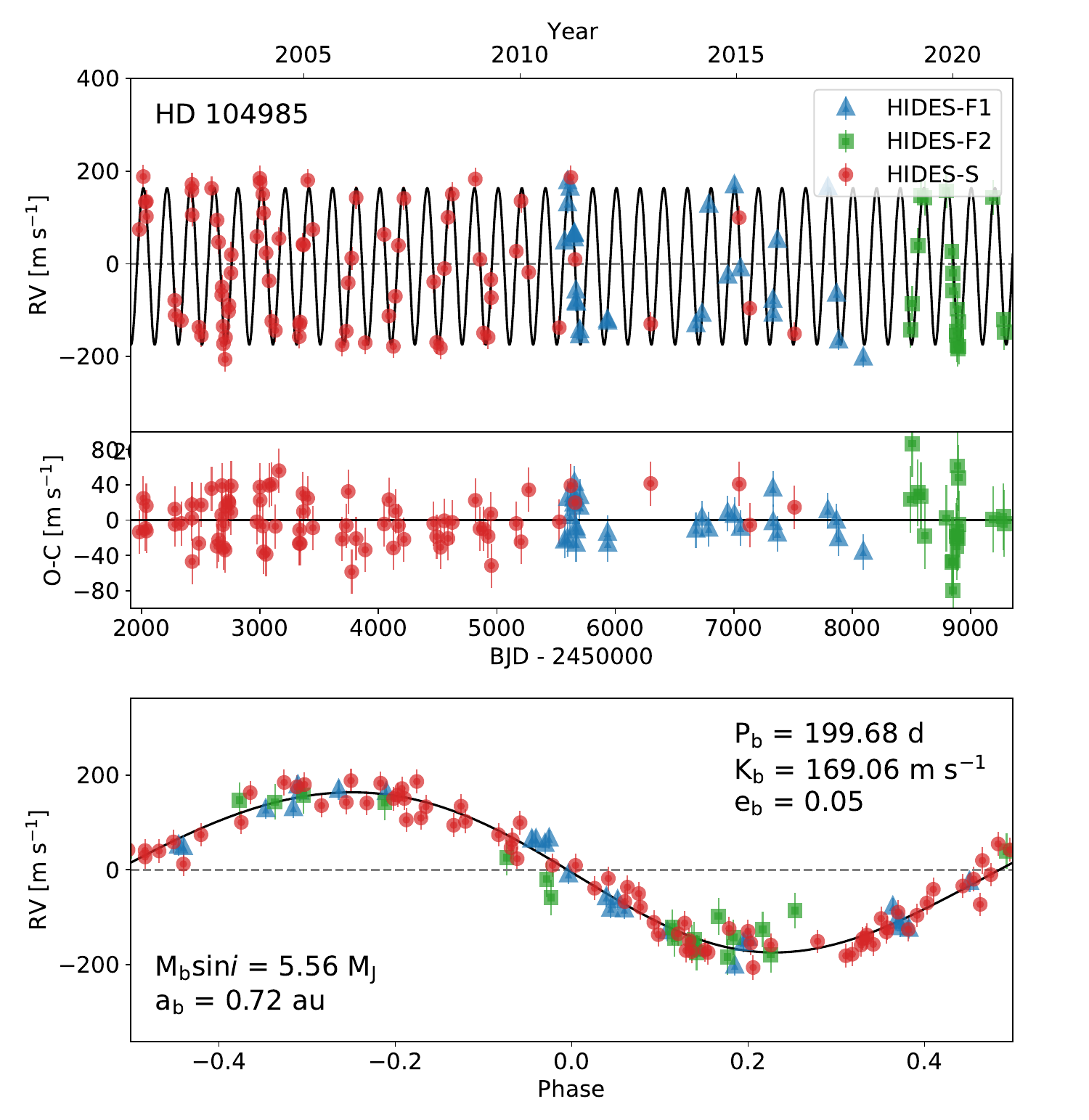}
\includegraphics[scale=0.25]{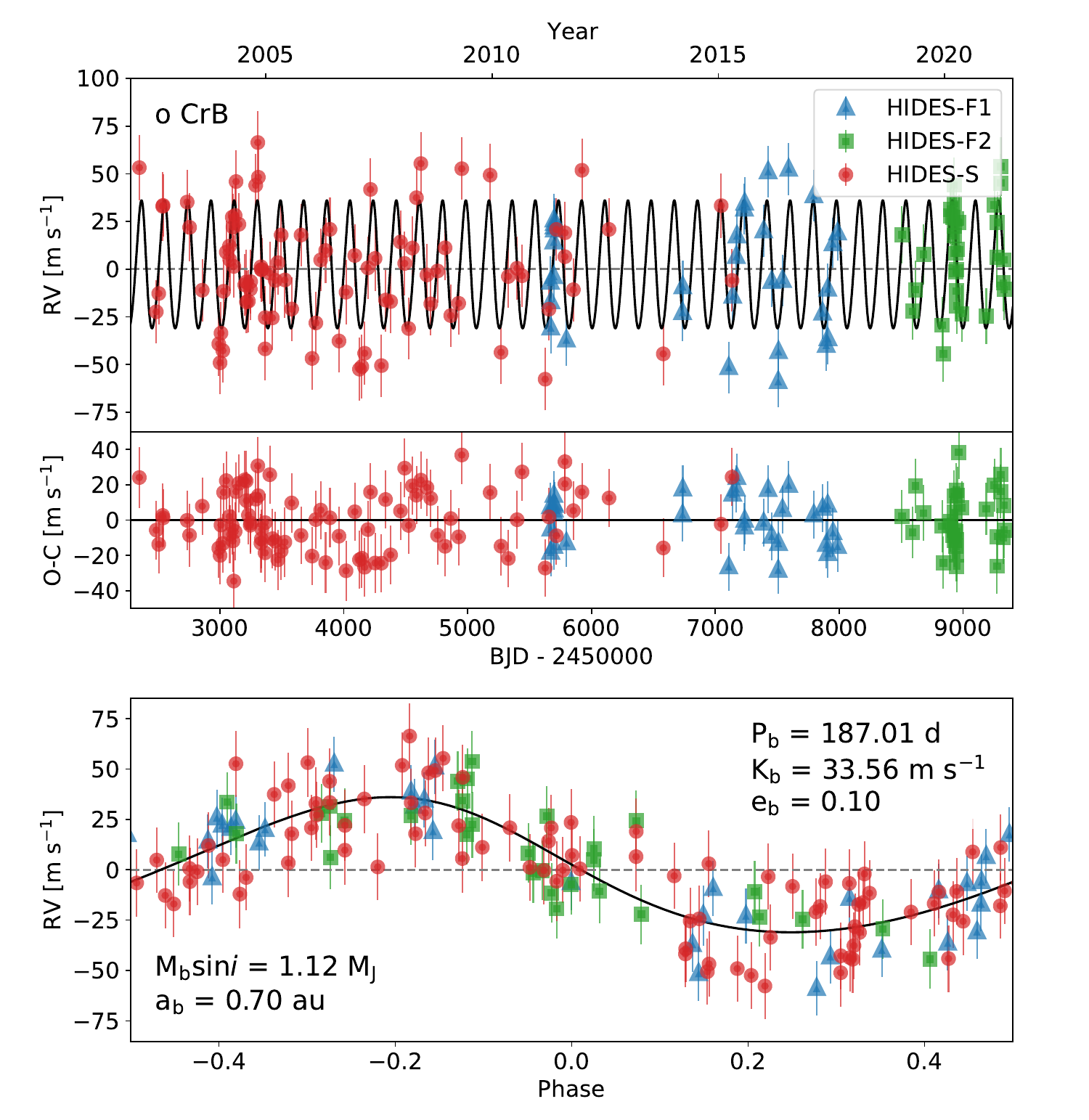}
\includegraphics[scale=0.25]{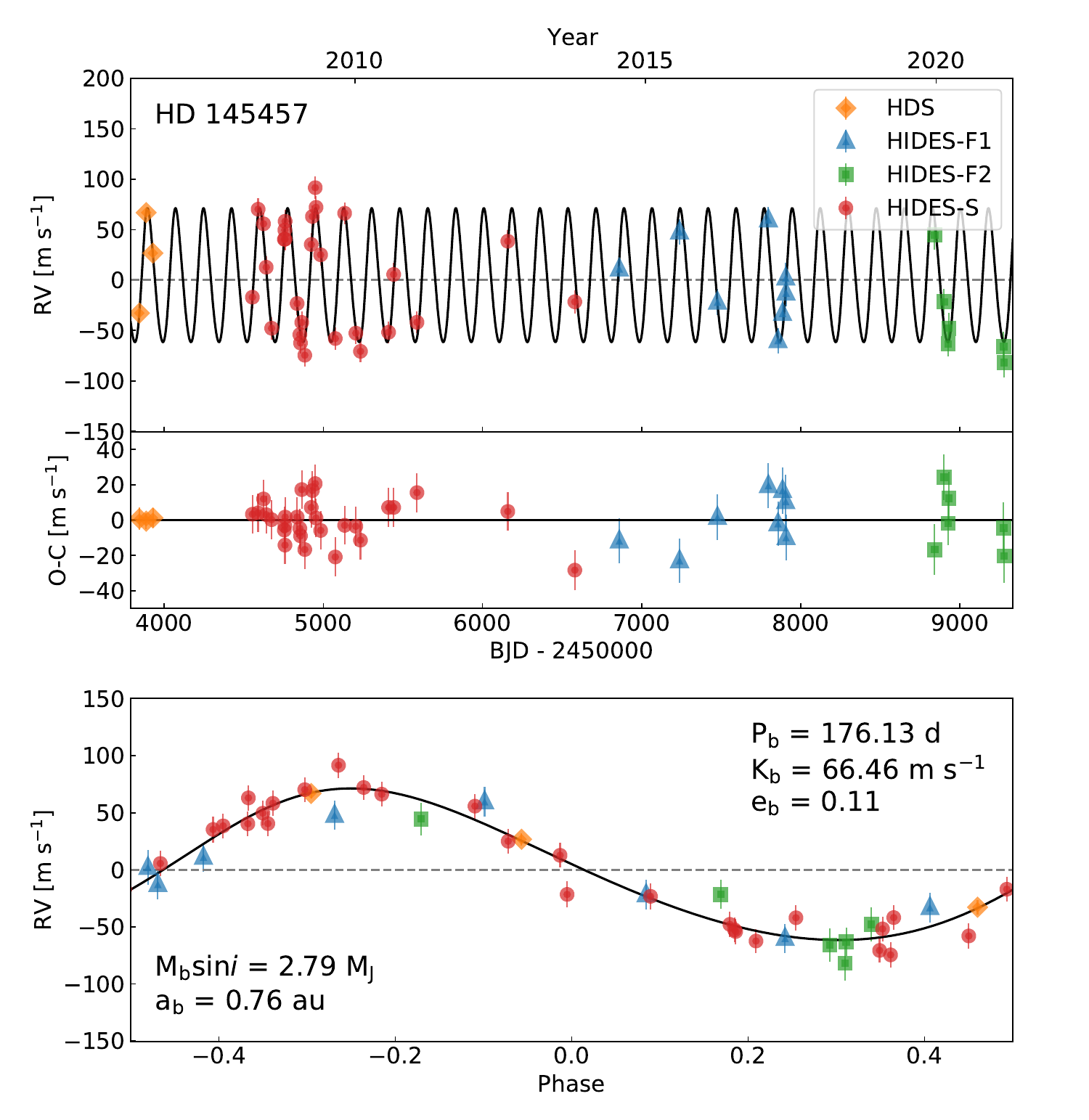}
\end{center}
\caption{
Keplerian orbital fit to o Uma, HD 104985, o CrB, and HD 145457. 
}\label{fig:HD104985_phase}
\end{figure*}

\begin{figure*}
\begin{center}
\includegraphics[scale=0.25]{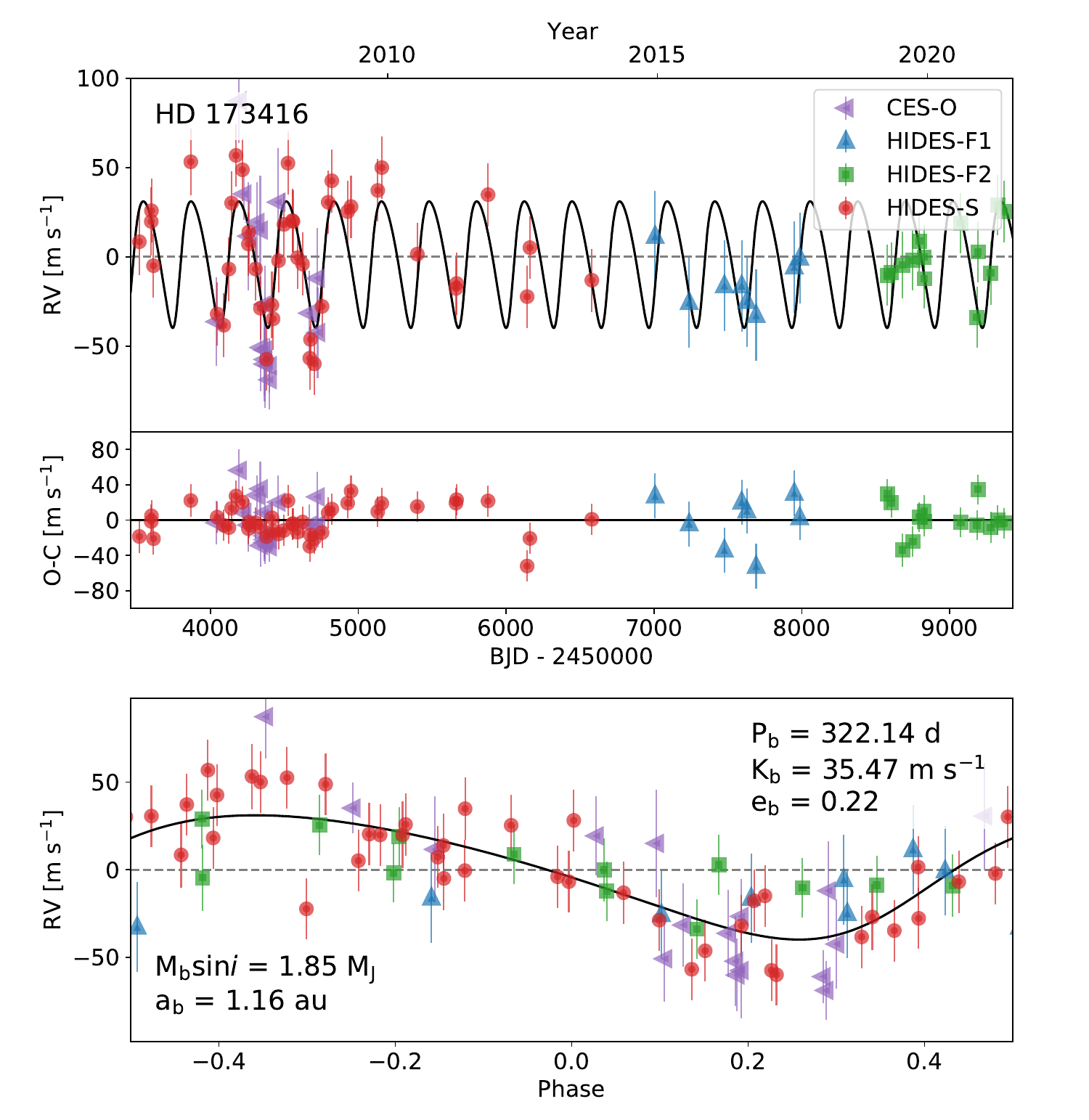}
\includegraphics[scale=0.25]{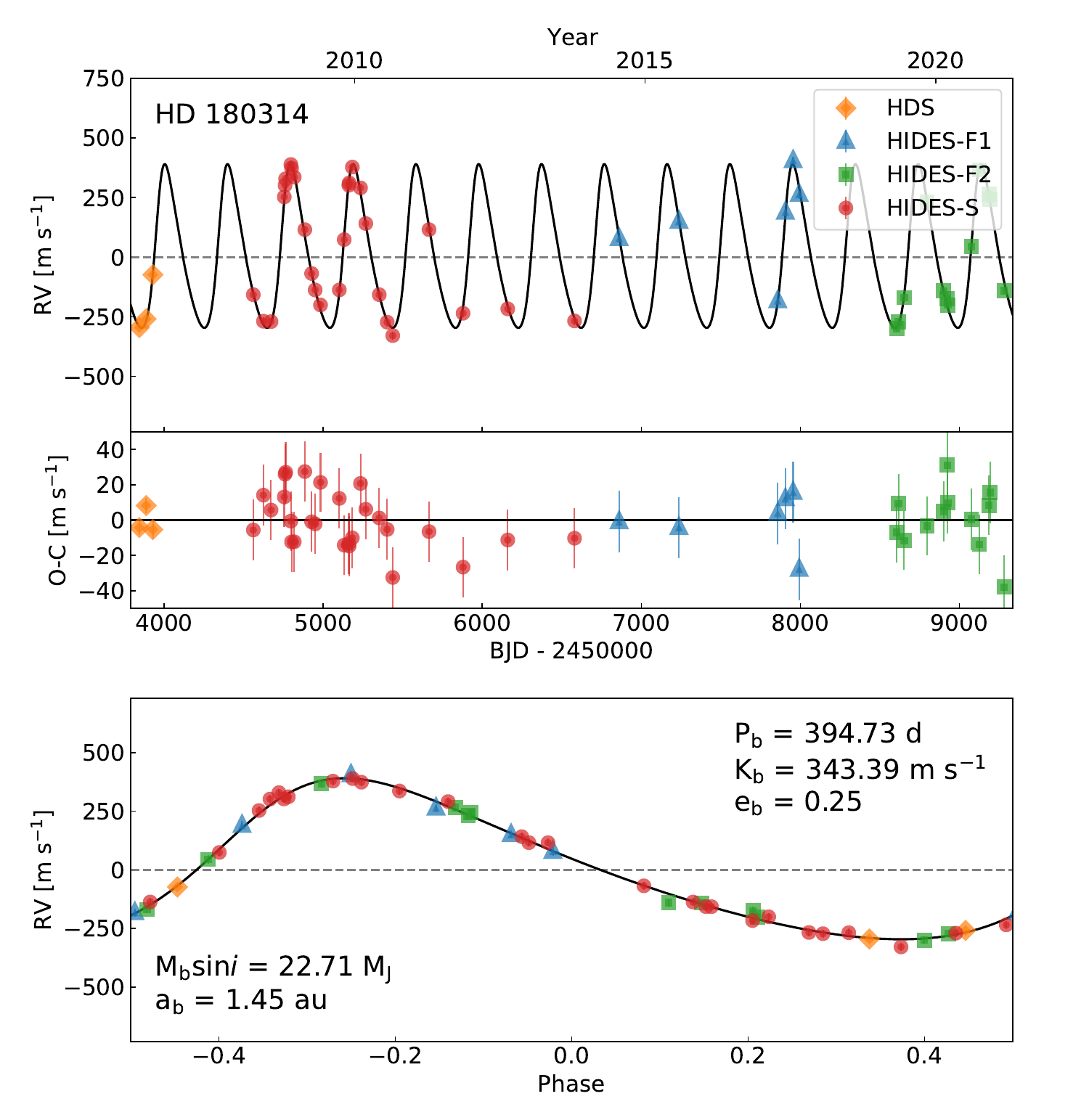}
\includegraphics[scale=0.25]{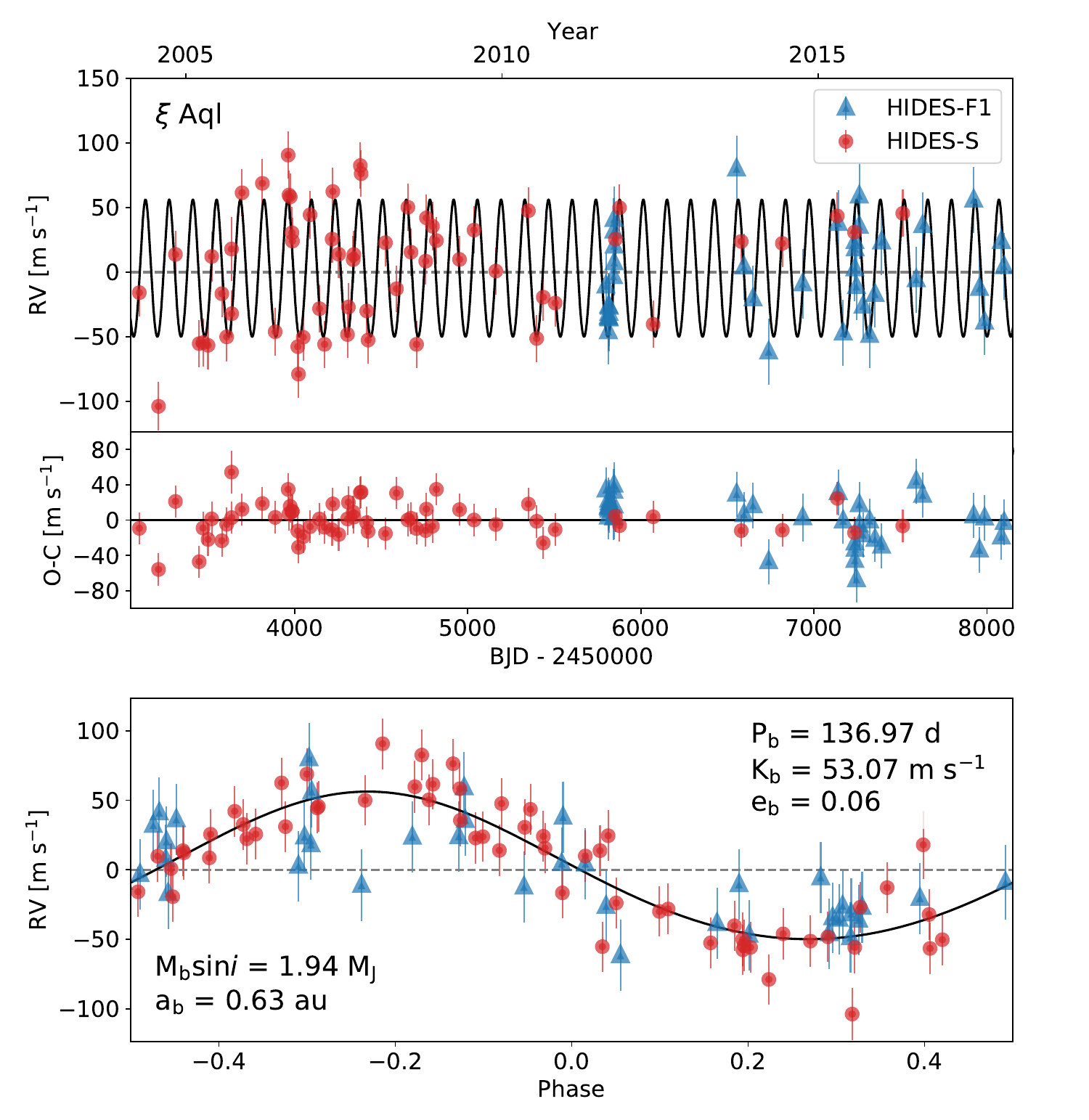}
\includegraphics[scale=0.25]{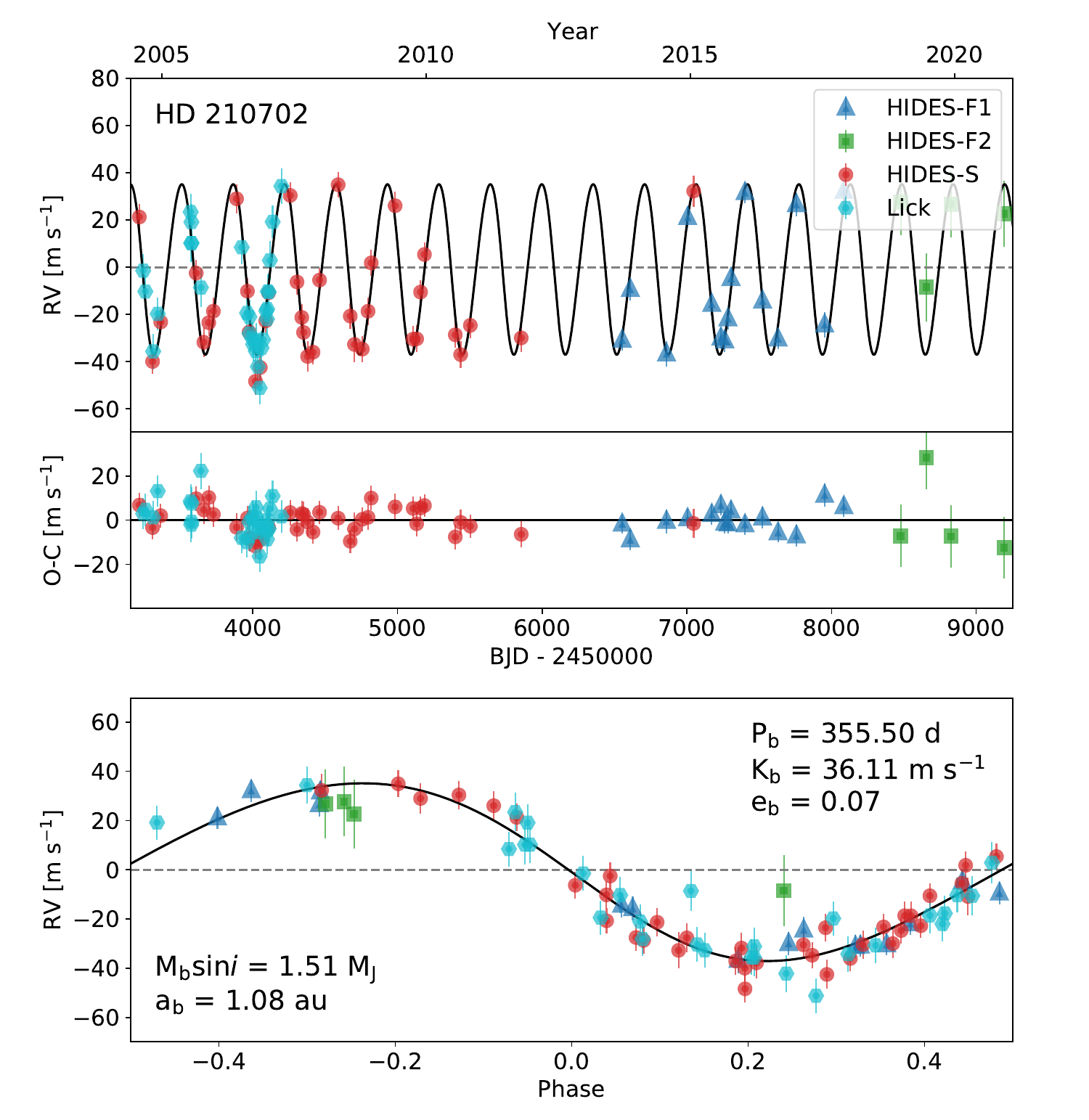}
\end{center}
\caption{
Keplerian orbital fit to HD 173416, HD 180314, $\xi$ Aql, and HD 210702.
}\label{fig:HD173416_phase}
\end{figure*}

\begin{figure*}
\begin{center}
\includegraphics[scale=0.25]{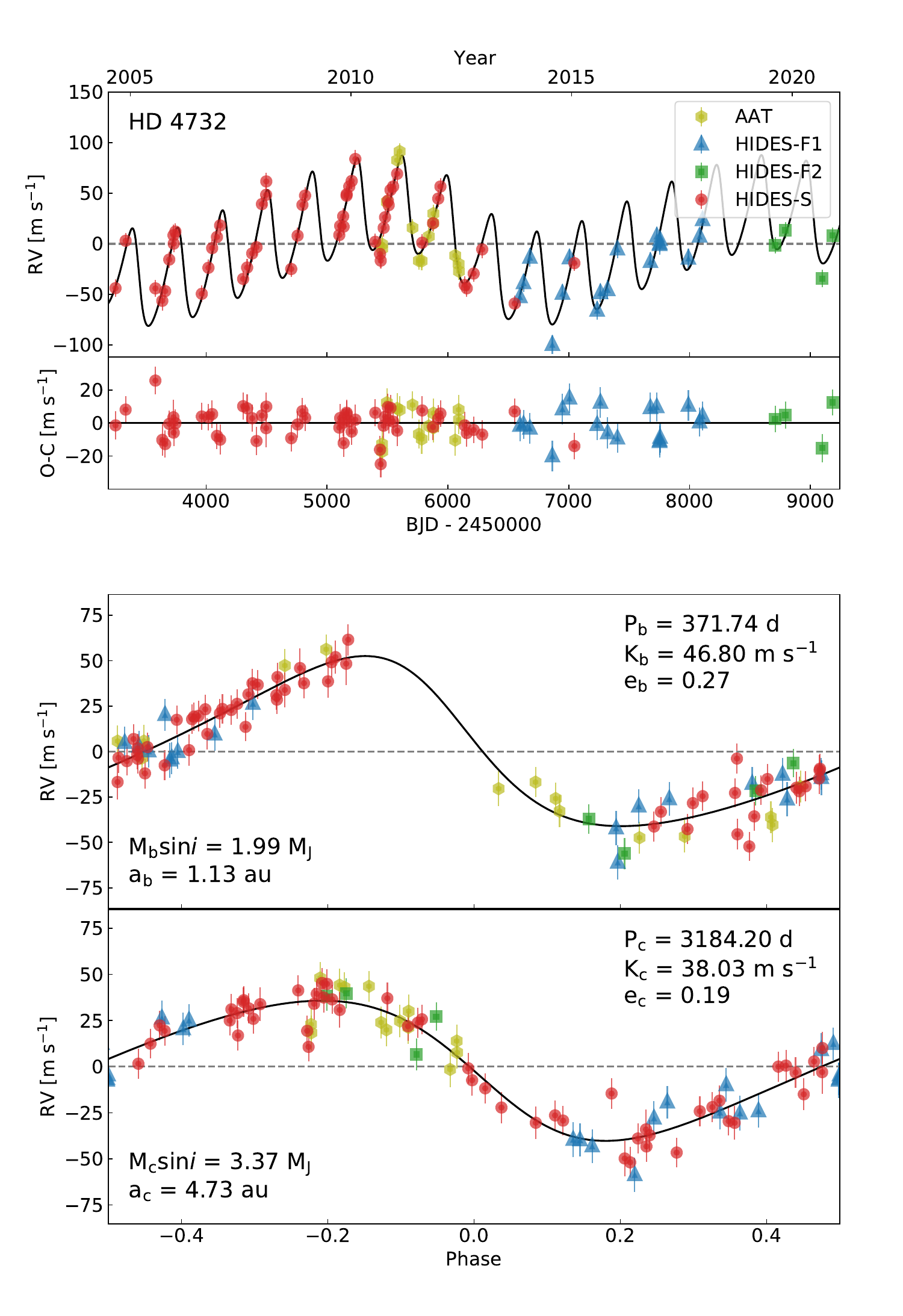}
\includegraphics[scale=0.25]{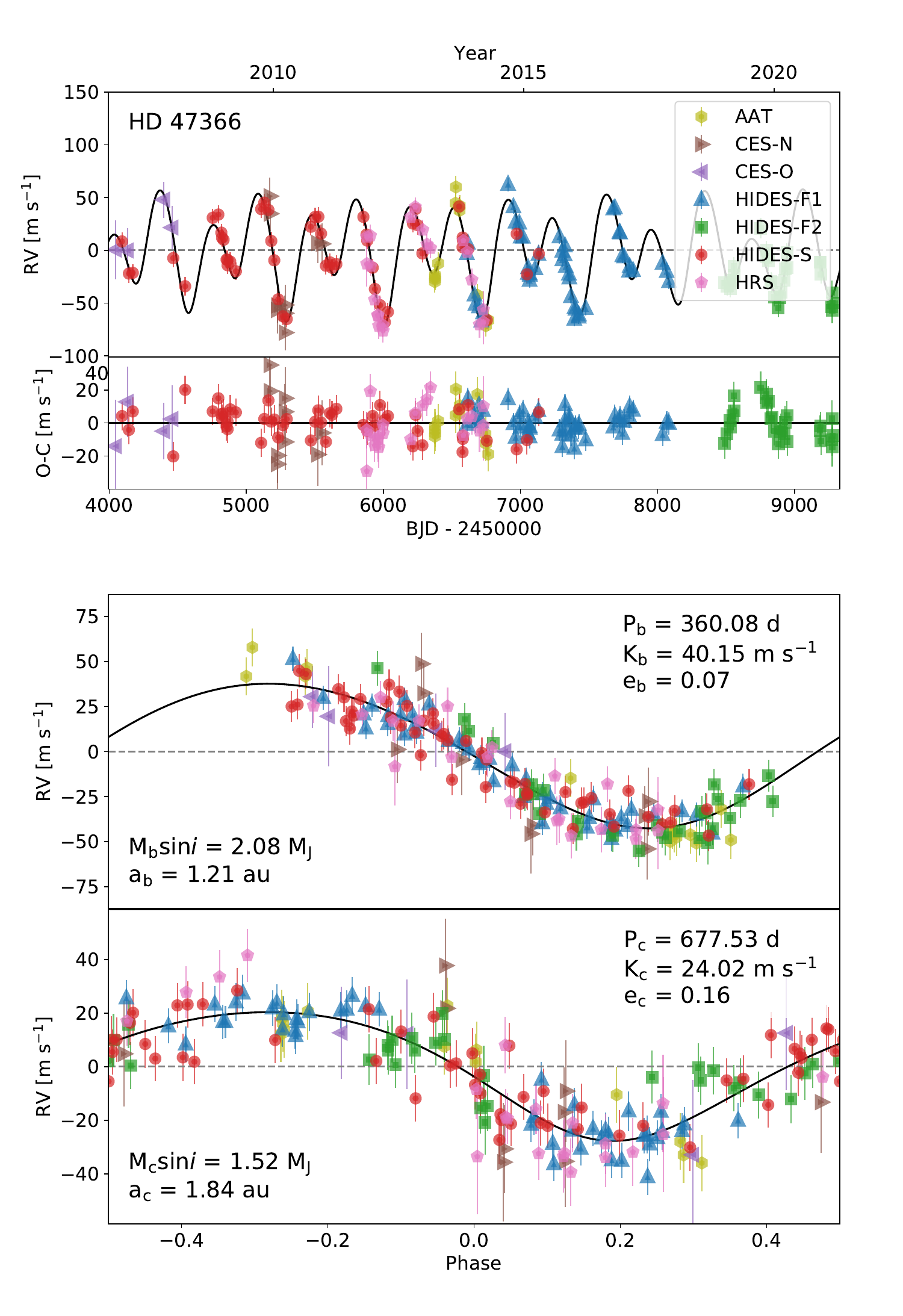}
\includegraphics[scale=0.25]{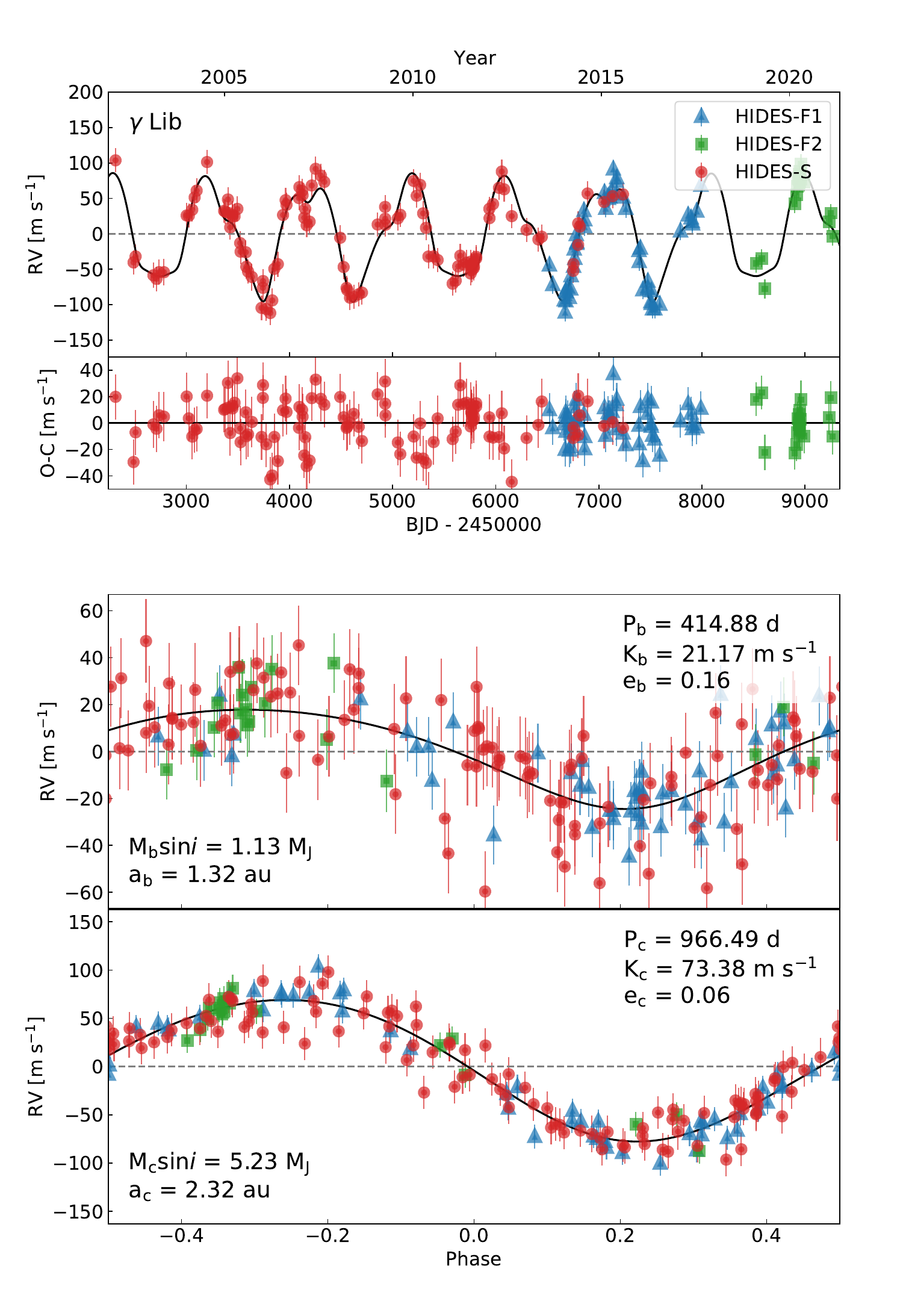}
\includegraphics[scale=0.25]{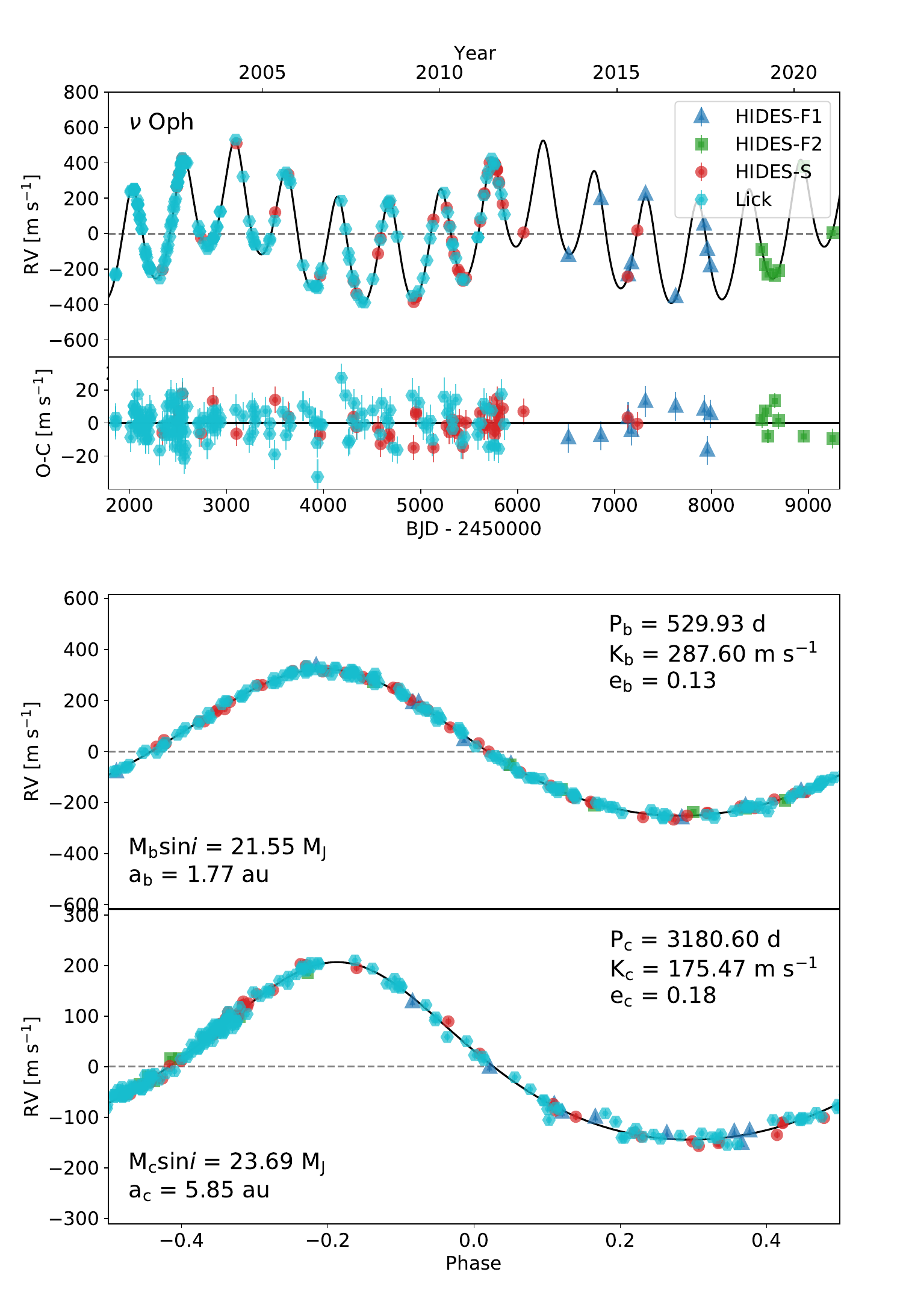}
\end{center}
\caption{
Keplerian orbital fit to HD 4732, HD 47366, $\gamma$ Lib, and $\nu$ Oph.
}\label{fig:HD4732_phase}
\end{figure*}

\begin{figure*}
\begin{center}
\includegraphics[scale=0.25]{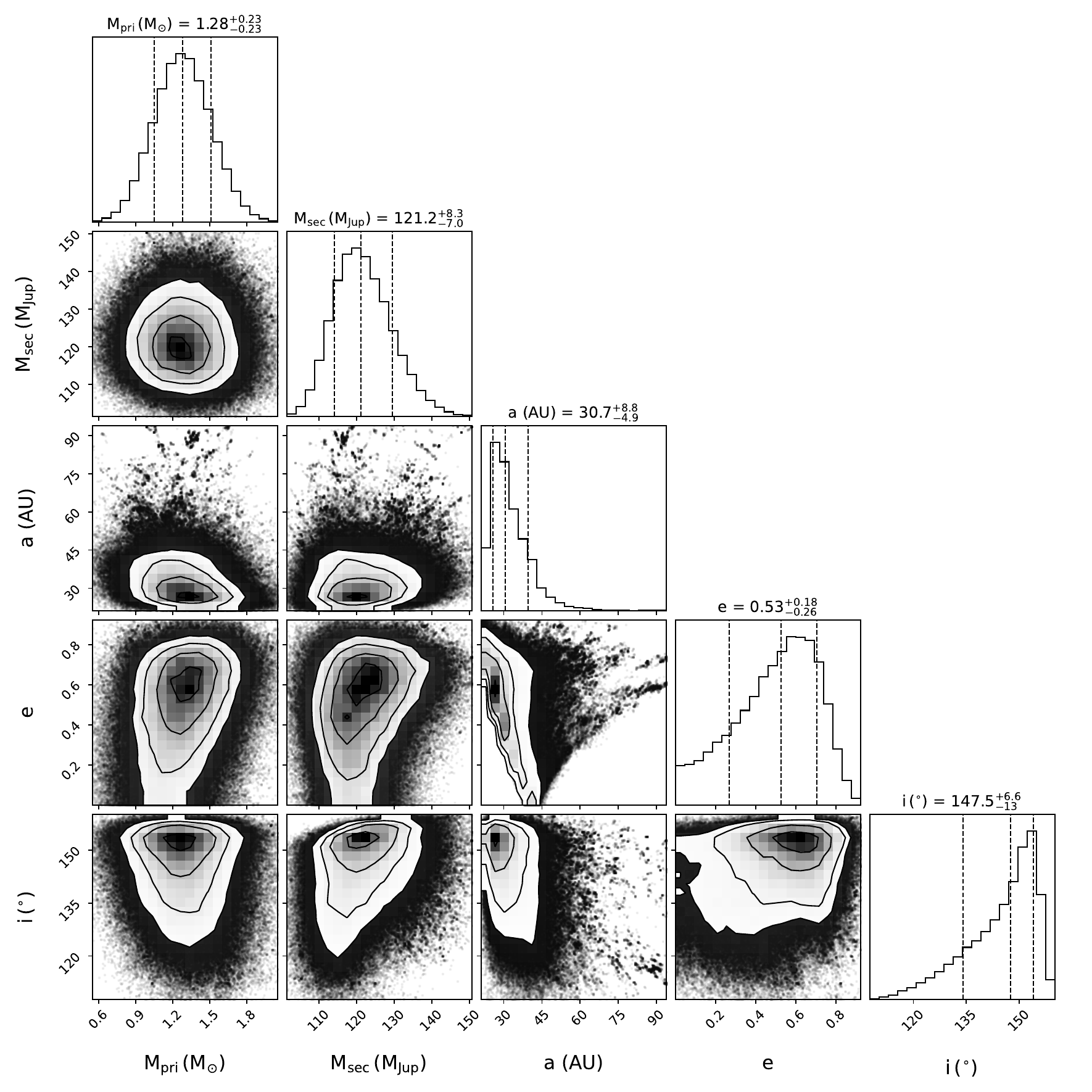}
\includegraphics[scale=0.25]{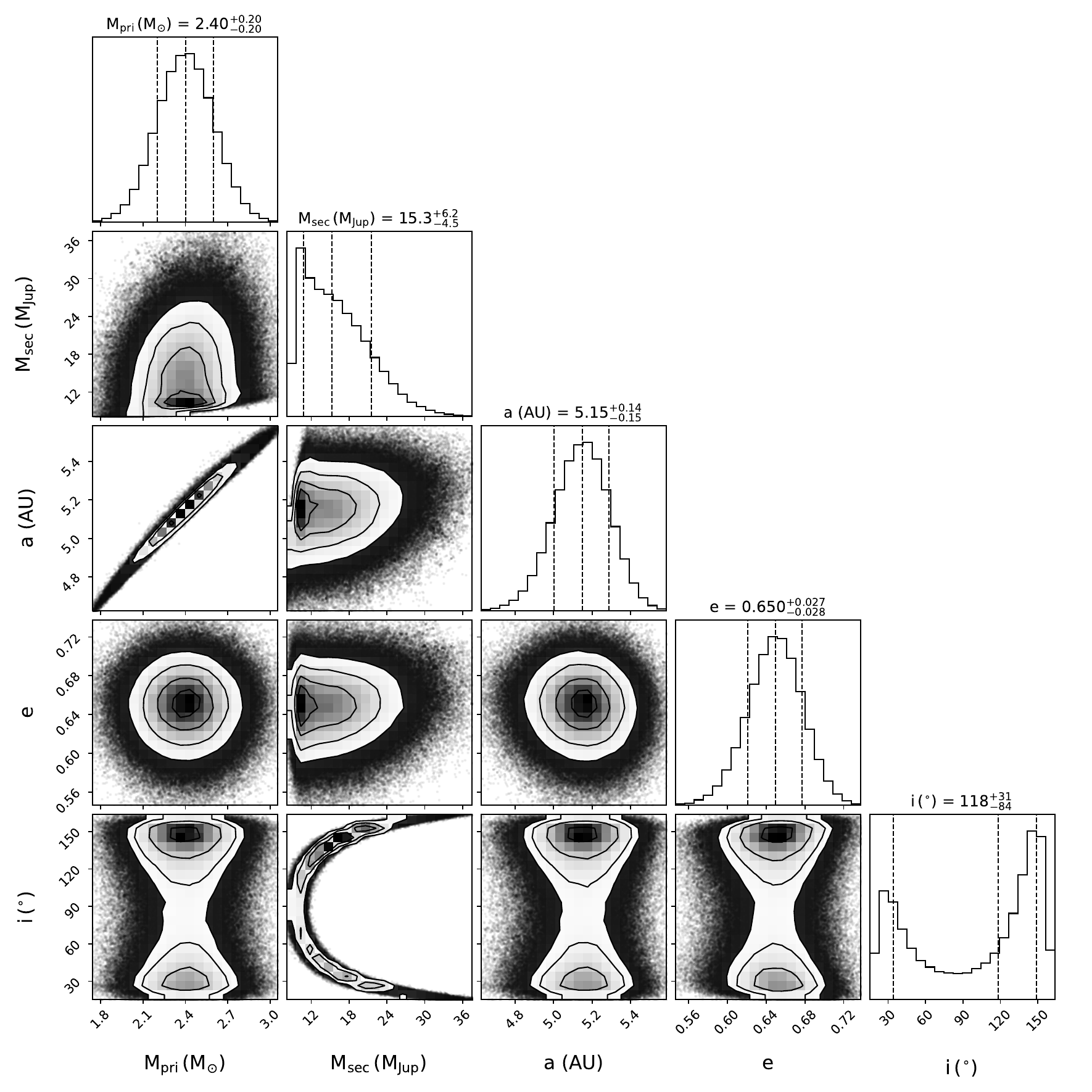}
\includegraphics[scale=0.25]{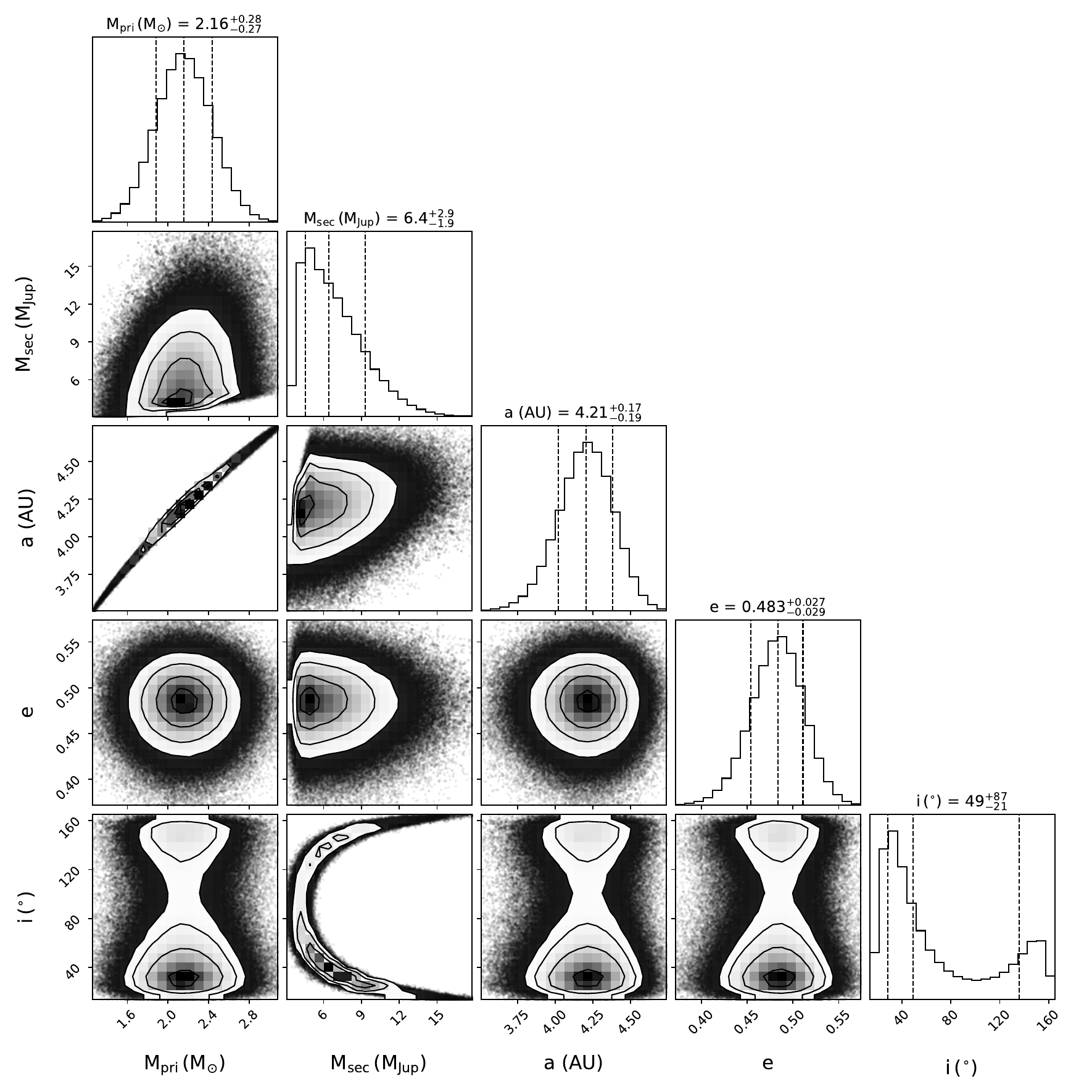}
\includegraphics[scale=0.25]{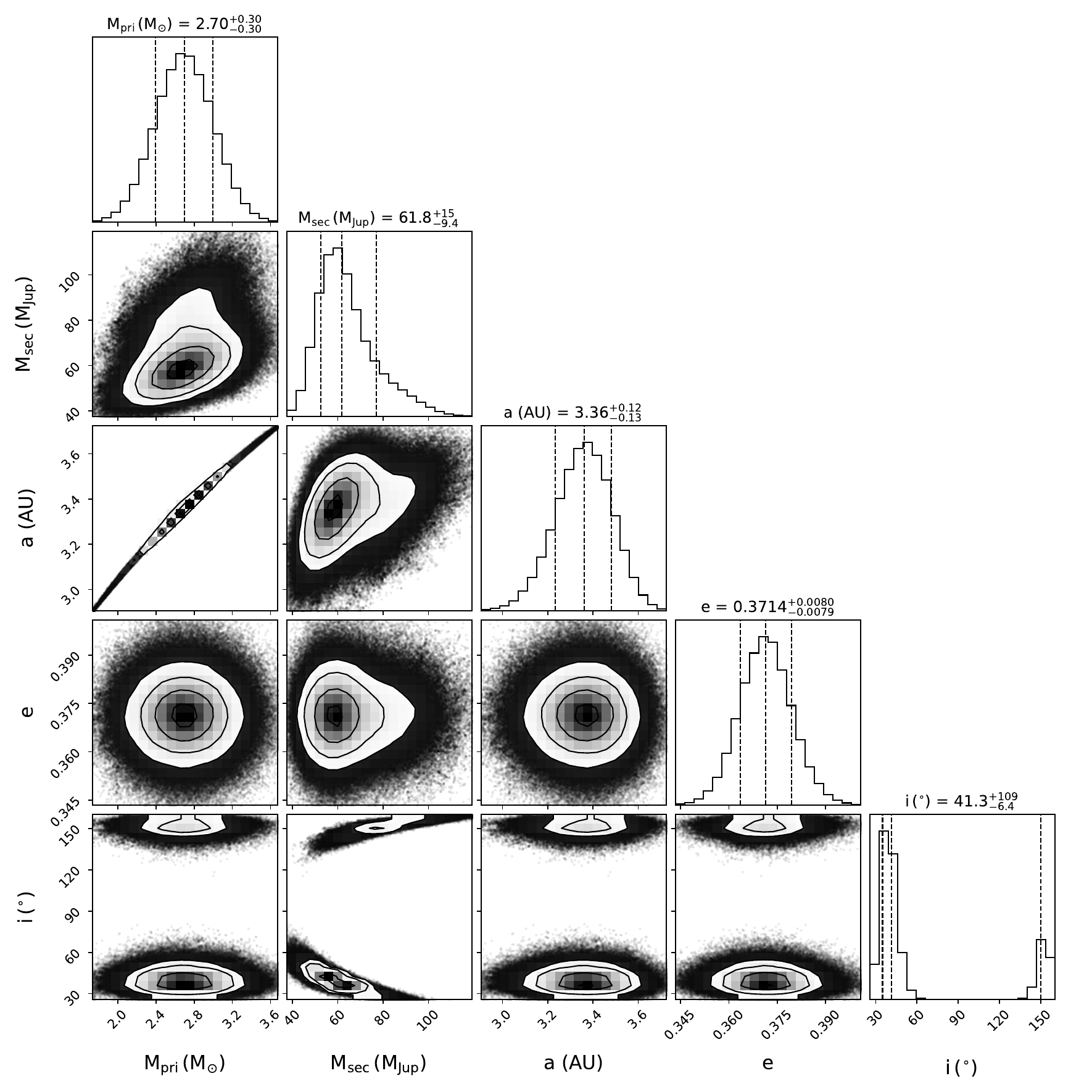}
\end{center}
\caption{
\texttt{orvara} corner plot of HD 5608, HD 14067, HD 120084, and HD 175679. 
}\label{fig:HD5608_orvara_corner}
\end{figure*}

In this section, we present the Keplerian orbital solutions to the normal stars, which are not concretely described in the main text (Section \ref{sec:res}, \ref{sec:interest} and \ref{sec:astrometry}).
In Figure \ref{fig:HD2952_phase}, \ref{fig:HD104985_phase}, and \ref{fig:HD173416_phase}, we present the orbits of single-planet systems, including HD 2952, HD 4732, HD 16400, 6 Lyn, 4 Uma, o Uma, HD 104985, o CrB, HD 145457, HD 173416, HD 180314, $\xi$ Aql, and HD 210702.
In Figure \ref{fig:HD4732_phase}, we present the orbits of double-planet systems, including HD 4732, HD 47366, $\gamma$ Lib, and $\nu$ Oph.
Finally in Figure \ref{fig:HD5608_orvara_corner}, we present the corner plots of HD 5608, HD 14067, HD 120084, and HD 175679 from \texttt{orvara} orbit fitting.

\end{document}